# Covariate influence in spatially autocorrelated occupancy and abundance data

David C Bardos[1*], Gurutzeta Guillera-Arroita[1] and Brendan A Wintle[1]

[1]*ARC Centre of Excellence for Environmental Decisions, School of Botany, The University of Melbourne, Parkville, Victoria 3010, Australia*

**Abstract.** The autologistic model and related auto-models, commonly applied as autocovariate regression, offer distinct advantages for analysing spatially autocorrelated species distribution or abundance data, allowing simple and direct modelling of the dependence of nearby observations. However, Carl and Kühn (Ecological Modelling, 2007, **207**, 159) and Dormann (Ecological Modelling, 2007, **207**, 234) questioned their validity: the former analysing simulated and empirical abundance data, the latter presenting theoretical arguments and analysis of simulated data. Further simulation studies were carried out by Dormann et al. (Ecography, 2007, **30**, 609) and Beale et al. (Ecology Letters, 2010, **13**, 246). These studies reached negative conclusions concerning autocovariate regression, based on numerical evidence of 'bias' comprising examples where autocovariate regression yielded much smaller covariate parameter magnitudes than associated linear regressions. In all but the first study, this numerical evidence is erroneous due to use of invalid neighbourhood weighting schemes and, in the auto-Poisson case, invalid application to cooperative interactions. Here we show that even when these technical errors are corrected, a more fundamental conceptual error remains: all four studies are founded on a mathematically incorrect notion of bias, involving direct comparison of parameter estimates across models differing in mathematical structure. We develop a set of simulation-based measures of covariate influence that are directly comparable across models. We apply these to examples from the abovementioned studies where covariate parameter estimates are much smaller for auto-models than for the associated linear models. We find that in these cases, the effect of smaller auto-model parameters is similar to (and consistent with) the corresponding linear model effects, due to a phenomenon within auto-models that we refer to as 'covariate amplification'. Thus, simple comparison of parameter magnitudes between structurally different models can be highly misleading. We demonstrate that the recent critique of auto-models is entirely unfounded. Correctly applied and interpreted, autocovariate regression provides a practical approach to inference for spatially autocorrelated species distribution or abundance data, while overcoming well-known limitations of generalized linear models.

Key words: abundance, autologistic, autocovariate, covariates, occupancy, species distribution model, SDM, spatial autocorrelation

*Corresponding author. E-mail: dcbardos@unimelb.edu.au

# Introduction

Species occupancy and abundance data usually exhibits strong spatial autocorrelation (SAC) due to some combination of (a) intrinsic processes causing dependence between nearby observations and (b) the influence of environmental covariates, usually incompletely known, that themselves have strong SAC. Consequently, analysis of such data usually requires going beyond the generalized linear model (GLM) framework. The class of auto-models, introduced by Besag (1974), is capable of modelling a wide variety of spatial phenomena and can be conveniently estimated using autocovariate regression, which is equivalent (Possolo 1986) to the pseudo-likelihood approximation (Besag 1975). Auto-models have been extensively used in ecology since Augustin, Mugglestone and Buckland (1996) applied an autologistic model to red deer census data.

However a series of papers has questioned the validity of auto-models, principally on the basis of examples where covariate parameter estimates obtained by autocovariate regression appear to be anomalous in the sense of being far smaller in magnitude than corresponding estimates in GLMs. Furthermore, in examples where the data is simulated, the anomalous estimates differ greatly from corresponding parameters in simulation models used to generate the data.

Carl and Kühn (2007) consider a range of simulated and empirical datasets; of particular interest is their analysis of a highly spatially autocorrelated presence-absence dataset for the plant species *Hydrocotyle vulgaris* in Germany, using altitude and temperature covariates that themselves exhibit strong SAC. They found that the autologistic model yielded covariate parameter estimates of much smaller magnitude than those obtained from a logistic model. This and similar results for simulated datasets led to their conclusion that the autologistic model performed very poorly in parameter estimation and could not be recommended as a method to account for spatial autocorrelation.

Dormann (2007) examines estimation of an autologistic model on artificial data intended to represent a strongly spatially autocorrelated population of 'snouter' on an island. The data is generated by deriving 'snouter' probability maps from a logistic model with a single highly spatially autocorrelated environmental covariate ("rain") and then applying distortions derived from spatially autocorrelated Gaussian noise. Estimation proceeds by including the autocovariate in a logistic model. Dormann *et al.* (2007) report on similar simulations and estimation procedures for the autologistic model, with the addition of simulated count and real-valued 'snouter' abundance data that are analysed using auto-Poisson and (uncentered) auto-normal models respectively. For each of these auto-models, Dormann (2007) and Dormann *et al.* (2007), hereafter 'Dormann et al.', found that parameter estimates for the "rain" covariate differed dramatically from the GLM parameters originally used to derive probability maps in the data generation procedure and from corresponding parameters in GLMs fitted to the data. These results led Dormann et al. to draw negative conclusions about auto-models.

Beale *et al.* (2010) compared parameter estimates across a range of models including linear and uncentered auto-normal models, using large numbers of simulated datasets in which both observations and covariates possess strong SAC. They described the overall performance of uncentered auto-normal models as "highly (downward) biased" and generally concurred with the assessments of Dormann et al.

We have shown elsewhere (Bardos, Guillera-Arroita & Wintle 2015) that a particular implementation error, namely the use of invalid neighbourhood weighting schemes, is widespread in the application of auto-models and that for the 'snouter' examples in Dormann et al., correcting implementation errors is sufficient to resolve the apparent anomalies in the sense that correct auto-model covariate parameter estimates turn out to be very close to covariate parameters driving the data-generating model and also similar to corresponding GLM estimates.

This similarity of auto-model, GLM and generating model parameters, whilst not an unusual occurrence, belies the complexity of auto-models: correct covariate parameter estimates for auto-models need *not* be similar to those for closely related GLMs. When simulated data with strong SAC is being fitted, correct covariate parameter estimates for auto-models can differ drastically from corresponding parameters used in a GLM as part of a simulation method for generating the data.

For a given dataset comprising observations and covariates, there is no *intrinsic parameter value* associated with a covariate $X$ that would be a correct estimate for *all* models fitted to the observations. Suppose we happen to have *exact* knowledge of all ecological processes giving rise to the data and can therefore specify the *exact* form of the probability distribution (i.e. the true model) from which the data was drawn, including all explanatory variables involved in the ecological processes and their associated true parameter values. Then if we fit the *exact* true model to the data, our parameter estimates are expected to approximate these true values to a degree determined by the size of the dataset and identifiability of the model.

In any realistic case, the true model structure and consequently the covariate set are unknown and any models we fit to the data are relatively crude structural approximations. The true parameter values for the exact model *need not* be correct estimates for parameters in approximate models. Disagreement between estimates in different models need not, therefore, imply mathematical error or faulty modelling of covariate influence on observations.

In this paper we re-examine apparent anomalies in auto-model estimation reported by Carl & Kuhn (2007) and Beale et al. (2010).

We begin with a brief discussion of the necessity for enforcing certain constraints on model parameters and neighbourhood weightings for correct implementation of auto-models. We summarise large estimation errors in the results of Beale *et al.* (2010) due to incorrect implementation.

When comparing structurally different models (even with identical covariate sets), simple comparisons of parameter magnitudes can be invalid. We therefore introduce three closely related measures of covariate influence in fitted models: the *covariate impact, covariate effect* and *standardized covariate effect*. These measures are evaluated using model predictions (i.e. predicted means) and can be compared across different models fitted to a common dataset.

This approach is feasible when pseudo-random simulated data can be drawn from the models of interest at their estimated parameter values, allowing predicted means to be obtained. We apply these methods to investigate apparent anomalies in (a) the *Hydrocotyle vulgaris* example (Carl & Kühn 2007) where apparent anomalies occur despite correct model implementation and (b) examples from Beale *et al.* (2010) where apparent anomalies are observed even after correcting implementation errors.

This paper focuses on the primary objection raised against auto-models in recent years, which concerns covariate parameter estimation. However, to complete our analysis of the recent critique of auto-models we need to consider a separate objection (Dormann 2007) concerning estimator bias in autologistic regression, based on a simulated-data example where the 'true' model contains no spatial autocorrelation. We provide an appendix re-examining this example both numerically and theoretically. The results presented here and elsewhere (Bardos, Guillera-Arroita & Wintle 2015) then allow general conclusions to be drawn on the validity of auto-model analysis of spatially autocorrelated ecological data.

# Correct implementation of auto-models

## Autologistic model

We consider a dataset comprising presence-absence observations $y_n$ at survey sites labelled $n = 1, 2, \ldots S,$ along with a set of $k$ covariates $X_i$, $i = 1, 2, \ldots, k$, writing $\mathbf{X}_n = (X_{1,n}, X_{2,n}, \ldots, X_{k,n})^T$ to denote the $k$-vector of covariate values at site $n$. We associate with each site a probability of occupation $p_n$ conditional on all other site observations, i.e. $p_n = \Pr(y_n = 1 \mid y_{-n})$, according to the model

$$\log\left(\frac{p_n}{1-p_n}\right) = \alpha + \boldsymbol{\beta}_{\text{cov}} \cdot \mathbf{X}_n + \beta_{\text{auto}} \text{ autocov}_n,$$
$$\text{autocov}_n = \sum_{m \in N_n} w_{nm} y_m, \tag{1}$$

where $\boldsymbol{\beta}_{\text{cov}}$ is a $k$-vector of covariate parameters and the *autocovariate* $\text{autocov}_n$ depends on observations in surrounding cells within the neighbourhood $N_n$ of site $n$, $\beta_{\text{auto}}$ is the associated *autocovariate parameter* and $w_{nm}$ are *neighbourhood weights* that determine the influence of neighbouring sites $n$ and $m$ on each other. For the model to be valid, the neighbourhood weights must obey the symmetry rule $w_{nm} = w_{mn}$ (Besag 1974; Bardos, Guillera-Arroita & Wintle 2015) .

## Uncentered auto-normal model

For datasets similar to the above but containing real-valued abundance observations $y_n$, we can define an analogous auto-normal model, with common variance $\sigma^2$ across all sites, as

$$y_n \sim N\left(\alpha + \boldsymbol{\beta}_{\text{cov}} \cdot \mathbf{X}_n + \beta_{\text{auto}} \text{ autocov}_n, \sigma^2\right),$$
$$\text{autocov}_n = \sum_{m \in N_n} w_{nm} y_m, \tag{2}$$

where the symmetry rule $w_{nm} = w_{mn}$ once again holds and $\beta_{\text{auto}}$ must be consistent with the positive-definiteness condition (Bardos, Guillera-Arroita & Wintle 2015) that can be violated for large enough $|\beta_{\text{auto}}|$; it can therefore be necessary to constrain this magnitude during model estimation.

## Validity of auto-models

A theoretical objection concerning the 'circularity' of auto-models was raised by Dormann (2007), arguing that "since autologistic regression is based on an *explanatory* variable being constructed from the non-independent values of a *response* variable one might suspect circularity in the method" (italics in original), and hence appears to suggest that a fundamental problem exists in the structure of autologistic models. Expressed in graphical terms, the autologistic model indeed contains undirected cycles, so that a site influences its neighbours and also depends on them in a 'circular' manner.

However, the main point of Besag's seminal 1974 paper was the resolution of whether (or for what parameter values) the collection of conditionally specified distributions that generate these cycles are consistent. Besag (1974) showed that the consistency requirement for auto-models is quite severe: auto-models *do not* lead to a valid joint distribution (i.e. a valid model likelihood) for *any* arbitrary parameters or arbitrary dependencies between neighbouring sites. Instead, for each particular type of auto-model, the parameters and dependencies must obey certain constraints, which must therefore be built into any implementation of the models.

**Implementation errors in comparative studies**

We refer to Bardos, Guillera-Arroita and Wintle (2015) for detailed background on auto-models and errors arising from their incorrect implementation, with extensive analysis of 'snouter' examples from Dormann et al., for which reported covariate anomalies are resolved by correcting the auto-model implementation.

Beale *et al.* (2010) fitted a variety of models, including linear and uncentered auto-normal models (described as 'Simple Autoregressive (AR)' in their nomenclature) to eight types of simulated data scenarios designed to have various types and strengths of spatial autocorrelation, which is partly due to covariates. The covariates, also simulated, are classified by the eight different types of autocorrelation structure employed in their generating models. For each scenario type, 1000 simulated real-valued datasets were analysed, for each of which a set of covariates (one of each of the eight types) and then observations were generated using Gaussian random fields (i.e. each dataset has a different covariate set).

Non-enforcement of the neighbourhood weighting symmetry rule $w_{nm} = w_{mn}$ in uncentered auto-normal models caused large estimation errors in many examples considered in Beale *et al.* (2010). This resulted from using a weighted-mean construction for their autocovariate, which breaks the symmetry rule near boundaries (Ord 1975; Bardos, Guillera-Arroita & Wintle 2015). We repeat their estimation of simulated datasets, using a weighted-sum construction that always obeys the symmetry rule and compare results with analyses using the incorrect (weighted-mean) autocovariates. For correct (weighted-sum) estimation we used standard unconstrained maximum-likelihood estimation of the autocovariate regression and tested the positive-definiteness condition (Bardos, Guillera-Arroita & Wintle 2015), which can be violated for large $\hat{\beta}_{auto}$ magnitudes. If the condition is violated we used a bisection procedure to determine the critical magnitude and used constrained maximum-likelihood to ensure that $\left|\hat{\beta}_{auto}\right|$ does not exceed it (see supplement for R code).

We see that the large apparent anomalies (differences between auto-normal and linear model covariate estimates) arising when using weighted means (eg for covariate $X_8$) are greatly reduced when weighted sums are used (Table 1). However, notwithstanding the average results, some large apparent anomalies still remain for particular simulated datasets, which we will examine subsequently.

## Comparing covariate influence across different models

To permit comparisons between structurally different models we introduce quantitative measures of covariate influence based on the effect of estimated parameters on model predictions, rather than comparison of parameter values.

Suppose we have estimated some model $A$ that includes a covariate of interest $X$ and an associated parameter $\beta_X$ with estimate $\hat{\beta}_X$. We would like to know what the value $\hat{\beta}_X$ implies about the covariate's importance in the estimated model or, equivalently, in simulations drawn from it (since the estimated model is a *probability distribution for observations*). We now introduce a systematic approach to asking this question in terms of estimated model predictions, i.e. the means of simulation draws.

For linear models, where the parameter $\beta_X$ is a linear regression coefficient, we will see that the question simply reduces to "what is the magnitude $|\hat{\beta}_X|$ of the regression coefficient". Thus the customary interpretation of linear regression coefficient magnitudes is recovered and no actual simulations need be conducted. For GLMs, the same outcome obtains by modifying the question so that we are asking about predictions transformed via an appropriate link function. However, for more general models, such as auto-models, there is no reduction to a simple function of parameter values; actual simulations must therefore be drawn from estimated models. We consider models fitted to observations $\mathbf{y} = (y_1, y_2, \ldots, y_n)$ and covariates $X_i$, $i = 1, 2, \ldots, k$. The models may be structurally different, yet some common structure is required for comparing the influence of covariates to make sense. We assume (i) dependence on covariates is expressed through a collection of terms $Z_j$ each of which can be any function of covariates (but all terms are distinct); examples include polynomials in $X_q$, interaction terms $X_q X_r$, or simply individual covariates (a basic linear model has $Z_j = X_j$); (ii) A parameter $\beta_j$ is associated with each term $Z_j$ (for GLMs $\beta_j$ are regression coefficients) such that setting $\beta_j = 0$ removes the influence of $Z_j$ from the model. Thus the joint probability distribution for observations $\mathbf{y}$ in model $A$ is

$$P_A(\mathbf{y} \mid \boldsymbol{\theta}_A) = f_A(\beta_1, \beta_2, \ldots; Z_1, Z_2, \ldots; \boldsymbol{\gamma}_A, \mathbf{y}), \qquad (3)$$

where $f_A$ is a function such that $f_A(\ldots, \beta_{j-1}, 0, \beta_{j+1}, \ldots; \ldots, Z_{j-1}, Z_j, Z_{j+1}, \ldots; \boldsymbol{\gamma}_A, \mathbf{y})$ does not vary with $Z_j$, $\boldsymbol{\theta}_A$ is the full parameter vector for model $A$ and $\boldsymbol{\gamma}_A$ represents parameters unrelated to covariates; for example if $A$ is an autologistic model (1), $\boldsymbol{\gamma}_A = \beta_{\text{auto}}$.

We now introduce measures of the influence of individual terms $Z_j$, focussing on those that are simply some given covariate $X$, i.e. on some term $Z_j = X$ with associated parameter $\beta_X$. In doing this we are examining the *direct* influence of $X$ and disregarding the influence of any other terms involving $X$, such as polynomial or interaction terms that might be present in a model.

*Covariate impact*

For direct comparisons between models, we can measure the influence of a given covariate $X$ by drawing simulations from models at their fitted parameters $\hat{\boldsymbol{\theta}}$ and at modifications of $\hat{\boldsymbol{\theta}}$. Suppose model $A$, including $X$ as a covariate, is fitted to a dataset, yielding a parameter estimate vector $\hat{\boldsymbol{\theta}}^A$. From this we derive the vector $\hat{\boldsymbol{\theta}}^A_{-X}$, identical to $\hat{\boldsymbol{\theta}}^A$ except that one of its elements, comprising the parameter estimate $\beta_X = \hat{\beta}_X$ for covariate $X$, has been replaced by zero, removing direct influence due to covariate $X$ from the model. Denoting by $\hat{p}_n$ the predicted (i.e. mean) response at site $n$, we define the *covariate impact* as the site-wise average $I^A_X$ of absolute differences between predictions by model $A$ at parameters $\hat{\boldsymbol{\theta}}^A$ and $\hat{\boldsymbol{\theta}}^A_{-X}$, i.e.

$$I^A_X = \frac{1}{S}\sum_{n=1}^{S}\left|\hat{p}_n(\hat{\boldsymbol{\theta}}^A) - \hat{p}_n(\hat{\boldsymbol{\theta}}^A_{-X})\right|, \qquad (4)$$
$$\approx \frac{1}{SR}\sum_{n=1}^{S}\left|\sum_{m=1}^{R}\left[y_{n,m}(\hat{\boldsymbol{\theta}}^A) - y_{n,m}(\hat{\boldsymbol{\theta}}^A_{-X})\right]\right|,$$

where $y_{n,m}(\hat{\boldsymbol{\theta}}^A)$ is the observation for site $n$ within the *m*-th simulation drawn from model $A$ with parameters $\hat{\boldsymbol{\theta}}^A$, $N$ is the number of sites and $R$ is a suitably large number of simulations. The same calculation can be repeated for any other model $B$ containing $X$ as a covariate. The results $I^A_X$, $I^B_X$ are directly comparable as measures of the influence (averaged across all sites) of covariate $X$ on the response variable in the two estimated models.

An advantage of the definition (4) is that all models are compared on an equal footing. However, for a linear model this definition reduces to a product of covariate parameter and averaged covariate magnitudes

$$I^{\text{linear}}_X = \left|\hat{\beta}_X\right|\frac{1}{S}\sum_{n=1}^{S}\left|X_n\right|. \qquad (5)$$

This shows that (4) can give much higher impacts for $X$ than for a centered version $X_C = X - \bar{X}$ of the covariate. That is, arbitrarily high covariate impact can be attained simply by adding a large enough constant to a covariate, so that impact magnitudes *for different covariates* may not be meaningfully comparable. For some models, covariates can be centered without affecting the fitted covariate parameters. In standard linear models, for example, covariate regression coefficients are invariant to shifts in covariates, which only affect the intercept. If all models under comparison are likewise invariant then we can center all covariates when calculating the impact (4) and impact magnitudes will then be more meaningfully comparable between different covariates.

However, some models are not invariant to shifts in covariates: centering changes the model. For example in linear models without intercept, centering of covariates usually changes the regression coefficients and there may be cogent reasons to prefer the uncentered model. Therefore we cannot always center the covariates, in which case comparison of impact magnitudes between different covariates may not be meaningful.

In a strict sense, this presents no problem for our purpose of model comparison: we are not comparing impacts for different covariates in the same model nor comparing the impact of $X$ in one model with the impact of $X_C$ in another, but instead comparing impacts of identical covariates in different models. Nevertheless, we now consider elaborations of (4) with less sensitivity to shifts in covariates.

*Covariate effect*

Rescaling the covariate impact by some measure of the covariate magnitude has intuitive appeal and the structure of (4) suggests that the mean covariate magnitude appearing in (5) is a reasonable choice. We therefore define the *covariate effect*

$$\varepsilon_X^A = \frac{\sum_{n=1}^{S} \left| \hat{p}_n(\hat{\boldsymbol{\theta}}^A) - \hat{p}_n(\hat{\boldsymbol{\theta}}_{-X}^A) \right|}{\sum_{n=1}^{S} |X_n|},$$

$$\approx \frac{\sum_{n=1}^{S} \left| \sum_{m=1}^{R} \left[ y_{n,m}(\hat{\boldsymbol{\theta}}^A) - y_{n,m}(\hat{\boldsymbol{\theta}}_{-X}^A) \right] \right|}{R \sum_{n=1}^{S} |X_n|},$$

(6)

which for linear models reduces to the regression coefficient magnitude,

$$\varepsilon_X^{\text{linear}} = \left| \hat{\beta}_X \right|,$$

(7)

sometimes called the *effect size* in a linear regression context, so that the covariate effect can be naturally interpreted as providing a comparison between linear models and all other models. By incorporating a link function, the definition (6) can be generalized to provide analogous comparisons in which a GLM takes the place of the linear model. Illustrating this for the logit link, we define the *covariate logit-effect*

$$_{\text{logit}}\varepsilon_X^A = \frac{\sum_{n=1}^{S} \left| \text{logit}\left( \hat{p}_n(\hat{\boldsymbol{\theta}}^A) \right) - \text{logit}\left( \hat{p}_n(\hat{\boldsymbol{\theta}}_{-X}^A) \right) \right|}{\sum_{n=1}^{S} |X_n|},$$

$$\approx \frac{\sum_{n=1}^{S} \left| \text{logit}\left( \frac{1}{R}\sum_{m=1}^{R} y_{n,m}(\hat{\boldsymbol{\theta}}^A) \right) - \text{logit}\left( \frac{1}{R}\sum_{m=1}^{R} y_{n,m}(\hat{\boldsymbol{\theta}}_{-X}^A) \right) \right|}{\sum_{n=1}^{S} |X_n|}.$$

(8)

For logistic models this reduces to

$$_{\text{logit}}\varepsilon_X^{\text{logistic}} = \left| \hat{\beta}_X \right|,$$

(9)

so that (8) facilitates their comparison with other models $A$ whose covariate effects $_{\text{logit}}\varepsilon_X^A$ can then be compared directly to logistic regression coefficients. The logit link requires that predictions $\hat{p}_n$ from $A$ fall within the $(0,1)$ domain of the logit function, which will usually be the case for models of presence-absence data. For other GLMs the appropriate link function $g$ can be substituted into (8) to define the corresponding covariate effect ${}_g\varepsilon_X^A$ applicable to models giving predictions compatible with $g$; substituting an identity link $g(x) = x$ (denoted $g = \mathrm{id}$) recovers (6), i.e. ${}_{\mathrm{id}}\varepsilon_X^A = \varepsilon_X^A$.

*Standardized Covariate effect*

Since the covariate effect ${}_g\varepsilon_X^A$ reduces to the regression coefficient magnitude $|\hat{\beta}_X|$ when $A$ is a GLM with link function $g$, but $\hat{\beta}_X$ is unstandardized and thus varies when the covariate $X$ is rescaled, then ${}_g\varepsilon_X^A$ is not generally invariant under rescaling of $X$. Notably, the covariate *impact* for a linear model (5) *is* invariant under rescaling, since the estimate $\hat{\beta}_X$ varies reciprocally with $X$ when the latter is rescaled; this invariance was lost in defining the covariate *effect* due to the denominators in (6) and (8). Accordingly, we define a *standardized covariate effect* ${}_g\tilde{\varepsilon}_X^A$ using the covariate standard deviation $\sigma_X$ as the standardisation factor:

$$ {}_g\tilde{\varepsilon}_X^A = \sigma_X \, {}_g\varepsilon_X^A, \tag{10} $$

which reduces, when $A$ is a GLM with link function $g$, to the magnitude of the standardized regression coefficient ${}_s\hat{\beta}_X$

$$ {}_g\tilde{\varepsilon}_X^{\mathrm{GLM}} = \left| {}_s\hat{\beta}_X \right| \equiv \sigma_X \left| \hat{\beta}_X \right|. \tag{11} $$

Note that ${}_s\hat{\beta}_X$ does not include any rescaling factor for the dependent variable, unlike some definitions of standardized coefficients for linear models, because it makes little sense to do so for GLMs such as the logistic model.

As measures of covariate influence for a class of models much wider than GLMs, ${}_g\varepsilon_X^A$ and ${}_g\tilde{\varepsilon}_X^A$ provide natural generalizations of the unstandardized and standardized GLM regression coefficients respectively. Unrestricted comparisons between models are enabled by avoiding any requirement for covariate centering. Nevertheless, centering can be used if all models being compared are invariant to it. We summarize the above definitions in Table 2, along with similar adaptations ${}_g\mathrm{I}_X^A$ of the covariate impact (4) for GLMs.

**Choice of link function** $g$

For any of the measures ${}_g\mathrm{I}_X^A$, ${}_g\varepsilon_X^A$ or ${}_g\tilde{\varepsilon}_X^A$ of covariate influence, a link function $g$ will be valid if its value is defined for all predictions from all models under comparison. Usually several obvious choices will be valid; the identity link is always valid but is not necessarily the most useful choice.

For presence-absence data, logit or probit links are natural choices; we use a logit link for investigating covariate influence in the *Hydrocotyle vulgaris* example of Carl and Kühn (2007), so that covariate effects $_{\text{logit}}\varepsilon_X^{\text{autologistic}}$ for the autologistic model can be directly compared to logistic regression coefficient magnitudes $\left|\hat{\beta}_X^{\text{logistic}}\right| = {_{\text{logit}}\varepsilon_X^{\text{logistic}}}$.

For abundance data the link $g(x) = \log(x)$ is a natural choice since it transforms the domain of meaningful (i.e. non-negative) abundance values to the extended real line. All covariate effects $_{\log}\varepsilon_X^A$ can then be directly compared with Poisson regression coefficient magnitudes $\left|\hat{\beta}_X^{\text{Poisson}}\right| = {_{\log}\varepsilon_X^{\text{Poisson}}}$. However, the logarithmic link will sometimes be invalid if a linear model is among those being compared: any linear model for abundance allows non-zero probability for negative abundance, since it has a normal error model. Simulation draws from it can therefore yield draws with negative abundance, i.e. outside the domain where $\log(x)$ is defined. Nevertheless a logarithmic link will be valid choice in Table 2 provided the linear-model *predictions* (i.e. means) at each site are positive, which is often (but not always) the case in practice. If some predictions are negative then the logarithmic link cannot be used, reflecting the fact that a linear model cannot (strictly) be a valid model for a positive quantity such as abundance.

If the main focus is to compare other models $A$ with a linear model, the identity link is convenient because all covariate effects $_{\text{id}}\varepsilon_X^A$ can then be directly compared with linear regression coefficient magnitudes $\left|\hat{\beta}_X^{\text{linear}}\right| = {_{\text{id}}\varepsilon_X^{\text{linear}}}$. The real-valued simulated datasets of Beale *et al.* (2010) are not specifically intended to represent abundance and are skewed negative, so a logarithmic link is invalid in that case. Our purpose in examining those datasets is comparison with a linear model, so we use the identity link for investigating covariate influence.

**Interactions and other complications**

As noted earlier, the above definitions are intended to quantify the *direct* influence of a covariate $X$, ignoring the influence of interaction terms and non-linear terms in $X$. It is, however, possible to modify the definitions to examine what could be called the 'total effect' of $X$, simply by redefining $\hat{\boldsymbol{\theta}}_{-X}^A$ such that all parameters associated with terms involving $X$ have been set to zero. Alternatively, for the 'total independent effect', including the influence of non-linear terms in $X$, we would redefine $\hat{\boldsymbol{\theta}}_{-X}^A$ to set to zero parameters for all terms involving $X$ without involving other covariates. Whatever definition is used, careful attention to the particular models in question is required when interpreting the results. In particular we note that depending on the structure of $f_A$ in (3), the model $A$ may depend non-linearly even on the linear terms.

# Amplification of covariate influence in auto-models

**Autologistic models: *Hydrocotyle vulgaris* data**

Carl and Kühn (2007) analysed presence-absence data for the water-plant *Hydrocotyle vulgaris* using two covariates (altitude and temperature) and comparing autocorrelation of residuals for various models, including logistic and autologistic models. The results (Carl & Kühn 2007, Fig. 5(b)) show that their autologistic model outperformed all other models in removing residual autocorrelation.

Nevertheless, they rejected the autologistic model due to what they described as poor parameter estimation: much smaller covariate parameters than were estimated in the corresponding logistic model. That rejection is based on the conceptual error we address in the present work: an implicit assumption that covariate parameters in structurally different models, fitted to the same data, should have similar values.

We reanalyse the *Hydrocotyle vulgaris* example, working with a dataset derived from Fig. 4 of Carl and Kühn (2007). This dataset approximates the original data and results in a similar pattern to that in Table 5 of Carl and Kühn (2007): autologistic covariate parameter estimates are small relative to corresponding logistic estimates. We investigate covariate influence in logistic and autologistic analyses that each use altitude as a covariate. In Fig. 1 we show presence-absence samples drawn from the two fitted models along with the observed data and receiver operator characteristic (ROC) plots (see Hoeting, Leecaster & Bowden 2000) for each model. Evidently the survey region is divided into two contrasting regions of good and poor habitat and both models approximate this division reasonably well using the highly informative altitude covariate. The autologistic model provides a better description of the observed clustering of presences within the 'poor' region and absences in the 'good' regions, since the logistic model cannot represent clustering at all unless it is present in the covariate. ROC plots indicate better overall predictive performance by the autologistic model due to much higher true-positive rates at small false-positive rates, although the logistic model outperforms at intermediate false-positive rates.

Estimates for the two fitted models, each with a single covariate (altitude), are given in Table 3 along with covariate effects calculated using predictions from the fitted models, which requires Gibbs sampling in the autologistic case (see Supplement for R code). We see that although the altitude parameter in the logistic model is five times the magnitude of its counterpart in the autologistic model, the covariate effect is only 24% larger. Thus the effect of a given covariate parameter value is strongly enhanced in the autologistic model; we refer to this phenomenon as *covariate amplification*.

Our results demonstrate that the influence of small covariate parameters can be greatly amplified within the autologistic model (relative to their influence in a corresponding logistic model) when the autocovariate parameter is large.

**Auto-normal models: Beale *et al.* (2010) simulations**

We investigate the apparent covariate estimation anomaly, concerning $\beta_{X_8}$ in simulated data scenario type 6 (see Table 1) from Beale *et al.* (2010). Linear model and uncentered auto-normal model estimates for $\beta_{X_8}$ are plotted in Fig. 2(a) for each of the 1000 simulated datasets of scenario type 6 and are summarized as histograms in Fig. 2(b). Clearly the linear model estimates are distributed symmetrically around the value of 0.5 used to generate the datasets. However, the small negative region of the linear-model estimate distribution has almost vanished in the auto-normal case, while the larger positive region appears to have been compressed unevenly toward zero, i.e. with a disproportionate shift of larger values. Since this would be consistent with a reduction in covariate parameter magnitudes due to covariate amplification, we investigate the dataset (Fig. 3) exhibiting the largest parameter shift. We show in Table 4 that for this dataset, the covariate effect (for covariate $X_8$) is only 18% lower for the auto-normal model than the linear model, despite a halving of the parameter estimate. We conclude that the lower $\beta_{X_8}$ estimates for the uncentered auto-normal model (averaged over 1000 datasets for scenario type 6) represent a smaller reduction in the influence of this covariate, relative to the linear model, than the parameter averages suggest.

**Mechanism of covariate amplification**

Linear models constitute a very small subcategory of auto-models in which there are no inter-site interactions. Therefore it is reasonable to expect auto-models to be capable of exotic behaviour by linear-model standards, covariate amplification being an example.

The Gibbs sampling process for generating simulated data from autologistic models provides a useful visualization of the mechanism of covariate amplification: strongly autocorrelated covariates tend to generate 'seeds' consisting of multiple neighbouring presences (i.e. clusters), which can then grow rapidly due to strong inter-site interactions if $\beta_{auto}$ is large. Whereas weakly spatially autocorrelated covariates will scatter presences more evenly, making neighbourhood interactions less likely, so the seeds are less likely to persist as simulations proceed.

Concomitantly, for parameter estimates to be consistent with the simulations (i.e. with model predictions), a valid estimation procedure must have a tendency to reduce magnitudes of covariate parameters for strongly autocorrelated covariates when estimating an auto-model on data that has strong autocorrelation beyond that explained by the covariates. In that circumstance, the auto-model will necessarily yield large $\beta_{auto}$ values, which if combined with covariate parameters at "full strength" (i.e. at values estimated in a GLM), would yield simulations from the estimated auto-model with excessive correlation to the covariates. Thus covariate amplification in predictive simulation from estimated auto-models must be accompanied by covariate *parameter attenuation* in auto-model estimation. From the above, we see that covariate parameter attenuation can occur when observations and informative covariates have strong SAC, but the strong SAC in observations is only partly due to covariates.

# Discussion

We have shown that simple comparison of parameter magnitudes between structurally different models can be highly misleading as a representation of covariate influence. In a phenomenon we call *covariate amplification*, small covariate parameters in auto-models can exert a large influence on model predictions, similar to the effect of much larger covariate parameters in the corresponding generalized linear model (the model obtained from the auto-model by setting all inter-site interactions to zero).

Consequently we cannot simply compare parameter values across models as a measure of covariate influence, nor interpret disparate parameter estimates for the same covariate (in different models) as "bias". Instead, measuring covariate influence requires obtaining predictions from estimated models (e.g. via simulation) and examining the change in predicted occupancy or abundance when covariate parameters are varied.

In the approach developed here, we define the *covariate impact, covariate effect* and *standardized covariate effect*, evaluated for each covariate by comparing predictions at the fitted parameter values with predictions at modified values that remove any influence due to the covariate of interest. When using this approach, we naturally expect that structurally different models will yield covariate effects that differ to some extent. For example, we would not be surprised to find a reduction in influence for a covariate when additional covariates are added to the model. In comparing auto-models with their associated GLMs, we found the auto-models assigned somewhat less influence to covariates, which is easily understood in view of the additional spatial aspect of these models, by analogy to the inclusion of an additional covariate. To make room for spatial effects, some reduction in covariate influence occurred, however this was very slight in relation to the reduction in covariate parameter magnitudes.

The results given here, in conjunction with our investigation into auto-model implementation errors (Bardos, Guillera-Arroita & Wintle 2015), demonstrate that assertions (Carl & Kühn 2007; Dormann 2007; Dormann *et al.* 2007; Beale *et al.* 2010) of covariate parameter estimation 'bias' in auto-models are incorrect.

We addressed a theoretical objection (Dormann 2007) concerning 'circularity' of auto-model estimation, pointing out that the issue of mathematical consistency of conditionally specified models was settled in Besag (1974), leading to the specific requirements given there for correct implementation.

Finally, in the Appendix we examine a different notion of bias, namely the standard definition of bias in parameter estimation, in the context of a separate objection from Dorman (2007), based on auto-covariate parameter estimates obtained by fitting an autologistic model to simple binomial data. This objection fails because the autologistic model has the correct asymptotic behaviour in the large data limit: the small estimator bias observed in Dorman (2007) tends to become smaller still with more data.

In summary, all objections raised against auto-models in Carl and Kühn (2007), Dormann (2007), Dormann *et al.* (2007) and Beale *et al.* (2010) have been systematically investigated and shown to be unfounded.

**Outlook**

When correctly implemented, auto-models provide an effective and relatively straightforward approach for analysing spatially autocorrelated ecological data such as species distribution and abundance data.

The measures of covariate influence introduced here provide for valid comparison of structurally different models using predictive simulations, allowing auto-models to be readily compared with GLMs.

# Appendix

**Bias in parameter estimation**

The term 'bias' has a precise meaning in non-Bayesian parameter estimation methodology, where one attempts to use an 'estimator', for example the Maximum Likelihood Estimator (MLE), to obtain estimates for the parameter vector $\theta$ of a model, given an empirical dataset $D_E$. An estimator is 'biased', in this sense, if its expected value over an infinite number of datasets drawn from the true model distribution differs from the true parameter vector $\theta_T$. That is, for a given true parameter $\theta_T$, the true model is a probability distribution for datasets $D$. Suppose we generate a large number $k$ of sample datasets drawn from this true model distribution. We then have a set of $k$ simulated datasets $D_i$, $i = 1, 2, ..., k$., each corresponding to an imaginary 'experiment' yielding data drawn from the model at its true parameter vector $\theta_T$. We use our estimator (eg. MLE) to estimate the model for each $D_i$, yielding estimate vectors $\hat{\theta}_i$; an approximation for the *bias of the estimator* is the difference between the average of these estimates and the true parameter

$$\text{bias} \approx k^{-1} \sum_{i=1}^{N} \hat{\theta}_i - \theta_T,$$

where this approximation is expected to converge to the exact value of the bias as $k \to \infty$. If the bias is zero the estimator is called unbiased. For many models in applied statistics, commonly used estimators *are* biased, but this does not render them impractical because their bias typically becomes smaller for larger datasets.

**Estimator bias in autologistic model estimation**

We performed bias calculations on a sets of 1000 simulated datasets comprising square grids of observations simulated as independent Bernoulli trials with all sites having common probabilities of presence $p$. Each dataset was analysed using an autologistic model without covariates, for various neighbourhood sizes, using the Maximum Pseudo Likelihood (MPL) estimator. Since the true model contains no SAC, the true value of the autocovariate parameter is zero. The MPL estimator has negative bias (Fig A1); the bias falls to zero strongly as the size of the observation grid increases and is very small for 50x50 grids.

Dormann (2007) reports results of similar calculations where autologistic models without covariates are fitted to 1000 simulated datasets, each comprising a 50x50 grid of observations simulated as independent Bernoulli trials with a common probability of presence $p = 0.25$. However, Dormann (2007) calculated a measure differing from the definition of bias given above, in that for each simulated dataset the neighbourhood size of the autologistic model was selected to minimize residual autocorrelation. Thus, the calculation does not measure the bias of any particular autologistic model, but is rather an analogous measure for a procedure that includes model selection as well as autologistic estimation. Consequently the numerical results in Dormann (2007) differ from our bias calculations in the details, but nevertheless they found similarly that estimates of the selected models were clustered around a mean value very close to the true value of zero. Dormann (2007) uses this small deviation from the true value as an argument against the validity of the model.

The problem with this argument is that the same argument would render invalid the use of many standard statistical models for which useful estimators are biased. For example the MLE for the variance of a normal distribution is biased. Such models are *not* summarily discarded, because as the amount of data increases, bias usually decreases in a probabilistic sense: estimators *converge in probability* to the correct value as the number of observations increases; this asymptotic property is called *consistency* and is usually required in practice for statistical models to be useful. In the case of the autologistic model, it has been shown that the MPL estimator is consistent (Geman & Graffigne 1986; Possolo 1986; Cressie 1991). Our results in Fig. A1 illustrate that, as expected, the bias reduces as the size of the dataset increases.

| Simulated scenario type | Covariate type | | | | | | | |
|---|---|---|---|---|---|---|---|---|
| | $X_1$ | $X_2$ | $X_3$ | $X_4$ | $X_5$ | $X_6$ | $X_7$ | $X_8$ |
| 1 | 0<br>0 | -0.0216<br>-0.0554 | -9<br>-23 | -2<br>-5 | -8<br>-19 | -0.0036<br>-0.0016 | | |
| 2 | 0<br>-1 | -0.0032<br>-0.0142 | -2<br>-8 | -1<br>-2 | -2<br>-6 | -0.0012<br>-0.0002 | | |
| 3 | 0<br>0 | -0.0037<br>-0.0172 | -2<br>-7 | -1<br>-2 | -2<br>-6 | -0.0009<br>-0.0004 | | |
| 4 | 0<br>0 | 0.0001<br>0.0004 | 0<br>0 | 0<br>0 | 0<br>0 | -0.0006<br>-0.0008 | | |
| 5 | 0.0285<br>0.1239 | | | | | | -9<br>-39 | -14<br>-78 |
| 6 | 0.0522<br>0.1247 | | | | | | -16<br>-39 | -29<br>-81 |
| 7 | | | 0.0263<br>0.3032 | | | | -3<br>-17 | -14<br>-147 |
| 8 | | | -2<br>-96 | | | | 0<br>-8 | -2<br>-69 |

**Table 1.** Differences (shaded) or percentage differences (unshaded) between average parameter estimates for uncentered auto-normal models and linear models. Means are calculated for each covariate type $X_i$, averaging over estimates fitted to sets of 1000 simulated real-valued datasets for each of eight scenario types. Differences $\underset{\text{datasets}}{\text{mean}}\left(\beta_{X_i}^{\text{auto-normal}}\right) - \underset{\text{datasets}}{\text{mean}}\left(\beta_{X_i}^{\text{linear}}\right)$ are then evaluated and are either (unshaded entries) expressed as percentages of $\underset{\text{datasets}}{\text{mean}}\left(\beta_{X_i}^{\text{linear}}\right)$, or else (shaded entries) given directly if $\underset{\text{datasets}}{\text{mean}}\left(\beta_{X_i}^{\text{linear}}\right) \approx 0$, which can occur when $X_i$ is absent from the data generation model. We report two values for each scenario and covariate type: the upper entry (black) based on correct estimation using the weighted-sum construction for autocovariates; the lower entry (red) based on incorrect estimation with the weighted-mean construction used in Beale et al. (2010). For 52 of the 8000 simulated datasets, the correct estimation procedure required constrained maximum-likelihood estimation to enforce the positive-definiteness condition for uncentered auto-normal models. This usually resulted in small changes to estimated covariate parameters (relative to estimates from unconstrained maximization that violate the condition) and the change, due to enforcement of this condition, was very small for averages calculated over 1000 datasets. Therefore the large reduction of differences between linear and auto-normal model averages, evident in the table when auto-normal estimation is corrected, was almost entirely due to enforcement of the symmetry rule via weighted-sum autocovariates. Estimation was performed in R (see Supplement) by extending and modifying the code given in Beale *et al.* (2010); all calculations for scenario types 5 and 6 include corrections to a coding error affecting neighbourhood sizes.

| Measure of covariate influence | Definition evaluating influence of covariate $X$ in model $A$; predictions are transformed using link function $g$ | Value when $A$ is a GLM with link function $g$ |
|---|---|---|
| Covariate impact | $_g\mathrm{I}_X^A = \dfrac{1}{S}\sum_{n=1}^{S}\left|g\left(\hat{p}_n(\hat{\boldsymbol{\theta}}^A)\right) - g\left(\hat{p}_n(\hat{\boldsymbol{\theta}}_{-X}^A)\right)\right|,$ | $_g\mathrm{I}_X^{GLM} = \left|\hat{\beta}_X\right|\dfrac{1}{S}\sum_{n=1}^{S}\left|X_n\right|$ |
| Covariate effect | $_g\varepsilon_X^A = \left(\dfrac{1}{S}\sum_{n=1}^{S}\left|X_n\right|\right)^{-1} {_g\mathrm{I}_X^A}$ | $_g\varepsilon_X^{GLM} = \left|\hat{\beta}_X\right|,$ |
| Standardized Covariate effect | $_g\tilde{\varepsilon}_X^A = \sigma_X \, _g\varepsilon_X^A,$ | $_g\tilde{\varepsilon}_X^{GLM} = \sigma_X \left|\hat{\beta}_X\right|,$ |

**Table 2.** Definitions for measures of covariate influence that apply to a wide class of models and are directly related to regression coefficients $\hat{\beta}_X$ when the model is a GLM.

| Model | Estimated parameters | | | Measures of covariate influence | | |
|---|---|---|---|---|---|---|
| | $\alpha$ | $\beta_{auto}$ | $\beta_{altitude}$ | $_{logit}I^A_{altitude}$ | $_{logit}\varepsilon^A_{altitude}$ | $_{logit}\tilde{\varepsilon}^A_{altitude}$ |
| logistic | 2.78 | | -0.792 | 3.37 | 0.792 | 2.30 |
| autologistic | -2.12 | 1.43 | -0.159 | 2.72 | 0.639 | 1.85 |

**Table 3.** Measures of covariate influence for the altitude covariate, evaluated using predictions from logistic and auto-logistic models (at the estimates shown) that were fitted to the *Hydrocotyle vulgaris* dataset.

| Model | Estimated parameters | | | | | Measures of covariate influence | | |
|---|---|---|---|---|---|---|---|---|
| | $\alpha$ | $\beta_{X_1}$ | $\beta_{X_7}$ | $\beta_{X_8}$ | $\beta_{auto}$ | $I^A_{X_8}$ | $\varepsilon^A_{X_8}$ | $\tilde{\varepsilon}^A_{X_8}$ |
| | | | | | | | | |
| linear | -2.00 | -0.062 | 0.509 | 1.15 | | 0.967 | 1.15 | 1.15 |
| uncentered auto-normal | -0.293 | -0.137 | 0.677 | 0.553 | 0.017 | 0.822 | 0.981 | 0.981 |

**Table 4.** Measures of covariate influence for the $X_8$ covariate from scenario type 6 of simulated dataset number 180 of Beale *et al.* (2010). For this dataset the linear model has an estimate $\beta_{X_8}$ approximately twice that of the auto-normal model, but the covariate effect is only 18% higher. The $X_8$ covariate is generated with unit variance in the simulations of Beale *et al.* (2010); consequently $\tilde{\varepsilon}^A_{X_8} = \varepsilon^A_{X_8}$.

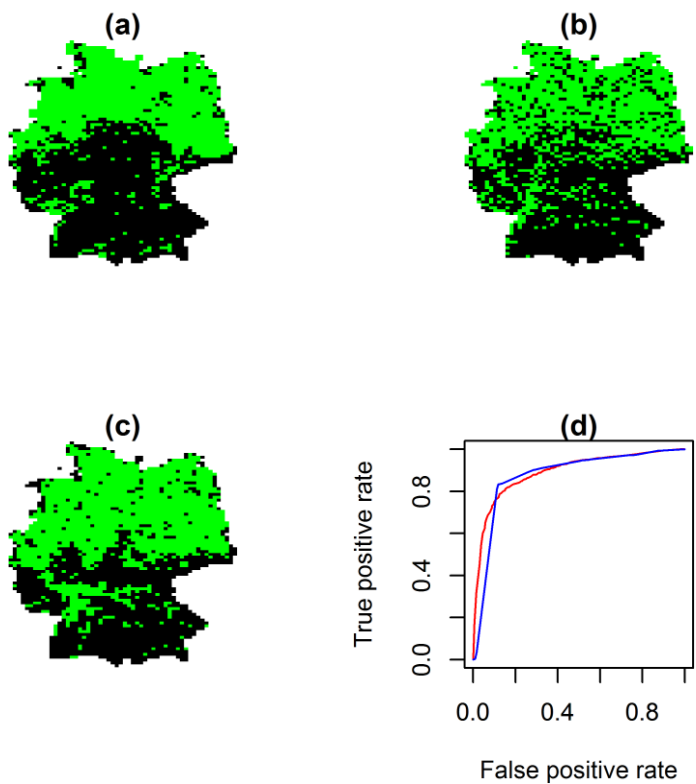

**Figure 1.** Reanalysis of the *Hydrocotyle vulgaris* example using data derived from Fig. 4 of Carl and Kühn (2007). Observed presence-absence data (a) is compared with simulated presence-absence maps drawn from (b) logistic and (c) autologistic models fitted to the observations, each using altitude as a covariate; (d) Receiver operator characteristics (ROC) for logistic (blue) and autologistic (red) predictions; corresponding area under curve (AUC) results are 0.876 and 0.892 respectively.

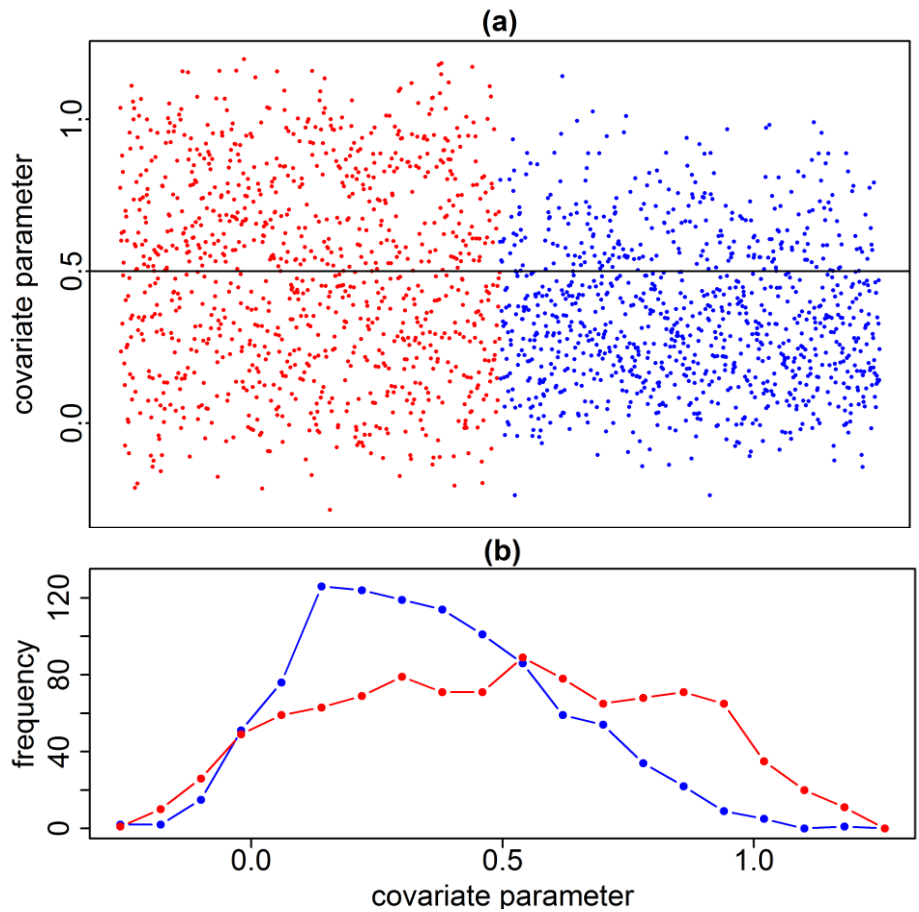

**Figure 2.** (a) Scatterplots of covariate parameter estimates $\beta_{X_8}$, obtained from fitting linear models (red, left) and uncentered auto-normal models (blue, right) to 1000 simulated datasets of scenario type 6 from Beale *et al.* (2010); (b) corresponding histograms for each model show the relative attenuation of covariate parameters in the auto-normal model.

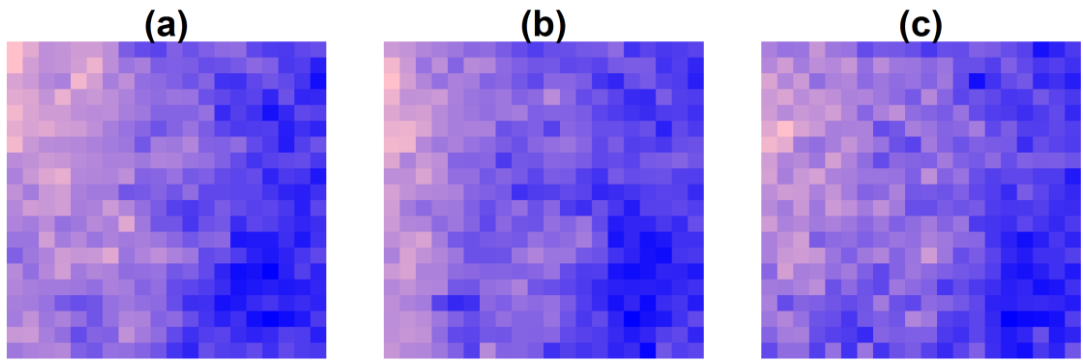

**Figure 3.** (a) Simulated dataset number 180 from scenario type 6 of Beale *et al.* (2010), to which a linear model and an uncentered auto-normal model were fitted. Simulations (b) and (c) were then drawn from the linear and auto-normal models respectively.

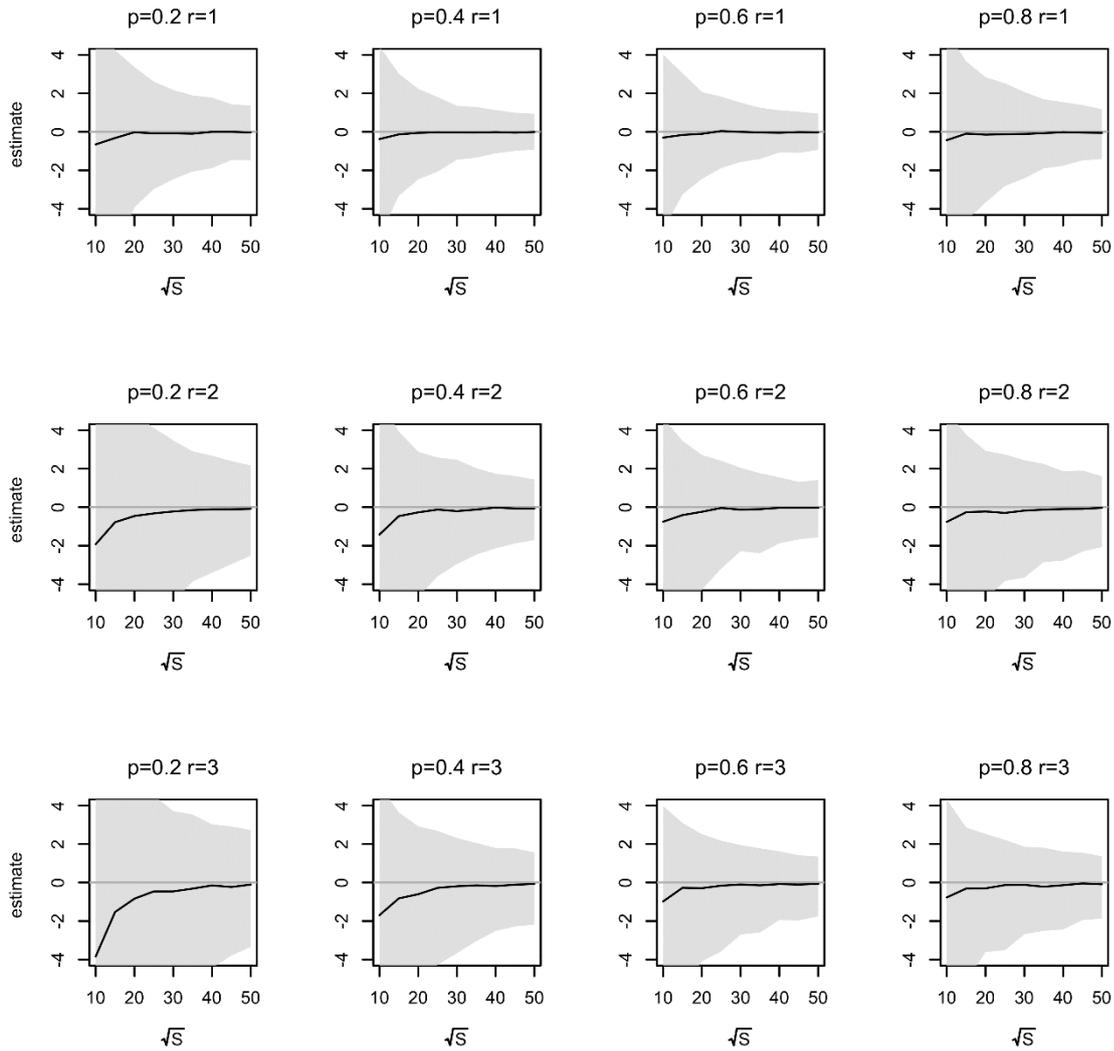

**Figure A1.** Bias as a function of the number $S$ of observed sites in autocovariate parameter estimation for an autologistic model fitted to simulated datasets generated without spatial autocorrelation. Solid lines show the mean of the estimated autocovariate values and shaded areas encompass estimates within the 2.5 and 97.5 percentiles. Plots show results for different grid sizes (from $S$ = 10x10 to $S$ = 50x50 sites), neighbourhood sizes (radius $r = 1$, 2 or 3) and species prevalence ($p$). Results obtained from 1000 simulated datasets per scenario, with models fitted using the maximum pseudo-likelihood approximation (i.e. incorporating an autocovariate to a logistic regression model). Bias decreases rapidly with grid size, as expected theoretically.

# Supplementary information

## R programs and an associated data file.

The first three programs below require additional R code from the supplementary information of Beale *et al.* (2010):

  http://onlinelibrary.wiley.com/doi/10.1111/j.1461-0248.2009.01422.x/suppinfo

The fourth program, 'Autologistic covariate effect.r', requires the data file 'Hydrocotyle.data.txt' reproduced below, containing data derived from Fig. 4 of Carl and Kühn (2007).

### UnconstrainedMPLE.R

```
# Auto-normal estimation:
# unconstrained MPLE for comparing covariate parameter estimates obtained
# using invalid  (weighted-mean) and valid (weighted-sum) autocovariates
# Author: David C Bardos
# Uses package RandomFields version 2.0.71 due to use of functions from
# Beale et al (2010) code; consequently it runs in R version 2.15.3
# Requires R file "ELE_1422_sm_HelperFunctions.r" from the
# supplementary information of Beale et al (2010) available at
# http://onlinelibrary.wiley.com/doi/10.1111/j.1461-0248.2009.01422.x/suppinfo
# (the main Beale at al (2010) code is also obtainable at that link)

# for 1000 iterations:

# 1. for each of 8 scenarios, simulates a dataset (covariates and observations)
#    using simulation code from Beale et al (2010);

# 2. for the 8 simulated datasets, runs unconstrained MPLE with
#    weighted-mean autocovariates, using code from Beale et al (2010)
#    with corrections to neighbourhood-size bug for scenarios 5 & 6

# 3. for the 8 simulated datasets, runs unconstrained MPLE with
#    weighted-sum autocovariates (code from Beale et al (2010) is
#    modified to use weighted sum autocovariates)

# 4. for the 8 simulated datasets, fits a linear model using
#    code from Beale et al (2010)

# For simulations generated at iteration i:

# the vectors:
# WeightedSumCoeffs1[i,], WeightedSumCoeffs2[i,], WeightedSumCoeffs3[i,] & WeightedSumCoeffs4[i,]
# give parameter estimates for scenarios 1 to 4, estimated using weighted sum autocovariates, with format:
# WeightedSumCoeffs1[i,]={intercept, beta_X1, beta_X2, beta_X3, beta_X4, beta_X5, beta_X6, beta_auto}

# the vectors:
# WeightedSumCoeffs5[i,], WeightedSumCoeffs6[i,]
# give parameter estimates for scenarios 5 & 6, estimated using weighted sum autocovariates, with format:
# WeightedSumCoeffs5[i,]={intercept, beta_X1, beta_X7, beta_X8, beta_auto}

# the vectors:
# WeightedSumCoeffs7[i,], WeightedSumCoeffs8[i,]
# give parameter estimates for scenarios 7 & 8, estimated using weighted sum autocovariates, with format:
# WeightedSumCoeffs5[i,]={intercept, beta_X3, beta_X7, beta_X8, beta_auto}

# The vectors WeightedMeanCoeffs1[i,], WeightedMeanCoeffs2[i,] ..etc give estimates obtained with
# weighted-mean autocovariates, with the same formatting as above.

# The vectors lmCoeffs1[i,], lmCoeffs2[i,] ..etc give linear model estimates in a similar format to
# the above except that there is no autocovariate parameter, so eg.
# lmCoeffs5[i,]={intercept, beta_X1, beta_X7, beta_X8}
```

```r
###### set up initial conditions and source additional functions
setwd("C:/CovariateEffect") #Change as required.
library(RandomFields)
library(MASS)
source("ELE_1422_sm_HelperFunctions.r")

x <- rep(1:20, times = 20)                                  # Set up coordinates.
y <- rep(1:20, each = 20)
dists <- as.matrix(dist(cbind(x,y), upper = T, diag = T))   # Calculate distance matrix.

MMsetUp <- matrixMoranSetUp(mat = matrix(0, ncol = 20, nrow = 20),   # Define matrix moran setup values
         equal.lag.dist = TRUE, nbins = 17)

ACdist <- function (mat1 = Y1, nbins1 = 17, useStartUp1 = MMsetUp) {
# Function to calculate distance at which Moran's I first becomes 0
# (spatial scale of pattern). Takes a matrix (mat1), the number of bins for
# Moran's I (nbins1) and the matrix moran startup values (useStartUp1).
    MM1 <- matrixMoran(mat = mat1, nbins = nbins1, useStartUp = useStartUp1,   # Calculates Moran's I for given bins.
         equal.lag.dist = T)
    first0 <- MM1[[1]][,2] < 0                   # Identifies negative I values
    first0 <- cumsum(first0)                     # Cumulitive sum of negative I values
    first0 <- min(which(first0 == 1))            # Identifies first negative I calue
    return(ceiling(MM1[[1]][first0,1]))          # Returns the distance corresponding to this bin.
}

###############################################################################

Nitr=1000;
ACmat=matrix(nrow=Nitr,ncol=8);
#load("ACmat(Nitr1000).RData")
WeightedMeanCoeffs1=matrix(nrow=Nitr,ncol=8);   #const + X1 + X2 + X3 + X4 + X5 + X6 + ary1
WeightedMeanCoeffs2=WeightedMeanCoeffs3=WeightedMeanCoeffs4=WeightedMeanCoeffs1

WeightedMeanCoeffs5=matrix(nrow=Nitr,ncol=5);   #const + X1 + X7 + X8 + ary5
WeightedMeanCoeffs6=WeightedMeanCoeffs7=WeightedMeanCoeffs8=WeightedMeanCoeffs5

WeightedSumCoeffs1=matrix(nrow=Nitr,ncol=8);    #const + X1 + X2 + X3 + X4 + X5 + X6 + ary1
WeightedSumCoeffs2=WeightedSumCoeffs3=WeightedSumCoeffs4=WeightedSumCoeffs1

WeightedSumCoeffs5=matrix(nrow=Nitr,ncol=5);    #const + X1 + X7 + X8 + ary5
WeightedSumCoeffs6=WeightedSumCoeffs7=WeightedSumCoeffs8=WeightedSumCoeffs5

lmCoeffs1=matrix(nrow=Nitr,ncol=7);  #const + X1 + X2 + X3 + X4 + X5 + X6
lmCoeffs2=lmCoeffs3=lmCoeffs4=lmCoeffs1

lmCoeffs5=matrix(nrow=Nitr,ncol=4);  #const + X1 + X7 + X8
lmCoeffs6=lmCoeffs7=lmCoeffs8=lmCoeffs5

### main loop which runs for Nitr iterations
for (i in 1:Nitr){
  print(i);
  set.seed(i)                                   # Set seed for repeatability.
    X1 <- multiPattern(n=2, dim=c(20,20), taus = c(.3,7), corr=0.8)    # Generates 2 correlated exponential fields
    X2 <- matrix(scale(as.vector(X1[[2]])), 20)                        # Scale fields to mean 0, var 1.
    X1 <- matrix(scale(as.vector(X1[[1]])), 20)
    X3 <- matrix(scale(as.vector(GaussRF (1:20,1:20,param = c(1,5,1,3), method =# Generate uncorrelated field with intermediate
         "cutoff CE", model = "exponential", grid = T))), 20)         #                    scale.
    X4 <- multiPattern(n=2, dim=c(20,20), taus = c(.3,7), corr=0.8)    # Repeat above for next two variables but adding
    X5 <- matrix( scale( as.vector(X4[[2]] + matrix( rnorm( 20*20, sd = 1),   #    white noise to reduce strength of
         ncol = 20))), 20)                     #         spatial autocorrelation.
    X4 <- matrix( scale( as.vector(X4[[1]] + matrix( rnorm( 20*20, sd = 1),
         ncol = 20))), 20)
    X6 <- matrix(scale(as.vector(GaussRF (1:20,1:20,param = c(1,5,1,3), method =
         "cutoff CE", model = "exponential", grid = T) +
         matrix(rnorm(20*20, sd = 1), ncol = 20))), 20)
    X7 <- X1 * seq(0, 1, len = 20) + X2 * seq(1, 0, len = 20)          # Generate Field with variable scale in AC
    X8 <- matrix(scale(as.vector(X3 + seq(min(X3), max(X3), len = 20))), 20)

    Y1E <- Y1 <- matrix(scale(as.vector(GaussRF (1:20,1:20,param = c(1,5,1,3),   # Generate 2 random fields to become the error
         method = "cutoff CE", model = "exponential", grid = T))), 20)   #       in the Y variables.
    Y2E <- Y2 <- matrix(scale(as.vector(GaussRF (1:20,1:20,param = c(1,5,1,0.7),
         method ="cutoff CE", model = "exponential", grid = T))), 20)
```

```r
    Y3 <- matrix(scale(as.vector(Y1 +  matrix(rnorm(20*20, sd = 2), ncol = 20))),# Add white noise to weaken AC in error term and
        20) - 2 + 0.5 * X1 + 0.5 * X3 + 0.5 * X4 + 0.5 * X5           #          add deterministic parts.
    Y3E <- Y3 + 2 - 0.5 * X1 - 0.5 * X3 - 0.5 * X4 - 0.5 * X5
    Y4 <- matrix(scale(as.vector(Y2 +  matrix(rnorm(20*20, sd = 2), ncol = 20))),
        20) - 2 + 0.5 * X1 + 0.5 * X3 + 0.5 * X4 + 0.5 * X5
    Y4E <- Y4 + 2 - 0.5 * X1 - 0.5 * X3 - 0.5 * X4 - 0.5 * X5
    Y1 <- Y1 - 2 + 0.5 * X1 + 0.5 * X3 + 0.5 * X4 + 0.5 * X5              # Add only deterministic parts.
    Y2 <- Y2 - 2 + 0.5 * X1 + 0.5 * X3 + 0.5 * X4 + 0.5 * X5
    Y5 <- matrix(scale(as.vector(GaussRF (1:20,1:20,param = c(1,5,1,1.5),   #   Fine scale error only
        method ="cutoff CE", model = "exponential", grid = T))), 20) -
        2 + 0.5 * X7 +0.5 * X8
    Y5E <- Y5 + 2 - 0.5 * X7 - 0.5 * X8
    Y6 <- matrix(scale(as.vector(GaussRF (1:20,1:20,param = c(1,5,1,10),    #   Medium scale error only
        method = "cutoff CE", model = "exponential", grid = T))), 20) -
        2 + 0.5 * X7 +0.5 * X8
    Y6E <- Y6 + 2 - 0.5 * X7 - 0.5 * X8
    X11 <- multiPattern(n=2, dim=c(20,20), taus = c(.3,7), corr=0.8)       # Generates 2 correlated exponential fields
    X12 <- matrix(scale(as.vector(X11[[2]])), 20)                # Scale fields to mean 0, var 1.
    X11 <- matrix(scale(as.vector(X11[[1]])), 20)
    Y7 <- X11 * seq(0, 1, len = 20) + X12 * seq(1, 0, len = 20) - 2 + 0.5 * X7 +
        0.5 * X8
    Y7E <- Y7 + 2 - 0.5 * X7 - 0.5 * X8
    Y8 <- matrix(scale(as.vector(X11 + seq(min(X11), max(X11), len = 20))), 20) -
        2 + 0.5 * X7 + 0.5 * X8
    Y8E <- Y8 + 2 - 0.5 * X7 - 0.5 * X8

    my.df <- data.frame(x = x, y = y, X1 =as.vector(X1), X2 =as.vector(X2), X3 = # Combine all X and Y variables and coordinates
        as.vector(X3), X4 =as.vector(X4), X5 =as.vector(X5), X6 =        #       into one data.frame
        as.vector(X6), X7 = as.vector(X7), X8 = as.vector(X8), Y1 =
        as.vector(Y1), Y2 =as.vector(Y2), Y3 = as.vector(Y3), Y4 =
        as.vector(Y4), Y5 = as.vector(Y5), Y6 = as.vector(Y6), Y7 =
        as.vector(Y7), Y8 = as.vector(Y8))
#print("Variables created")

    AC1 <- ACdist(mat1 = Y1, nbins1 = 17, useStartUp1 = MMsetUp)  # Calculate Autocorrelation distance for Y
    AC2 <- ACdist(mat1 = Y2, nbins1 = 17, useStartUp1 = MMsetUp)  #                variables.
    AC3 <- ACdist(mat1 = Y3, nbins1 = 17, useStartUp1 = MMsetUp)
    AC4 <- ACdist(mat1 = Y4, nbins1 = 17, useStartUp1 = MMsetUp)
    AC5 <- ACdist(mat1 = Y5, nbins1 = 17, useStartUp1 = MMsetUp)
    AC6 <- ACdist(mat1 = Y6, nbins1 = 17, useStartUp1 = MMsetUp)
    AC7 <- ACdist(mat1 = Y7, nbins1 = 17, useStartUp1 = MMsetUp)
    AC8 <- ACdist(mat1 = Y8, nbins1 = 17, useStartUp1 = MMsetUp)

# print("Autocorrelation distances done")
    ACmat[i,1]=AC1; ACmat[i,2]=AC2; ACmat[i,3]=AC3; ACmat[i,4]=AC4;
    ACmat[i,5]=AC5; ACmat[i,6]=AC6; ACmat[i,7]=AC7; ACmat[i,8]=AC8;

    ary1 <- ary2 <- ary3 <- ary4 <- ary5 <- ary6 <- ary7 <- ary8 <- numeric(400)   # Create vector for autoregressive Y variables.
    for (j in 1:400) {                         # For each cell...
      ns1 <- dists[j,] < AC1 & dists[j,] != 0                   # Identify cells within autocorrelation range.
      ns2 <- dists[j,] < AC2 & dists[j,] != 0
      ns3 <- dists[j,] < AC3 & dists[j,] != 0
      ns4 <- dists[j,] < AC4 & dists[j,] != 0
      ns5 <- dists[j,] < AC5 & dists[j,] != 0
      ns6 <- dists[j,] < AC6 & dists[j,] != 0
      ns7 <- dists[j,] < AC7 & dists[j,] != 0
      ns8 <- dists[j,] < AC8 & dists[j,] != 0

      ary1[j] <- weighted.mean(as.vector(Y1)[ns1],1/dists[j,ns1]) # Calculate mean Y values weighted by 1/distance
      ary2[j] <- weighted.mean(as.vector(Y2)[ns2],1/dists[j,ns2]) #            for cells with AC range.
      ary3[j] <- weighted.mean(as.vector(Y3)[ns3],1/dists[j,ns3])
      ary4[j] <- weighted.mean(as.vector(Y4)[ns4],1/dists[j,ns4])
#     ary5[j] <- weighted.mean(as.vector(Y5)[ns1],1/dists[j,ns1])
#     ary6[j] <- weighted.mean(as.vector(Y6)[ns2],1/dists[j,ns2])
      ary5[j] <- weighted.mean(as.vector(Y5)[ns5],1/dists[j,ns5]) # correction of bug in Beale et al (2010)
      ary6[j] <- weighted.mean(as.vector(Y6)[ns6],1/dists[j,ns6]) # affecting neighbourhood sizes in ary5 and ary6
      ary7[j] <- weighted.mean(as.vector(Y7)[ns7],1/dists[j,ns7])
      ary8[j] <- weighted.mean(as.vector(Y8)[ns8],1/dists[j,ns8])
    }
#print("weighted means calculated")

    ar1.y1 <- lm(Y1 ~ X1 + X2 + X3 + X4 + X5 + X6 + ary1, data = my.df) # unconstrained autocovariate regression
```

```r
ar1.y2 <- lm(Y2 ~ X1 + X2 + X3 + X4 + X5 + X6 + ary2, data = my.df)  # using weighted-mean autocovariates
ar1.y3 <- lm(Y3 ~ X1 + X2 + X3 + X4 + X5 + X6 + ary3, data = my.df)
ar1.y4 <- lm(Y4 ~ X1 + X2 + X3 + X4 + X5 + X6 + ary4, data = my.df)
ar1.y5 <- lm(Y5 ~ X1 + X7 + X8 + ary5, data = my.df)
ar1.y6 <- lm(Y6 ~ X1 + X7 + X8 + ary6, data = my.df)
ar1.y7 <- lm(Y7 ~ X3 + X7 + X8 + ary7, data = my.df)
ar1.y8 <- lm(Y8 ~ X3 + X7 + X8 + ary8, data = my.df)
#print("weighted means auto-regression done")
WeightedMeanCoeffs1[i,]=unname(ar1.y1$coefficients);WeightedMeanCoeffs2[i,]=unname(ar1.y2$coefficients);
WeightedMeanCoeffs3[i,]=unname(ar1.y3$coefficients);WeightedMeanCoeffs4[i,]=unname(ar1.y4$coefficients);
WeightedMeanCoeffs5[i,]=unname(ar1.y5$coefficients);WeightedMeanCoeffs6[i,]=unname(ar1.y6$coefficients);
WeightedMeanCoeffs7[i,]=unname(ar1.y7$coefficients);WeightedMeanCoeffs8[i,]=unname(ar1.y8$coefficients);

#print("weighted-mean coeffs extracted")

    for (j in 1:400) {                     # For each cell...

      ns1 <- dists[j,] < AC1 & dists[j,] != 0   # Identify cells within autocorrelation range.
      ns2 <- dists[j,] < AC2 & dists[j,] != 0
      ns3 <- dists[j,] < AC3 & dists[j,] != 0
      ns4 <- dists[j,] < AC4 & dists[j,] != 0
      ns5 <- dists[j,] < AC5 & dists[j,] != 0
      ns6 <- dists[j,] < AC6 & dists[j,] != 0
      ns7 <- dists[j,] < AC7 & dists[j,] != 0
      ns8 <- dists[j,] < AC8 & dists[j,] != 0

      ary1[j] <- sum(as.vector(Y1)[ns1]/dists[j,ns1])    # sum Y values weighted by 1/distance
      ary2[j] <- sum(as.vector(Y2)[ns2]/dists[j,ns2])    #           for cells with AC range.
      ary3[j] <- sum(as.vector(Y3)[ns3]/dists[j,ns3])
      ary4[j] <- sum(as.vector(Y4)[ns4]/dists[j,ns4])
      ary5[j] <- sum(as.vector(Y5)[ns5]/dists[j,ns5])
      ary6[j] <- sum(as.vector(Y6)[ns6]/dists[j,ns6])
      ary7[j] <- sum(as.vector(Y7)[ns7]/dists[j,ns7])
      ary8[j] <- sum(as.vector(Y8)[ns8]/dists[j,ns8])
    }
#print("weighted sums calculated")

  ar1.y1 <- lm(Y1 ~ X1 + X2 + X3 + X4 + X5 + X6 + ary1, data = my.df) # unconstrained autocovariate regression
  ar1.y2 <- lm(Y2 ~ X1 + X2 + X3 + X4 + X5 + X6 + ary2, data = my.df) # using weighted-sum autocovariates
  ar1.y3 <- lm(Y3 ~ X1 + X2 + X3 + X4 + X5 + X6 + ary3, data = my.df)
  ar1.y4 <- lm(Y4 ~ X1 + X2 + X3 + X4 + X5 + X6 + ary4, data = my.df)
  ar1.y5 <- lm(Y5 ~ X1 + X7 + X8 + ary5, data = my.df)
  ar1.y6 <- lm(Y6 ~ X1 + X7 + X8 + ary6, data = my.df)
  ar1.y7 <- lm(Y7 ~ X3 + X7 + X8 + ary7, data = my.df)
  ar1.y8 <- lm(Y8 ~ X3 + X7 + X8 + ary8, data = my.df)
#print("weighted-sum auto-regression done")

WeightedSumCoeffs1[i,]=unname(ar1.y1$coefficients);WeightedSumCoeffs2[i,]=unname(ar1.y2$coefficients);
WeightedSumCoeffs3[i,]=unname(ar1.y3$coefficients);WeightedSumCoeffs4[i,]=unname(ar1.y4$coefficients);
WeightedSumCoeffs5[i,]=unname(ar1.y5$coefficients);WeightedSumCoeffs6[i,]=unname(ar1.y6$coefficients);
WeightedSumCoeffs7[i,]=unname(ar1.y7$coefficients);WeightedSumCoeffs8[i,]=unname(ar1.y8$coefficients);
#print("weighted-sum coeffs extracted")

  r1.y1 <- lm(Y1 ~ X1 + X2 + X3 + X4 + X5 + X6, data = my.df) # linear regression
  r1.y2 <- lm(Y2 ~ X1 + X2 + X3 + X4 + X5 + X6, data = my.df)
  r1.y3 <- lm(Y3 ~ X1 + X2 + X3 + X4 + X5 + X6, data = my.df)
  r1.y4 <- lm(Y4 ~ X1 + X2 + X3 + X4 + X5 + X6, data = my.df)
  r1.y5 <- lm(Y5 ~ X1 + X7 + X8, data = my.df)
  r1.y6 <- lm(Y6 ~ X1 + X7 + X8, data = my.df)
  r1.y7 <- lm(Y7 ~ X3 + X7 + X8, data = my.df)
  r1.y8 <- lm(Y8 ~ X3 + X7 + X8, data = my.df)
#print("simple lm done")

lmCoeffs1[i,]=unname(r1.y1$coefficients);lmCoeffs2[i,]=unname(r1.y2$coefficients);
lmCoeffs3[i,]=unname(r1.y3$coefficients);lmCoeffs4[i,]=unname(r1.y4$coefficients);
lmCoeffs5[i,]=unname(r1.y5$coefficients);lmCoeffs6[i,]=unname(r1.y6$coefficients);
lmCoeffs7[i,]=unname(r1.y7$coefficients);lmCoeffs8[i,]=unname(r1.y8$coefficients);
#print("simple lm coeffs extracted")
}

save(ACmat,file ="ACtable.RData")
save.image("UnconstrainedMPLE.RData")
```

# ConstrainedMPLE.R

```r
# Auto-normal estimation:
# constrained MPLE using weighted-sum autocovariates
# Author: David C Bardos
# Uses package RandomFields version 2.0.71 due to use of functions from
# Beale et al (2010) code; consequently it runs in R version 2.15.3
# Requires R file "ELE_1422_sm_HelperFunctions.r" from the
# supplementary information of Beale et al (2010) available at
# http://onlinelibrary.wiley.com/doi/10.1111/j.1461-0248.2009.01422.x/suppinfo
# (the main Beale at al (2010) code is also obtainable at that link)

# for 1000 iterations:

# 1. generates 8 simulated datasets using data generation code from Beale et al (2010);

# 2. runs unconstrained MPLE for each simulated dataset
#    (code from Beale et al (2010) is modified to use weighted sum autocovariates)

# 3. tests for positive-definiteness; if test fails, runs constrained MPLE

# For simulations generated at iteration i:

# the vectors:
# WeightedSumCCoeffs1[i,], WeightedSumCCoeffs2[i,], WeightedSumCCoeffs3[i,] & WeightedSumCCoeffs4[i,]
# give parameter estimates for scenarios 1 to 4, estimated using weighted sum autocovariates and
# (where necessary) constrained maximization, with format:
# WeightedSumCCoeffs1[i,]={intercept, beta_X1, beta_X2, beta_X3, beta_X4, beta_X5, beta_X6, beta_auto}

# corresponding estimates for scenarios 5 & 6 are stored in
# WeightedSumCCoeffs5[i,], WeightedSumCCoeffs6[i,] with format:
# WeightedSumCCoeffs5[i,]={intercept, beta_X1, beta_X7, beta_X8, beta_auto}

# corresponding estimates for scenarios 7 & 8 are stored in
# WeightedSumCCoeffs7[i,], WeightedSumCCoeffs8[i,] with format:
# WeightedSumCCoeffs5[i,]={intercept, beta_X3, beta_X7, beta_X8, beta_auto}

###### set up initial conditions and source additional functions
setwd("C:/CovariateEffect") #Change as required.
library(RandomFields)
library(MASS);library(lattice);
source("ELE_1422_sm_HelperFunctions.r")

x <- rep(1:20, times = 20)                              # Set up coordinates.
y <- rep(1:20, each = 20)
dists <- as.matrix(dist(cbind(x,y), upper = T, diag = T))         # Calculate distance matrix.

################################################################################

Nitr=1000;
# load autocorrelation distances
load("ACtable.RData")
# unconstrained MPLE code must be run 1st to create this file

WeightedSumCCoeffs1=matrix(nrow=Nitr,ncol=8);  #const + X1 + X2 + X3 + X4 + X5 + X6 + ary1
WeightedSumCCoeffs2=WeightedSumCCoeffs3=WeightedSumCCoeffs4=WeightedSumCCoeffs1

WeightedSumCCoeffs5=matrix(nrow=Nitr,ncol=5);  #const + X1 + X7 + X8 + ary5
WeightedSumCCoeffs6=WeightedSumCCoeffs7=WeightedSumCCoeffs8=WeightedSumCCoeffs5

Ysig=matrix(0,nrow=Nitr,ncol=8);lwr=matrix(-1,nrow=Nitr,ncol=8);convrg=matrix(NA,nrow=Nitr,ncol=8);

### main loop which runs for Nitr iterations
for (i in 1:Nitr){

  print(i);
  set.seed(i)                                     # Set seed for repeatability.
```

```r
X1 <- multiPattern(n=2, dim=c(20,20), taus = c(.3,7), corr=0.8)        # Generates 2 correlated exponential fields
X2 <- matrix(scale(as.vector(X1[[2]])), 20)                             # Scale fields to mean 0, var 1.
X1 <- matrix(scale(as.vector(X1[[1]])), 20)
X3 <- matrix(scale(as.vector(GaussRF (1:20,1:20,param = c(1,5,1,3), method =# Generate uncorrelated field with intermediate
       "cutoff CE", model = "exponential", grid = T))), 20)             #                    scale.
X4 <- multiPattern(n=2, dim=c(20,20), taus = c(.3,7), corr=0.8)         # Repeat above for next two variables but adding
X5 <- matrix( scale( as.vector(X4[[2]] + matrix( rnorm( 20*20, sd = 1),  #         white noise to reduce strength of
       ncol = 20))), 20)                                                #         spatial autocorrelation.
X4 <- matrix( scale( as.vector(X4[[1]] + matrix( rnorm( 20*20, sd = 1),
       ncol = 20))), 20)
X6 <- matrix(scale(as.vector(GaussRF (1:20,1:20,param = c(1,5,1,3), method =
       "cutoff CE", model = "exponential", grid = T) +
       matrix(rnorm(20*20, sd = 1), ncol = 20))), 20)
X7 <- X1 * seq(0, 1, len = 20) + X2 * seq(1, 0, len = 20)               # Generate Field with variable scale in AC
X8 <- matrix(scale(as.vector(X3 + seq(min(X3), max(X3), len = 20))), 20)

Y1E <- Y1 <- matrix(scale(as.vector(GaussRF (1:20,1:20,param = c(1,5,1,3),  # Generate 2 random fields to become the error
       method = "cutoff CE", model = "exponential", grid = T))), 20)    #            in the Y variables.
Y2E <- Y2 <- matrix(scale(as.vector(GaussRF (1:20,1:20,param = c(1,5,1,0.7),
       method ="cutoff CE", model = "exponential", grid = T))), 20)
Y3 <- matrix(scale(as.vector(Y1 +  matrix(rnorm(20*20, sd = 2), ncol = 20)),# Add white noise to weaken AC in error term and
       20) - 2 + 0.5 * X1 + 0.5 * X3 + 0.5 * X4 + 0.5 * X5             #            add deterministic parts.
Y3E <- Y3 + 2 - 0.5 * X1 - 0.5 * X3 - 0.5 * X4 - 0.5 * X5
Y4 <- matrix(scale(as.vector(Y2 +  matrix(rnorm(20*20, sd = 2), ncol = 20))),
       20) - 2 + 0.5 * X1 + 0.5 * X3 + 0.5 * X4 + 0.5 * X5
Y4E <- Y4 + 2 - 0.5 * X1 - 0.5 * X3 - 0.5 * X4 - 0.5 * X5
Y1 <- Y1 - 2 + 0.5 * X1 + 0.5 * X3 + 0.5 * X4 + 0.5 * X5                # Add only deterministic parts.
Y2 <- Y2 - 2 + 0.5 * X1 + 0.5 * X3 + 0.5 * X4 + 0.5 * X5
Y5 <- matrix(scale(as.vector(GaussRF (1:20,1:20,param = c(1,5,1,1.5),    #    Fine scale error only
       method ="cutoff CE", model = "exponential", grid = T))), 20) -
       2 + 0.5 * X7 +0.5 * X8
Y5E <- Y5 + 2 - 0.5 * X7 - 0.5 * X8
Y6 <- matrix(scale(as.vector(GaussRF (1:20,1:20,param = c(1,5,1,10),     #    Medium scale error only
       method = "cutoff CE", model = "exponential", grid = T))), 20) -
       2 + 0.5 * X7 +0.5 * X8
Y6E <- Y6 + 2 - 0.5 * X7 - 0.5 * X8
X11 <- multiPattern(n=2, dim=c(20,20), taus = c(.3,7), corr=0.8)         # Generates 2 correlated exponential fields
X12 <- matrix(scale(as.vector(X11[[2]])), 20)                             # Scale fields to mean 0, var 1.
X11 <- matrix(scale(as.vector(X11[[1]])), 20)
Y7 <- X11 * seq(0, 1, len = 20) + X12 * seq(1, 0, len = 20) - 2 + 0.5 * X7 +
       0.5 * X8
Y7E <- Y7 + 2 - 0.5 * X7 - 0.5 * X8
Y8 <- matrix(scale(as.vector(X11 + seq(min(X11), max(X11), len = 20))), 20) -
       2 + 0.5 * X7 + 0.5 * X8
Y8E <- Y8 + 2 - 0.5 * X7 - 0.5 * X8

my.df <- data.frame(x = x, y = y, X1 =as.vector(X1), X2 =as.vector(X2), X3 =  # Combine all X and Y variables and coordinates
       as.vector(X3), X4 =as.vector(X4), X5 =as.vector(X5), X6 =       #            into one data.frame
       as.vector(X6), X7 = as.vector(X7), X8 = as.vector(X8), Y1 =
       as.vector(Y1), Y2 =as.vector(Y2), Y3 = as.vector(Y3), Y4 =
       as.vector(Y4), Y5 = as.vector(Y5), Y6 = as.vector(Y6), Y7 =
       as.vector(Y7), Y8 = as.vector(Y8))
#print("Variables created")

# Autocorrelation distances
AC1=ACmat[i,1];AC2=ACmat[i,2];AC3=ACmat[i,3];AC4=ACmat[i,4];
AC5=ACmat[i,5];AC6=ACmat[i,6];AC7=ACmat[i,7];AC8=ACmat[i,8];

ary1 <- ary2 <- ary3 <- ary4 <- ary5 <- ary6 <- ary7 <- ary8 <- numeric(400)       # Create vector for autoregressive Y variables.

for (j in 1:400) {                                 # For each cell...

  ns1 <- dists[j,] < AC1 & dists[j,] != 0                                # Identify cells within autocorrelation range.
  ns2 <- dists[j,] < AC2 & dists[j,] != 0
  ns3 <- dists[j,] < AC3 & dists[j,] != 0
  ns4 <- dists[j,] < AC4 & dists[j,] != 0
  ns5 <- dists[j,] < AC5 & dists[j,] != 0
  ns6 <- dists[j,] < AC6 & dists[j,] != 0
  ns7 <- dists[j,] < AC7 & dists[j,] != 0
  ns8 <- dists[j,] < AC8 & dists[j,] != 0

  ary1[j] <- sum(as.vector(Y1)[ns1]/dists[j,ns1])          # sum Y values weighted by 1/distance
```

```
      ary2[j] <- sum(as.vector(Y2)[ns2]/dists[j,ns2])           #            for cells with AC range.
      ary3[j] <- sum(as.vector(Y3)[ns3]/dists[j,ns3])
      ary4[j] <- sum(as.vector(Y4)[ns4]/dists[j,ns4])
      ary5[j] <- sum(as.vector(Y5)[ns5]/dists[j,ns5])
      ary6[j] <- sum(as.vector(Y6)[ns6]/dists[j,ns6])
      ary7[j] <- sum(as.vector(Y7)[ns7]/dists[j,ns7])
      ary8[j] <- sum(as.vector(Y8)[ns8]/dists[j,ns8])
   }
#print("weighted sums calculated")

   ar1.y1 <- lm(Y1 ~ X1 + X2 + X3 + X4 + X5 + X6 + ary1, data = my.df)      # unconstrained autocovariate regression
   ar1.y2 <- lm(Y2 ~ X1 + X2 + X3 + X4 + X5 + X6 + ary2, data = my.df)
   ar1.y3 <- lm(Y3 ~ X1 + X2 + X3 + X4 + X5 + X6 + ary3, data = my.df)
   ar1.y4 <- lm(Y4 ~ X1 + X2 + X3 + X4 + X5 + X6 + ary4, data = my.df)
   ar1.y5 <- lm(Y5 ~ X1 + X7 + X8 + ary5, data = my.df)
   ar1.y6 <- lm(Y6 ~ X1 + X7 + X8 + ary6, data = my.df)
   ar1.y7 <- lm(Y7 ~ X3 + X7 + X8 + ary7, data = my.df)
   ar1.y8 <- lm(Y8 ~ X3 + X7 + X8 + ary8, data = my.df)
#print("weighted sums auto-covariate regression done")
Ysig[i,1]=sd(residuals(ar1.y1));Ysig[i,2]=sd(residuals(ar1.y2));Ysig[i,3]=sd(residuals(ar1.y3));Ysig[i,4]=sd(residuals(ar1.y4));
Ysig[i,5]=sd(residuals(ar1.y5));Ysig[i,6]=sd(residuals(ar1.y6));Ysig[i,7]=sd(residuals(ar1.y7));Ysig[i,8]=sd(residuals(ar1.y8));

WeightedSumCCoeffs1[i,]=unname(ar1.y1$coefficients);WeightedSumCCoeffs2[i,]=unname(ar1.y2$coefficients);
WeightedSumCCoeffs3[i,]=unname(ar1.y3$coefficients);WeightedSumCCoeffs4[i,]=unname(ar1.y4$coefficients);
WeightedSumCCoeffs5[i,]=unname(ar1.y5$coefficients);WeightedSumCCoeffs6[i,]=unname(ar1.y6$coefficients);
WeightedSumCCoeffs7[i,]=unname(ar1.y7$coefficients);WeightedSumCCoeffs8[i,]=unname(ar1.y8$coefficients);
#print("weighted sums coeffs extracted")

# the following code corrects the estimates where necessary, using constrained MPLE to ensure positive-definiteness

# Y1 positive-definiteness
wsc=WeightedSumCCoeffs1[i,]; #1
alphaWS=wsc[1]+wsc[2]*X1+wsc[3]*X2+wsc[4]*X3+wsc[5]*X4+wsc[6]*X5+wsc[7]*X6;
sd=Ysig[i,1]; #1
alpha.vec=as.vector(alphaWS);
beta=matrix(0,400,400);
for (m in 1:400) for(n in 1:400){dmn=dists[m,n];if (dmn<AC1 && dmn>0) beta[m,n]=wsc[8]/dmn;} #1 #wsc
B=diag(400) - beta;
Binv=solve(B)
mu.vec=Binv%*%alpha.vec
Sigma=(sd^2)*Binv;
eval=eigen(Sigma)$values;
if(min(eval)<10^(-6)) {           # test for positive definiteness
   upper=1;lower=0.0;           # bisection search for max b.auto consistent with positive definiteness
   for(k in 1:20){mid=0.5*(upper+lower);beta.1=beta*mid;B=diag(400)-beta.1;
                  Binv=solve(B);Sigma=(sd^2)*Binv;eval=eigen(Sigma)$values;
                  if(min(eval)<10^(-6)) upper=mid else lower=mid
                  print(c(i,lower,upper));
                  }
   b.auto.max=wsc[8]*lower; # use bisection result to set max b.auto for constrained ML estimation
   # construct -log(pseudo-likelihood) function
   llike=function(coeffs){sum((Y1-coeffs[1]-coeffs[2]*X1-coeffs[3]*X2-coeffs[4]*X3-coeffs[5]*X4-
                  coeffs[6]*X5-coeffs[7]*X6-coeffs[8]*ary1)^2);} #1
   #constrained ML estimation
   cls=constrOptim(c(wsc[1:7],0.9*b.auto.max),llike,NULL,ui=rbind(c(0,0,0,0,0,0,0,-1)),ci=-b.auto.max,control=list(fnscale=1));
   WeightedSumCCoeffs1[i,]=cls$par;    #1
   convrg[i,1]=cls$convergence;      #1
   lwr[i,1]=lower; #1
            } #endif pos-def

# Y2 positive-definiteness
wsc=WeightedSumCCoeffs2[i,]; #2
alphaWS=wsc[1]+wsc[2]*X1+wsc[3]*X2+wsc[4]*X3+wsc[5]*X4+wsc[6]*X5+wsc[7]*X6;
sd=Ysig[i,2]; #2
alpha.vec=as.vector(alphaWS);
beta=matrix(0,400,400);
for (m in 1:400) for(n in 1:400){dmn=dists[m,n];if (dmn<AC2 && dmn>0) beta[m,n]=wsc[8]/dmn;} #2 #wsc
B=diag(400) - beta;
Binv=solve(B)
mu.vec=Binv%*%alpha.vec
Sigma=(sd^2)*Binv;
eval=eigen(Sigma)$values;
```

```r
if(min(eval)<10^(-6)) {             # test for positive definiteness
   upper=1;lower=0.0;                # bisection search for max b.auto consistent with positive definiteness
   for(k in 1:20){mid=0.5*(upper+lower);beta.1=beta*mid;B=diag(400)-beta.1;
                  Binv=solve(B);Sigma=(sd^2)*Binv;eval=eigen(Sigma)$values;
                  if(min(eval)<10^(-6)) upper=mid else lower=mid
                  print(c(i,lower,upper));
                  }
   b.auto.max=wsc[8]*lower;  # use bisection result to set max b.auto for constrained ML estimation
   # construct -log(pseudo-likelihood) function
   llike=function(coeffs){sum((Y2-coeffs[1]-coeffs[2]*X1-coeffs[3]*X2-coeffs[4]*X3-coeffs[5]*X4-
                    coeffs[6]*X5-coeffs[7]*X6-coeffs[8]*ary2)^2);} #2
   #constrained ML estimation
   cls=constrOptim(c(wsc[1:7],0.9*b.auto.max),llike,NULL,ui=rbind(c(0,0,0,0,0,0,0,-1)),ci=-b.auto.max,control=list(fnscale=1));
   WeightedSumCCoeffs2[i,]=cls$par;   #2
   convrg[i,2]=cls$convergence;     #2
   lwr[i,2]=lower; #2
             } #endif pos-def

# Y3 positive-definiteness
wsc=WeightedSumCCoeffs3[i,]; #3
alphaWS=wsc[1]+wsc[2]*X1+wsc[3]*X2+wsc[4]*X3+wsc[5]*X4+wsc[6]*X5+wsc[7]*X6;
sd=Ysig[i,3]; #3
alpha.vec=as.vector(alphaWS);
beta=matrix(0,400,400);
for (m in 1:400) for(n in 1:400){dmn=dists[m,n];if (dmn<AC3 && dmn>0) beta[m,n]=wsc[8]/dmn;}  #3 #wsc
B=diag(400) - beta;
Binv=solve(B)
mu.vec=Binv%*%alpha.vec
Sigma=(sd^2)*Binv;
eval=eigen(Sigma)$values;
if(min(eval)<10^(-6)) {             # test for positive definiteness
   upper=1;lower=0.0;                # bisection search for max b.auto consistent with positive definiteness
   for(k in 1:20){mid=0.5*(upper+lower);beta.1=beta*mid;B=diag(400)-beta.1;
                  Binv=solve(B);Sigma=(sd^2)*Binv;eval=eigen(Sigma)$values;
                  if(min(eval)<10^(-6)) upper=mid else lower=mid
                  print(c(i,lower,upper));
                  }
   b.auto.max=wsc[8]*lower;  # use bisection result to set max b.auto for constrained ML estimation
   # construct -log(pseudo-likelihood) function
   llike=function(coeffs){sum((Y3-coeffs[1]-coeffs[2]*X1-coeffs[3]*X2-coeffs[4]*X3-coeffs[5]*X4-
                    coeffs[6]*X5-coeffs[7]*X6-coeffs[8]*ary3)^2);} #3
   #constrained ML estimation
   cls=constrOptim(c(wsc[1:7],0.9*b.auto.max),llike,NULL,ui=rbind(c(0,0,0,0,0,0,0,-1)),ci=-b.auto.max,control=list(fnscale=1));
   WeightedSumCCoeffs3[i,]=cls$par;   #3
   convrg[i,3]=cls$convergence;     #3
   lwr[i,3]=lower; #3
             } #endif pos-def

# Y4 positive-definiteness
wsc=WeightedSumCCoeffs4[i,]; #4
alphaWS=wsc[1]+wsc[2]*X1+wsc[3]*X2+wsc[4]*X3+wsc[5]*X4+wsc[6]*X5+wsc[7]*X6;
sd=Ysig[i,4]; #4
alpha.vec=as.vector(alphaWS);
beta=matrix(0,400,400);
for (m in 1:400) for(n in 1:400){dmn=dists[m,n];if (dmn<AC4 && dmn>0) beta[m,n]=wsc[8]/dmn;}  #4 #wsc
B=diag(400) - beta;
Binv=solve(B)
mu.vec=Binv%*%alpha.vec
Sigma=(sd^2)*Binv;
eval=eigen(Sigma)$values;
if(min(eval)<10^(-6)) {             # test for positive definiteness
   upper=1;lower=0.0;                # bisection search for max b.auto consistent with positive definiteness
   for(k in 1:20){mid=0.5*(upper+lower);beta.1=beta*mid;B=diag(400)-beta.1;
                  Binv=solve(B);Sigma=(sd^2)*Binv;eval=eigen(Sigma)$values;
                  if(min(eval)<10^(-6)) upper=mid else lower=mid
                  print(c(i,lower,upper));
                  }
   b.auto.max=wsc[8]*lower;  # use bisection result to set max b.auto for constrained ML estimation
   # construct -log(pseudo-likelihood) function
   llike=function(coeffs){sum((Y4-coeffs[1]-coeffs[2]*X1-coeffs[3]*X2-coeffs[4]*X3-coeffs[5]*X4-
                    coeffs[6]*X5-coeffs[7]*X6-coeffs[8]*ary4)^2);} #4
   #constrained ML estimation
   cls=constrOptim(c(wsc[1:7],0.9*b.auto.max),llike,NULL,ui=rbind(c(0,0,0,0,0,0,0,-1)),ci=-b.auto.max,control=list(fnscale=1));
   WeightedSumCCoeffs4[i,]=cls$par;   #4
   convrg[i,4]=cls$convergence;     #4
```

```r
       lwr[i,4]=lower; #4
                  } #endif pos-def

# Y5 positive-definiteness
wsc=WeightedSumCCoeffs5[i,]; #5
alphaWS=wsc[1]+wsc[2]*X1+wsc[3]*X7+wsc[4]*X8; # !!!!!
sd=Ysig[i,5]; #5
alpha.vec=as.vector(alphaWS);
beta=matrix(0,400,400);
for (m in 1:400) for(n in 1:400){dmn=dists[m,n];if (dmn<AC5 && dmn>0) beta[m,n]=wsc[5]/dmn;} #5 #wsc
B=diag(400) - beta;
Binv=solve(B)
mu.vec=Binv%*%alpha.vec
Sigma=(sd^2)*Binv;
eval=eigen(Sigma)$values;
if(min(eval)<10^(-6)) {           # test for positive definiteness
   upper=1;lower=0.0;             # bisection search for max b.auto consistent with positive definiteness
   for(k in 1:20){mid=0.5*(upper+lower);beta.1=beta*mid;B=diag(400)-beta.1;
                  Binv=solve(B);Sigma=(sd^2)*Binv;eval=eigen(Sigma)$values;
                  if(min(eval)<10^(-6)) upper=mid else lower=mid
                  print(c(i,lower,upper));
                                }
   b.auto.max=wsc[5]*lower;  # use bisection result to set max b.auto for constrained ML estimation
   # construct -log(pseudo-likelihood) function
   llike=function(coeffs){sum((Y5-coeffs[1]-coeffs[2]*X1-coeffs[3]*X7-coeffs[4]*X8-coeffs[5]*ary5)^2);} #5
   #constrained ML estimation
   cls=constrOptim(c(wsc[1:4],0.9*b.auto.max),llike,NULL,ui=rbind(c(0,0,0,0,-1)),ci=-b.auto.max,control=list(fnscale=1));
   WeightedSumCCoeffs5[i,]=cls$par;   #5
   convrg[i,5]=cls$convergence;      #5
   lwr[i,5]=lower; #5
                  } #endif pos-def

# Y6 positive-definiteness
wsc=WeightedSumCCoeffs6[i,]; #6
alphaWS=wsc[1]+wsc[2]*X1+wsc[3]*X7+wsc[4]*X8; # !!!!!
sd=Ysig[i,6]; #6
alpha.vec=as.vector(alphaWS);
beta=matrix(0,400,400);
for (m in 1:400) for(n in 1:400){dmn=dists[m,n];if (dmn<AC6 && dmn>0) beta[m,n]=wsc[5]/dmn;} #6 #wsc
B=diag(400) - beta;
Binv=solve(B)
mu.vec=Binv%*%alpha.vec
Sigma=(sd^2)*Binv;
eval=eigen(Sigma)$values;
if(min(eval)<10^(-6)) {           # test for positive definiteness
   upper=1;lower=0.0;             # bisection search for max b.auto consistent with positive definiteness
   for(k in 1:20){mid=0.5*(upper+lower);beta.1=beta*mid;B=diag(400)-beta.1;
                  Binv=solve(B);Sigma=(sd^2)*Binv;eval=eigen(Sigma)$values;
                  if(min(eval)<10^(-6)) upper=mid else lower=mid
                  print(c(i,lower,upper));
                                }
   b.auto.max=wsc[5]*lower;  # use bisection result to set max b.auto for constrained ML estimation
   # construct -log(pseudo-likelihood) function
   llike=function(coeffs){sum((Y6-coeffs[1]-coeffs[2]*X1-coeffs[3]*X7-coeffs[4]*X8-coeffs[5]*ary6)^2);} #6
   #constrained ML estimation
   cls=constrOptim(c(wsc[1:4],0.9*b.auto.max),llike,NULL,ui=rbind(c(0,0,0,0,-1)),ci=-b.auto.max,control=list(fnscale=1));
   WeightedSumCCoeffs6[i,]=cls$par;   #6
   convrg[i,6]=cls$convergence;      #6
   lwr[i,6]=lower;  #6
                  } #endif pos-def

# Y7 positive-definiteness
wsc=WeightedSumCCoeffs7[i,]; #7
alphaWS=wsc[1]+wsc[2]*X3+wsc[3]*X7+wsc[4]*X8; # !!!!!
sd=Ysig[i,7];
alpha.vec=as.vector(alphaWS);
beta=matrix(0,400,400);
for (m in 1:400) for(n in 1:400){dmn=dists[m,n];if (dmn<AC7 && dmn>0) beta[m,n]=wsc[5]/dmn;} #7 #wsc
B=diag(400) - beta;
Binv=solve(B)
mu.vec=Binv%*%alpha.vec
Sigma=(sd^2)*Binv;
eval=eigen(Sigma)$values;
if(min(eval)<10^(-6)) {           # test for positive definiteness
```

```r
        upper=1;lower=0.0;           # bisection search for max b.auto consistent with positive definiteness
        for(k in 1:20){mid=0.5*(upper+lower);beta.1=beta*mid;B=diag(400)-beta.1;
                       Binv=solve(B);Sigma=(sd^2)*Binv;eval=eigen(Sigma)$values;
                       if(min(eval)<10^(-6)) upper=mid else lower=mid
                       print(c(i,lower,upper));
                                         }
        b.auto.max=wsc[5]*lower;  # use bisection result to set max b.auto for constrained ML estimation
        # construct -log(pseudo-likelihood) function
        llike=function(coeffs){sum((Y7-coeffs[1]-coeffs[2]*X3-coeffs[3]*X7-coeffs[4]*X8-coeffs[5]*ary7)^2);} #7
        #constrained ML estimation
        cls=constrOptim(c(wsc[1:4],0.9*b.auto.max),llike,NULL,ui=rbind(c(0,0,0,0,-1)),ci=-b.auto.max,control=list(fnscale=1));
        WeightedSumCCoeffs7[i,]=cls$par;    #7
        convrg[i,7]=cls$convergence;        #7
        lwr[i,7]=lower; #7
                    } #endif pos-def

# Y8 positive-definiteness
wsc=WeightedSumCCoeffs8[i,]; #8
alphaWS=wsc[1]+wsc[2]*X3+wsc[3]*X7+wsc[4]*X8; # !!!!!
sd=Ysig[i,8]; #8
alpha.vec=as.vector(alphaWS);
beta=matrix(0,400,400);
for (m in 1:400) for(n in 1:400){dmn=dists[m,n];if (dmn<AC8 && dmn>0) beta[m,n]=wsc[5]/dmn;} #8 #wsc
B=diag(400) - beta;
Binv=solve(B)
mu.vec=Binv%*%alpha.vec
Sigma=(sd^2)*Binv;
eval=eigen(Sigma)$values;
if(min(eval)<10^(-6)) {            # test for positive definiteness
    upper=1;lower=0.0;           # bisection search for max b.auto consistent with positive definiteness
    for(k in 1:20){mid=0.5*(upper+lower);beta.1=beta*mid;B=diag(400)-beta.1;
                   Binv=solve(B);Sigma=(sd^2)*Binv;eval=eigen(Sigma)$values;
                   if(min(eval)<10^(-6)) upper=mid else lower=mid
                   print(c(i,lower,upper));
                                     }
    b.auto.max=wsc[5]*lower;  # use bisection result to set max b.auto for constrained ML estimation
    # construct -log(pseudo-likelihood) function
    llike=function(coeffs){sum((Y8-coeffs[1]-coeffs[2]*X3-coeffs[3]*X7-coeffs[4]*X8-coeffs[5]*ary8)^2);} #8
    #constrained ML estimation
    cls=constrOptim(c(wsc[1:4],0.9*b.auto.max),llike,NULL,ui=rbind(c(0,0,0,0,-1)),ci=-b.auto.max,control=list(fnscale=1));
    WeightedSumCCoeffs8[i,]=cls$par;    #8
    convrg[i,8]=cls$convergence;        #8
    lwr[i,8]=lower; #8
                } #endif pos-def

} # end main loop (i)

save.image("ConstrainedMPLE.RData");
```

# auto-normal covariate effect.R

```r
# Evaluates covariate impact, covariate effect and standardized covariate effect for
# covariate X8 in simulated data example i=180 for scenario 6 from Beale et al (2010)
# This example from Beale et al (2010) exhibits the strongest covariate amplification
# and therefore the strongest covariate parameter attenuation.
# Since auto-normal models are multivariate normal, simulations at estimated
# parameters (required for evaluating covariate influence) are obtained directly.
# (Gibbs sampling can be used but is not necessary for this model).
# Author: David C Bardos
# Uses package RandomFields version 2.0.71 due to use of functions from
# Beale et al (2010) code; consequently it runs in R version 2.15.3
# Requires R file "ELE_1422_sm_HelperFunctions.r" from the
# supplementary information of Beale et al (2010) available at
# http://onlinelibrary.wiley.com/doi/10.1111/j.1461-0248.2009.01422.x/suppinfo
# (the main Beale at al (2010) code is also obtainable at that link)

# This code:
# 1. generates the dataset (covariates and observations) for scenario 6, i=180
#    from Beale et al (2010); using simulation code from Beale et al (2010);

# 2. runs unconstrained MPLE for the dataset, with weighted-sum autocovariates (code
#    from Beale et al (2010) is modified to use weighted sum autocovariates and a
#    neighbourhood-size bug is corrected). Note: constrained MPLE is not required
#    for this example.

# 3. fits a linear model to the dataset, using code from Beale et al (2010)

# 4. runs simulations at the estimated parameters, which are used to evaluate
#    measures of covariate influence for the X8 covariate in the auto-normal model;
#    measures of covariate influence are also evaluated in the linear model.

library(lattice);library(MASS);

###### set up initial conditions and source additional functions
setwd("C:/CovariateEffect") #Change as required.
library(RandomFields)
library(MASS)
source("ELE_1422_sm_HelperFunctions.r")

x <- rep(1:20, times = 20)                              # Set up coordinates.
y <- rep(1:20, each = 20)
dists <- as.matrix(dist(cbind(x,y), upper = T, diag = T))    # Calculate distance matrix.
xy=cbind(x,y);

################################################################################
key=180;  # i=180 example for scenario 6
nbr=matrix(F,400,400)

Nitr=1000;
load("ACtable.RData");

WeightedSumCoeffs1=matrix(nrow=Nitr,ncol=8);  #const + X1 + X2 + X3 + X4 + X5 + X6 + ary1
WeightedSumCoeffs2=WeightedSumCoeffs3=WeightedSumCoeffs4=WeightedSumCoeffs1

WeightedSumCoeffs5=matrix(nrow=Nitr,ncol=5);  #const + X1 + X7 + X8 + ary5
WeightedSumCoeffs6=WeightedSumCoeffs7=WeightedSumCoeffs8=WeightedSumCoeffs5

lmCoeffs1=matrix(nrow=Nitr,ncol=7);  #const + X1 + X2 + X3 + X4 + X5 + X6
lmCoeffs2=lmCoeffs3=lmCoeffs4=lmCoeffs1

lmCoeffs5=matrix(nrow=Nitr,ncol=4);  #const + X1 + X7 + X8
lmCoeffs6=lmCoeffs7=lmCoeffs8=lmCoeffs5

Y6figs=list(); Y6Efigs=list();X1figs=list();X7figs=list(); X8figs=list();
Y6sig=rep(0,Nitr);

i=key;  # loop removed, single iteration
  print(i);
  set.seed(i)                                # Set seed for repeatability.
  X1 <- multiPattern(n=2, dim=c(20,20), taus = c(.3,7), corr=0.8)     # Generates 2 correlated exponential fields
  X2 <- matrix(scale(as.vector(X1[[2]])), 20)           # Scale fields to mean 0, var 1.
  X1 <- matrix(scale(as.vector(X1[[1]])), 20)
```

```r
X3 <- matrix(scale(as.vector(GaussRF (1:20,1:20,param = c(1,5,1,3), method =# Generate uncorrelated field with intermediate
    "cutoff CE", model = "exponential", grid = T))), 20)           #                    scale.
X4 <- multiPattern(n=2, dim=c(20,20), taus = c(.3,7), corr=0.8)         # Repeat above for next two variables but adding
X5 <- matrix( scale( as.vector(X4[[2]] + matrix( rnorm( 20*20, sd = 1),    #      white noise to reduce strength of
    ncol = 20))), 20)                                 #              spatial autocorrelation.
X4 <- matrix( scale( as.vector(X4[[1]] + matrix( rnorm( 20*20, sd = 1),
    ncol = 20))), 20)
X6 <- matrix(scale(as.vector(GaussRF (1:20,1:20,param = c(1,5,1,3), method =
    "cutoff CE", model = "exponential", grid = T) +
    matrix(rnorm(20*20, sd = 1), ncol = 20))), 20)
X7 <- X1 * seq(0, 1, len = 20) + X2 * seq(1, 0, len = 20)       # Generate Field with variable scale in AC
X8 <- matrix(scale(as.vector(X3 + seq(min(X3), max(X3), len = 20))), 20)

Y1E <- Y1 <- matrix(scale(as.vector(GaussRF (1:20,1:20,param = c(1,5,1,3),  # Generate 2 random fields to become the error
    method = "cutoff CE", model = "exponential", grid = T))), 20)     #       in the Y variables.
Y2E <- Y2 <- matrix(scale(as.vector(GaussRF (1:20,1:20,param = c(1,5,1,0.7),
    method ="cutoff CE", model = "exponential", grid = T))), 20)
Y3 <- matrix(scale(as.vector(Y1 +  matrix(rnorm(20*20, sd = 2), ncol = 20))),# Add white noise to weaken AC in error term and
    20) - 2 + 0.5 * X1 + 0.5 * X3 + 0.5 * X4 + 0.5 * X5       #             add deterministic parts.
Y3E <- Y3 + 2 - 0.5 * X1 - 0.5 * X3 - 0.5 * X4 - 0.5 * X5
Y4 <- matrix(scale(as.vector(Y2 +  matrix(rnorm(20*20, sd = 2), ncol = 20))),
    20) - 2 + 0.5 * X1 + 0.5 * X3 + 0.5 * X4 + 0.5 * X5
Y4E <- Y4 + 2 - 0.5 * X1 - 0.5 * X3 - 0.5 * X4 - 0.5 * X5
Y1 <- Y1 - 2 + 0.5 * X1 + 0.5 * X3 + 0.5 * X4 + 0.5 * X5         # Add only deterministic parts.
Y2 <- Y2 - 2 + 0.5 * X1 + 0.5 * X3 + 0.5 * X4 + 0.5 * X5
Y5 <- matrix(scale(as.vector(GaussRF (1:20,1:20,param = c(1,5,1,1.5),   #     Fine scale error only
    method ="cutoff CE", model = "exponential", grid = T))), 20) -
    2 + 0.5 * X7 +0.5 * X8
Y5E <- Y5 + 2 - 0.5 * X7 - 0.5 * X8
Y6 <- matrix(scale(as.vector(GaussRF (1:20,1:20,param = c(1,5,1,10),    #     Medium scale error only
    method = "cutoff CE", model = "exponential", grid = T))), 20) -
    2 + 0.5 * X7 +0.5 * X8
Y6E <- Y6 + 2 - 0.5 * X7 - 0.5 * X8
X11 <- multiPattern(n=2, dim=c(20,20), taus = c(.3,7), corr=0.8)        # Generates 2 correlated exponential fields
X12 <- matrix(scale(as.vector(X11[[2]])), 20)              # Scale fields to mean 0, var 1.
X11 <- matrix(scale(as.vector(X11[[1]])), 20)
Y7 <- X11 * seq(0, 1, len = 20) + X12 * seq(1, 0, len = 20) - 2 + 0.5 * X7 +
    0.5 * X8
Y7E <- Y7 + 2 - 0.5 * X7 - 0.5 * X8
Y8 <- matrix(scale(as.vector(X11 + seq(min(X11), max(X11), len = 20))), 20) -
    2 + 0.5 * X7 + 0.5 * X8
Y8E <- Y8 + 2 - 0.5 * X7 - 0.5 * X8

my.df <- data.frame(x = x, y = y, X1 =as.vector(X1), X2 =as.vector(X2), X3 =  # Combine all X and Y variables and coordinates
    as.vector(X3), X4 =as.vector(X4), X5 =as.vector(X5), X6 =        #             into one data.frame
    as.vector(X6), X7 = as.vector(X7), X8 = as.vector(X8), Y1 =
    as.vector(Y1), Y2 =as.vector(Y2), Y3 = as.vector(Y3), Y4 =
    as.vector(Y4), Y5 = as.vector(Y5), Y6 = as.vector(Y6), Y7 =
    as.vector(Y7), Y8 = as.vector(Y8))
#print("Variables created")

Y6figs[[i]]=levelplot(Y6);  X1figs[[i]]=levelplot(X1);X7figs[[i]]=levelplot(X7); X8figs[[i]]=levelplot(X8);
Y6Efigs[[i]]=levelplot(Y6E);

AC1=ACmat[i,1];AC2=ACmat[i,2];AC3=ACmat[i,3];AC4=ACmat[i,4];
AC5=ACmat[i,5];AC6=ACmat[i,6];AC7=ACmat[i,7];AC8=ACmat[i,8];
# print("Autocorrelation distances done")

ary1 <- ary2 <- ary3 <- ary4 <- ary5 <- ary6 <- ary7 <- ary8 <- numeric(400)    # Create vector for autoregressive Y variables.

for (j in 1:400) {                          # For each cell...

ns1 <- dists[j,] < AC1 & dists[j,] != 0               # Identify cells within autocorrelation range.
ns2 <- dists[j,] < AC2 & dists[j,] != 0
ns3 <- dists[j,] < AC3 & dists[j,] != 0
ns4 <- dists[j,] < AC4 & dists[j,] != 0
ns5 <- dists[j,] < AC5 & dists[j,] != 0
ns6 <- dists[j,] < AC6 & dists[j,] != 0
ns7 <- dists[j,] < AC7 & dists[j,] != 0
ns8 <- dists[j,] < AC8 & dists[j,] != 0

nbr[,j]=ns6;

ary1[j] <- sum(as.vector(Y1)[ns1]/dists[j,ns1])         # sum Y values weighted by 1/distance
```

```r
      ary2[j] <- sum(as.vector(Y2)[ns2]/dists[j,ns2])          #           for cells with AC range.
      ary3[j] <- sum(as.vector(Y3)[ns3]/dists[j,ns3])
      ary4[j] <- sum(as.vector(Y4)[ns4]/dists[j,ns4])
      ary5[j] <- sum(as.vector(Y5)[ns5]/dists[j,ns5])
      ary6[j] <- sum(as.vector(Y6)[ns6]/dists[j,ns6])
      ary7[j] <- sum(as.vector(Y7)[ns7]/dists[j,ns7])
      ary8[j] <- sum(as.vector(Y8)[ns8]/dists[j,ns8])
   }
#print("weighted sums calculated")

  ar1.y1 <- lm(Y1 ~ X1 + X2 + X3 + X4 + X5 + X6 + ary1, data = my.df)   # unconstrained autocovariate regression
  ar1.y2 <- lm(Y2 ~ X1 + X2 + X3 + X4 + X5 + X6 + ary2, data = my.df)   # note:constrained MPLE not required in i=180 example
  ar1.y3 <- lm(Y3 ~ X1 + X2 + X3 + X4 + X5 + X6 + ary3, data = my.df)
  ar1.y4 <- lm(Y4 ~ X1 + X2 + X3 + X4 + X5 + X6 + ary4, data = my.df)
  ar1.y5 <- lm(Y5 ~ X1 + X7 + X8 + ary5, data = my.df)
  ar1.y6 <- lm(Y6 ~ X1 + X7 + X8 + ary6, data = my.df)
  ar1.y7 <- lm(Y7 ~ X3 + X7 + X8 + ary7, data = my.df)
  ar1.y8 <- lm(Y8 ~ X3 + X7 + X8 + ary8, data = my.df)
#print("weighted sums auto-regression done")
Y6sig[i]=sd(residuals(ar1.y6));

WeightedSumCoeffs1[i,]=unname(ar1.y1$coefficients);WeightedSumCoeffs2[i,]=unname(ar1.y2$coefficients);
WeightedSumCoeffs3[i,]=unname(ar1.y3$coefficients);WeightedSumCoeffs4[i,]=unname(ar1.y4$coefficients);
WeightedSumCoeffs5[i,]=unname(ar1.y5$coefficients);WeightedSumCoeffs6[i,]=unname(ar1.y6$coefficients);
WeightedSumCoeffs7[i,]=unname(ar1.y7$coefficients);WeightedSumCoeffs8[i,]=unname(ar1.y8$coefficients);
#print("weighted sums coeffs extracted")

  r1.y1 <- lm(Y1 ~ X1 + X2 + X3 + X4 + X5 + X6, data = my.df)          # linear regression
  r1.y2 <- lm(Y2 ~ X1 + X2 + X3 + X4 + X5 + X6, data = my.df)
  r1.y3 <- lm(Y3 ~ X1 + X2 + X3 + X4 + X5 + X6, data = my.df)
  r1.y4 <- lm(Y4 ~ X1 + X2 + X3 + X4 + X5 + X6, data = my.df)
  r1.y5 <- lm(Y5 ~ X1 + X7 + X8, data = my.df)
  r1.y6 <- lm(Y6 ~ X1 + X7 + X8, data = my.df)
  r1.y7 <- lm(Y7 ~ X3 + X7 + X8, data = my.df)
  r1.y8 <- lm(Y8 ~ X3 + X7 + X8, data = my.df)
#print("simple lm done")

lmCoeffs1[i,]=unname(r1.y1$coefficients);lmCoeffs2[i,]=unname(r1.y2$coefficients);
lmCoeffs3[i,]=unname(r1.y3$coefficients);lmCoeffs4[i,]=unname(r1.y4$coefficients);
lmCoeffs5[i,]=unname(r1.y5$coefficients);lmCoeffs6[i,]=unname(r1.y6$coefficients);
lmCoeffs7[i,]=unname(r1.y7$coefficients);lmCoeffs8[i,]=unname(r1.y8$coefficients);
#print("simple lm coeffs extracted")

#########################################################################
#######   evaluate covariate influence
#########################################################################

wsc=WeightedSumCoeffs6[i,];lmc=lmCoeffs6[i,];
alphaWS=wsc[1]+wsc[2]*X1+wsc[3]*X7+wsc[4]*X8;
alphaWS.zero=wsc[1]+wsc[2]*X1+wsc[3]*X7
alphaLM=lmc[1]+lmc[2]*X1+lmc[3]*X7+lmc[4]*X8;
alphaLM.zero=lmc[1]+lmc[2]*X1+lmc[3]*X7
sd=Y6sig[i];

alpha.vec=as.vector(alphaWS);
beta=matrix(0,400,400);
for (i in 1:400) for(j in 1:400){dij=dists[i,j];if (dij<AC6 && dij>0) beta[i,j]=wsc[5]/dij;}
B=diag(400) - beta;
Binv=solve(B)
mu.vec=Binv%*%alpha.vec
Sigma=(sd^2)*Binv;
Ndraws=1000
mvndraw=mvrnorm(Ndraws,mu.vec,Sigma,tol=1e-10)
mvnhat=matrix(colMeans(mvndraw),20,20)
levelplot(mvnhat)

alpha.vec=as.vector(alphaWS.zero);
beta=matrix(0,400,400);
for (i in 1:400) for(j in 1:400){dij=dists[i,j];if (dij<AC6 && dij>0) beta[i,j]=wsc[5]/dij;}
B=diag(400) - beta;
```

```
Binv=solve(B)
mu.vec=Binv%*%alpha.vec
Sigma=(sd^2)*Binv;
Ndraws=1000
mvndraw.zero=mvrnorm(Ndraws,mu.vec,Sigma,tol=1e-10)
mvnhat.zero=matrix(colMeans(mvndraw.zero),20,20)
levelplot(mvnhat.zero)

( covariate.impact.autonormal=sum(abs(mvnhat-mvnhat.zero))/400 )
( covariate.impact.LM=sum(abs(alphaLM-alphaLM.zero))/400 )
meanAbsX8=mean(abs(X8));sdX=sd(as.vector(X8));

( covariate.effect.autonormal=covariate.impact.autonormal/meanAbsX8 )
( standardized.covariate.effect.autonormal=covariate.effect.autonormal*sdX )

( covariate.effect.LM=covariate.impact.LM/meanAbsX8 )
(standardized.covariate.effect.LM=covariate.effect.LM*sdX )
# covariate X8 has variance=1, so standardizing makes no difference.
```

# Autologistic covariate effect.R

```
# Influence of altitude covariate in auto-logistic and logistic analyses of Hydrocotyle vulgaris presence-absence data
# Author: David C Bardos

# The first section performs autologistic estimation by maximum pseudo-likelihood; the code allows for Gibbs sampling
# of missing data, with alternating Gibbs sweeps and MPLE steps, i.e. using a hybrid estimation approach similar to
# that of Augustin, Mugglestone and Buckland (1996), but implemented to ensure valid neighbourhood weightings. This
# algorithm is fully described elsewhere (Bardos, D.C., Guillera-Arroita, G. & Wintle, B.A. (2015) Valid auto-models for
# spatially autocorrelated occupancy and abundance data. arXiv:1501.06529 [q-bio.QM]) and is applied there for an
# example with missing data. However, the Hydrocotyle data analysed here is fully observed and does not require Gibbs
# sampling during estimation.

# The second section evaluates the 'covariate impact', covariate effect' and 'standardized covariate effect' of
# the altitude covariate in the autologistic model and in the associated GLM (i.e. the logistic model obtained by
# setting beta_auto=0 in the autologistic model). For the auto-logistic case, this evaluation requires generating
# simulated draws from the estimated auto-logistic model, which is performed by Gibbs sampling.
# Results are obtained for both identity and logistic link functions.

library(lattice);set.seed(999);
setwd("C:/CovariateEffect") #change as required

nML=3L;# number of MPLE phases (must have nML >2)
Tmin=2;  # start running averages (must have Tmin >1)
Tmin=min(Tmin,nML-1); # ensure Tmin < nML
t.R=2:3; # iteration numbers to retain full set of y_n maps
pa.maps=list(); # store presence-absence maps here
nMiss=1L; #number of Gibbs cycles after each MPLE phase

covCentFlag=F; # center covariates or not

obs.df=read.table("Hydrocotyle.data.txt",header=T,sep="");
# data file must have columns in the format: X, Y, covariate1, covariate2, ... , observations
# observations can be 0, 1 or 'NA'

radius=1;   # neighbourhood radius r
lattice.ratio=1; #ratio of lattice spacings,i.e. lattice.ratio=a.y/a.x where a.x & a.y are lattice spacing in the x & y directions
# for square lattice set lattice.ratio=1. Putting radius=1.5 with lattice.ratio=1 gives 2nd-order (8-cell) neighbourhood
nbimage=matrix(0L,ncol=2*radius+3,nrow=2*radius+3);
l.x=as.integer(ceiling(radius));l.y=as.integer(ceiling(radius/lattice.ratio));lr2=lattice.ratio^2;
nbimage=matrix(0L,ncol=2*l.y+3,nrow=2*l.x+3);
nbmat=matrix(,ncol=2,nrow=0,byrow=T); # start with empty list
for (i in (-l.x:l.x)) for (j in (-l.y:l.y))
    if ((i^2 + lr2*j^2<=radius^2)&&  (!identical(c(i,j),c(0L,0L))) ){ nbmat=rbind(nbmat,t(c(i,j)));nbimage[i+l.x+2,j+l.y+2]=1L; }
Nnb=length(nbmat[,1]);levelplot(nbimage)

power=2.0;range=2.5; # power law and exponential decay parameters
r.nb= sqrt(nbmat[,1]^2+lr2*nbmat[,2]^2);
# 4 otions for specifying weight vs distance:
# wvec=r.nb^(-power); # weights w,  power-law decline vs. central distance
# wvec=exp(-r.nb/range);# weights w,  exponential decline vs. central distance
#wvec=r.nb^(-power)*exp(-r.nb/range);# weights w,  exponential and powerlaw decline vs. central distance
wvec=rep(1,Nnb) # or just use uniform weights

dx=max(abs(nbmat[,1]));dy=max(abs(nbmat[,2]));border=max(dx,dy); # minimum border requirements
parameters.n=2;covflag=F; # start with autocovariate and constant term

obs.len=length(obs.df[,1]);obs.width=length(obs.df[1,]);covariate.data.n=obs.width-3;# number of columns aside from x,y and observation.

# user specifies covariates included in estimation:
#covariates.estimated=c("altitude","temperature");
covariates.estimated=c("altitude");
#covariates.estimated=c("temperature");

if(covariate.data.n > 0) {
  covariates.list=colnames(obs.df)[3:(obs.width-1)];
  if(length(which(covariates.estimated%in%covariates.list))==length(covariates.estimated)) {
  covindex=match(covariates.estimated,colnames(obs.df));covindex=unique(covindex);
  covflag=T;covariate.param.n=length(covindex);parameters.n=2+covariate.param.n} else warning("Covariate mismatch");
}
```

```r
obs.table=matrix(0L,nrow=obs.len,ncol=obs.width);
for (j in 1:obs.width) obs.table[,j]=obs.df[,j];
nsites=obs.len;xmin=min(obs.table[,1]);ymin=min(obs.table[,2]);
xmax=max(obs.table[,1]);ymax=max(obs.table[,2]);Length=ymax-ymin+1;width=xmax-xmin+1;
observed=missing=1L[-1]; #initially empty index vectors
datamat=matrix(0L,nrow=width+2*border,ncol=Length+2*border); # use 1st array index for x (width)
p.hat=p.hat.sum=p.hat.mean=p.sim=0.0*datamat;p.hat=p.hat-1; p.sim=p.sim-1;#  occupation probabilities
for (k in t.R) pa.maps[[k]]=datamat-1L;
if(covCentFlag) for (l in covindex) obs.table[,l]=obs.table[,l]-mean(obs.table[,l]); # center covariates

obs.table[,1]=obs.table[,1]-(xmin-1)+border;  obs.table[,2]=obs.table[,2]-(ymin-1)+border;
for (k in 1:obs.len) {if(is.na(obs.table[k,obs.width])) missing=append(missing,k) else
               {observed=append(observed,k);datamat[obs.table[k,1],obs.table[k,2]]=as.integer(obs.table[k,obs.width]);}
               #x is 1st coord in matrix
}# k loop

observed.n=length(observed);missing.n=length(missing);
observed.xy=matrix(0L,nrow=observed.n,ncol=2);missing.xy=matrix(0L,nrow=missing.n,ncol=2);
observed.pa=as.integer(obs.table[observed,obs.width]);missing.pa=as.integer(obs.table[missing,obs.width]);
#init to NA so we can check all values are set

if(observed.n>0) { for (k in 1:observed.n) {# generate observed.xy using "observed" to give sequence of entries in data file
  observed.xy[k,1]=as.integer(obs.table[observed[k],1]);observed.xy[k,2]=as.integer(obs.table[observed[k],2]);}# end k loop
} #endif observed.n>0

if(missing.n>0) {for (k in 1:missing.n) {# generate missing.xy using "missing" to give sequences of entries in data file
  missing.xy[k,1]=as.integer(obs.table[missing[k],1]);missing.xy[k,2]=as.integer(obs.table[missing[k],2]);}# end k loop
} #endif missing.n>0

observed.covariates=as.matrix(obs.table[observed,covindex]);
# as.matrix so %*% product with parameter vector is OK when only 1 covariate.
missing.covariates=as.matrix(obs.table[missing,covindex]);#nb: it is observations that are missing, these are the associated covariates
h.observed=h.observed.proposed=rep(0,observed.n);h.missing=h.missing.proposed=rep(0,missing.n);

theta=matrix(0,nrow=nML,ncol=parameters.n);

f1=glm(observed.pa~observed.covariates,family=binomial);c1=as.vector(coefficients(f1));
theta[1,]=c(c1,0);

if(covflag){ # initialize objects depending on parameters
covariate.params=theta[1,2:(1+covariate.param.n)];auto=theta[1,parameters.n];
h.observed[]=observed.covariates%*%covariate.params+theta[1,1]; # X_i.beta + alpha
h.missing[]=missing.covariates%*%covariate.params+theta[1,1]} else{
auto=theta[1,parameters.n]; h.observed[]=h.missing[]=theta[1,1];}

full.n=observed.n+missing.n;
full.xy=rbind(observed.xy,missing.xy);
observed.ac=rep(0,observed.n);

for(n in 2L:nML){  # n='t' in the pseudocode

# update missing data via gibbs sampling
if(missing.n>0){ pvec=runif(nMiss*missing.n,0,1);pcount=0L; # Bernoulli draws for nMiss Gibbs cycles

for(k in 1:nMiss) {
for(m in 1:missing.n)  {

pcount=pcount+1;
i=missing.xy[m,1]; j=missing.xy[m,2];
autocov=0;
for(l in 1:Nnb) {autocov=autocov+datamat[i+nbmat[l,1],j+nbmat[l,2]]*wvec[l] ;}

prob=exp(h.missing[m]+auto*autocov);
prob=prob/(1+prob);

if(pvec[pcount]<prob) {datamat[i,j]=1L;missing.pa[m]=1L;} else {datamat[i,j]=0L;missing.pa[m]=0L;}

} # m loop
} # k loop
```

```r
    } # if(missing.n>0)

#ML step

observed.ac[]=0;
for(m in 1:observed.n)  {

i=observed.xy[m,1]; j=observed.xy[m,2];
for(l in 1:Nnb) {observed.ac[m]=observed.ac[m]+datamat[i+nbmat[l,1],j+nbmat[l,2]]*wvec[l] ;}
             }

f1=glm(observed.pa~observed.covariates+observed.ac,family=binomial);c1=as.vector(coefficients(f1));
theta[n,]=c1;

print("n=");print(n);
print("theta[n,]=");print(theta[n,]);
if(n %in% t.R) for(m in 1:full.n)  {i=full.xy[m,1]; j=full.xy[m,2]; pa.maps[[n]][i,j]=datamat[i,j];}

if(covflag){ # calculate objects depending on parameters
covariate.params=theta[n,2:(1+covariate.param.n)];auto=theta[n,parameters.n];
h.observed[]=observed.covariates%*%covariate.params+theta[n,1]; # X_i.beta + alpha
h.missing[]=missing.covariates%*%covariate.params+theta[n,1]} else{
auto=theta[n,parameters.n]; h.observed[]=h.missing[]=theta[n,1];}

if(n>Tmin)   {h.full=c(h.observed,h.missing);
for(m in 1:full.n)  {
i=full.xy[m,1]; j=full.xy[m,2];
autocov=0;
for(l in 1:Nnb) {autocov=autocov+datamat[i+nbmat[l,1],j+nbmat[l,2]]*wvec[l] ;}
prob=exp(h.full[m]+auto*autocov);
prob=prob/(1+prob); p.hat[i,j]=prob;
} # m loop
p.hat.sum=p.hat.sum+p.hat;
         }# if(n>Tmin)

} #end main loop (n)

p.hat.mean=p.hat.sum/(nML-Tmin)

( theta.hat=colMeans(theta[Tmin:nML,]) );
levelplot(pa.maps[[max(t.R)]]) #show last retained presence-absence map
sum(datamat) # number of occupied sites at final iteration
levelplot(p.hat.mean)   # show map of estimated occupation probabilities

#########################################################
########  covariate effect in autologistic model  ######
#########################################################

theta.sim=theta.hat    # calculate covariate effect for autologistic fit
nGibbs=100;burnin=50;  #Gibbs sampling parameters

if(covflag){ # calculate objects depending on parameters
covariate.params=theta.sim[2:(1+covariate.param.n)];auto=theta.sim[parameters.n];
h.observed[]=observed.covariates%*%covariate.params+theta.sim[1]; # X_i.beta + alpha
h.missing[]=missing.covariates%*%covariate.params+theta.sim[1]} else{
auto=theta.sim[parameters.n]; h.observed[]=h.missing[]=theta.sim[1];}

h.full=c(h.observed,h.missing);

# note: Gibbs sampling starts with datamat from last estimation iterate as initial state for simulations
p.sum.est=0.0*datamat;
pvec=runif(nGibbs*full.n,0,1);pcount=0L; # Bernoulli draws for nGibbs full-scene Gibbs cycles
for(k in 1:nGibbs) { # number of Gibbs scans
for(m in 1:full.n)  {
pcount=pcount+1;
i=full.xy[m,1]; j=full.xy[m,2];
autocov=0;
for(l in 1:Nnb) {autocov=autocov+datamat[i+nbmat[l,1],j+nbmat[l,2]]*wvec[l] ;}
prob=exp(h.full[m]+auto*autocov);
prob=prob/(1+prob); p.sim[i,j]=prob;
if(pvec[pcount]<prob) {datamat[i,j]=1L;} else {datamat[i,j]=0L;}
} # m loop
```

```
if(k>burnin) p.sum.est=p.sum.est+p.sim;
} # k loop
p.mean.est=p.sum.est/(nGibbs-burnin)
levelplot(datamat);
PA.sample=0*datamat-1
for(m in 1:full.n)  {i=full.xy[m,1]; j=full.xy[m,2];PA.sample[i,j]=datamat[i,j]}
save(PA.sample,file="fitted PA sample.alt only AUTO.RData")

theta.sim[2]=0; #turn off 1st covariate

if(covflag){ # calculate objects depending on parameters
covariate.params=theta.sim[2:(1+covariate.param.n)];auto=theta.sim[parameters.n];
h.observed[]=observed.covariates%*%covariate.params+theta.sim[1]; # X_i.beta + alpha
h.missing[]=missing.covariates%*%covariate.params+theta.sim[1]} else{
auto=theta.sim[parameters.n]; h.observed[]=h.missing[]=theta.sim[1];}

h.full=c(h.observed,h.missing);

# note: Gibbs sampling starts with datamat from last Gibbs iterate (at estimates) as initial state for simulations
p.sum.zero=0.0*datamat;
pvec=runif(nGibbs*full.n,0,1);pcount=0L; # Bernoulli draws for nGibbs full-scene Gibbs cycles
for(k in 1:nGibbs) { # number of Gibbs scans
for(m in 1:full.n)  {
pcount=pcount+1;
i=full.xy[m,1]; j=full.xy[m,2];
autocov=0;
for(l in 1:Nnb) {autocov=autocov+datamat[i+nbmat[l,1],j+nbmat[l,2]]*wvec[l] ;}
prob=exp(h.full[m]+auto*autocov);
prob=prob/(1+prob); p.sim[i,j]=prob;
if(pvec[pcount]<prob) {datamat[i,j]=1L;} else {datamat[i,j]=0L;}
} # m loop
if(k>burnin) p.sum.zero=p.sum.zero+p.sim;
} # k loop
p.mean.zero=p.sum.zero/(nGibbs-burnin)
levelplot(datamat);

MeanAbsX= mean(abs(observed.covariates[,1]));   # takes the first covariate
sdX= sd(observed.covariates[,1]);   # takes the first covariate

## using identity link function ##
( Covariate.impact.Autologistic= sum(abs(p.mean.est-p.mean.zero))/nsites          )
( Covariate.effect.Autologistic= Covariate.impact.Autologistic/MeanAbsX )
( Standardized.Covariate.effect.Autologistic= Covariate.effect.Autologistic*sdX )

## using logit link function ##
Covariate.logit.impact.Autologistic=0;
for(m in 1:full.n)  {i=full.xy[m,1]; j=full.xy[m,2];
Covariate.logit.impact.Autologistic=Covariate.logit.impact.Autologistic+abs(qlogis(p.mean.est[i,j]) - qlogis(p.mean.zero[i,j])); }
(Covariate.logit.impact.Autologistic=Covariate.logit.impact.Autologistic/nsites)
( Covariate.logit.effect.Autologistic=Covariate.logit.impact.Autologistic/(MeanAbsX ) )
( Standardized.Covariate.logit.effect.Autologistic=Covariate.logit.effect.Autologistic*sdX  )

#save(p.mean.est,file="p.mean.est.alt only AUTO.RData")

###########################################
########  covariate effect in GLM   ########
###########################################

theta.sim=theta[1,];  # calculate covariate effect for GLM fit

if(covflag){ # calculate objects depending on parameters
covariate.params=theta.sim[2:(1+covariate.param.n)];auto=theta.sim[parameters.n];
h.observed[]=observed.covariates%*%covariate.params+theta.sim[1]; # X_i.beta + alpha
h.missing[]=missing.covariates%*%covariate.params+theta.sim[1]} else{
auto=theta.sim[parameters.n]; h.observed[]=h.missing[]=theta.sim[1];}

h.full=c(h.observed,h.missing);

for(m in 1:full.n)  {
i=full.xy[m,1]; j=full.xy[m,2];
prob=exp(h.full[m]);
prob=prob/(1+prob); p.sim[i,j]=prob;
```

```
} # m loop
p.mean.est=p.sim;

theta.sim[2]=0; #turn off 1st covariate

if(covflag){ # calculate objects depending on parameters
covariate.params=theta.sim[2:(1+covariate.param.n)];auto=theta.sim[parameters.n];
h.observed[]=observed.covariates%*%covariate.params+theta.sim[1]; # X_i.beta + alpha
h.missing[]=missing.covariates%*%covariate.params+theta.sim[1]} else{
auto=theta.sim[parameters.n]; h.observed[]=h.missing[]=theta.sim[1];}

h.full=c(h.observed,h.missing);

for(m in 1:full.n)  {
i=full.xy[m,1]; j=full.xy[m,2];
prob=exp(h.full[m]);
prob=prob/(1+prob); p.sim[i,j]=prob;
} # m loop
p.mean.zero=p.sim;

## using identity link function ##
( Covariate.impact.GLM=sum(abs(p.mean.est-p.mean.zero))/nsites )
( Covariate.effect.GLM= Covariate.impact.GLM/MeanAbsX )
( Standardized.Covariate.effect.GLM=Covariate.effect.GLM*sdX )

## using logit link function ##
Covariate.logit.impact.GLM=0;
for(m in 1:full.n)  {i=full.xy[m,1]; j=full.xy[m,2];
Covariate.logit.impact.GLM=Covariate.logit.impact.GLM+abs(qlogis(p.mean.est[i,j]) - qlogis(p.mean.zero[i,j])); }
(Covariate.logit.impact.GLM=Covariate.logit.impact.GLM/nsites)
( Covariate.logit.effect.GLM=Covariate.logit.impact.GLM/(MeanAbsX ) )
( Standardized.Covariate.logit.effect.GLM=Covariate.logit.effect.GLM*sdX )
theta[1,2];  #compare with GLM estimate (should agree except sign may differ)
#save(p.mean.est,file="p.mean.est.alt only.RData")
```

## Hydrocotyle.data.txt

```
"X" "Y" "altitude" "temperature" "obs"
"1" 1 16 1.71138775510204 8.26649746192893 1
"2" 2 15 1.86020408163265 8.154822335025381
"3" 2 16 1.78579591836735 8.0989847715736 1
"4" 2 17 1.78579591836735 8.0989847715736 0
"5" 2 18 1.78579591836735 8.0989847715736 0
"6" 2 19 1.86020408163265 8.0989847715736 1
"7" 3 15 1.86020408163265 8.154822335025381
"8" 3 16 1.78579591836735 8.154822335025381
"9" 3 17 1.78579591836735 8.154822335025380
"10" 3 18 1.78579591836735 8.154822335025381
"11" 3 19 1.78579591836735 8.154822335025381
"12" 3 20 1.78579591836735 8.0989847715736 1
"13" 3 21 1.78579591836735 8.154822335025381
"14" 3 22 1.78579591836735 8.154822335025381
"15" 3 23 1.78579591836735 8.154822335025381
"16" 3 24 1.86020408163265 8.154822335025381
"17" 4 15 1.86020408163265 8.154822335025381
"18" 4 16 1.78579591836735 8.21065989847716 1
"19" 4 17 1.78579591836735 8.21065989847716 1
"20" 4 18 1.78579591836735 8.21065989847716 1
"21" 4 19 1.78579591836735 8.21065989847716 1
"22" 4 20 1.78579591836735 8.21065989847716 1
"23" 4 21 1.78579591836735 8.21065989847716 1
"24" 4 22 1.78579591836735 8.21065989847716 1
"25" 4 23 1.78579591836735 8.21065989847716 1
"26" 4 24 1.86020408163265 8.21065989847716 1
"27" 4 25 1.71138775510204 8.26649746192893 1
"28" 4 26 0.223224489795918 8.21065989847716 1
"29" 5 15 1.71138775510204 8.154822335025381
"30" 5 16 1.86020408163265 8.04314720812183 1
"31" 5 17 1.71138775510204 8.0989847715736 1
"32" 5 18 1.78579591836735 8.0989847715736 0
"33" 5 19 1.78579591836735 8.0989847715736 1
```

```
"34" 5 20 1.78579591836735 8.0989847715736 1
"35" 5 21 1.78579591836735 8.0989847715736 1
"36" 5 22 1.78579591836735 8.0989847715736 1
"37" 5 23 1.78579591836735 8.0989847715736 1
"38" 5 24 1.78579591836735 8.0989847715736 1
"39" 5 25 1.78579591836735 8.0989847715736 1
"40" 5 26 1.78579591836735 8.15482233502538 1
"41" 5 44 0 8.21065989847716 0
"42" 5 45 1.78579591836735 8.0989847715736 1
"43" 5 46 1.78579591836735 8.0989847715736 1
"44" 6 17 1.78579591836735 8.15482233502538 0
"45" 6 18 1.78579591836735 8.15482233502538 0
"46" 6 19 1.78579591836735 8.15482233502538 1
"47" 6 20 1.78579591836735 8.15482233502538 1
"48" 6 21 1.78579591836735 8.15482233502538 1
"49" 6 22 1.78579591836735 8.15482233502538 1
"50" 6 23 1.78579591836735 8.15482233502538 1
"51" 6 24 1.78579591836735 8.15482233502538 1
"52" 6 25 1.78579591836735 8.15482233502538 1
"53" 6 26 1.78579591836735 8.15482233502538 1
"54" 6 32 1.78579591836735 8.0989847715736 1
"55" 6 33 0 8.04314720812183 1
"56" 6 44 1.71138775510204 8.0989847715736 1
"57" 6 45 1.78579591836735 8.15482233502538 1
"58" 6 46 1.78579591836735 8.0989847715736 1
"59" 6 47 1.86020408163265 8.0989847715736 1
"60" 6 48 0 8.04314720812183 0
"61" 7 17 0 8.21065989847716 0
"62" 7 18 1.71138775510204 8.21065989847716 0
"63" 7 19 1.78579591836735 8.21065989847716 0
"64" 7 20 1.78579591836735 8.21065989847716 1
"65" 7 21 1.78579591836735 8.21065989847716 1
"66" 7 22 1.78579591836735 8.21065989847716 1
"67" 7 23 1.78579591836735 8.21065989847716 1
"68" 7 24 1.78579591836735 8.21065989847716 1
"69" 7 25 1.78579591836735 8.21065989847716 1
"70" 7 26 1.78579591836735 8.15482233502538 1
"71" 7 27 1.86020408163265 8.21065989847716 1
"72" 7 28 1.71138775510204 8.21065989847716 1
"73" 7 32 1.78579591836735 8.32233502538071 1
"74" 7 33 1.78579591836735 8.21065989847716 1
"75" 7 40 1.78579591836735 8.15482233502538 1
"76" 7 41 1.78579591836735 8.15482233502538 1
"77" 7 42 1.78579591836735 8.15482233502538 1
"78" 7 43 1.78579591836735 8.15482233502538 1
"79" 7 44 1.78579591836735 8.0989847715736 1
"80" 7 45 1.78579591836735 8.21065989847716 1
"81" 7 46 1.78579591836735 8.21065989847716 1
"82" 7 47 1.78579591836735 8.37817258883249 1
"83" 8 17 1.78579591836735 8.15482233502538 1
"84" 8 18 1.78579591836735 8.0989847715736 1
"85" 8 19 1.78579591836735 8.0989847715736 0
"86" 8 20 1.78579591836735 8.0989847715736 1
"87" 8 21 1.78579591836735 8.0989847715736 1
"88" 8 22 1.78579591836735 8.0989847715736 1
"89" 8 23 1.78579591836735 8.0989847715736 1
"90" 8 24 1.78579591836735 8.0989847715736 1
"91" 8 25 1.78579591836735 8.0989847715736 1
"92" 8 26 1.78579591836735 8.0989847715736 1
"93" 8 27 1.78579591836735 8.0989847715736 1
"94" 8 28 1.78579591836735 8.0989847715736 1
"95" 8 29 1.78579591836735 8.21065989847716 1
"96" 8 30 1.78579591836735 8.0989847715736 1
"97" 8 31 1.78579591836735 8.0989847715736 1
"98" 8 32 1.78579591836735 8.0989847715736 1
"99" 8 40 1.78579591836735 8.15482233502538 1
"100" 8 41 1.78579591836735 8.0989847715736 1
"101" 8 42 1.78579591836735 8.0989847715736 1
"102" 8 43 1.78579591836735 8.0989847715736 1
"103" 8 44 1.78579591836735 8.0989847715736 1
"104" 8 45 1.78579591836735 8.0989847715736 1
"105" 8 46 1.78579591836735 8.15482233502538 1
"106" 8 47 1.78579591836735 8.04314720812183 1
"107" 8 48 1.78579591836735 8.15482233502538 1
"108" 9 17 1.71138775510204 8.15482233502538 1
```

```
"109" 9 18 1.78579591836735 8.0989847715736 1
"110" 9 19 1.78579591836735 8.0989847715736 0
"111" 9 20 1.78579591836735 8.0989847715736 1
"112" 9 21 1.78579591836735 8.0989847715736 1
"113" 9 22 1.78579591836735 8.0989847715736 1
"114" 9 23 1.78579591836735 8.0989847715736 1
"115" 9 24 1.78579591836735 8.0989847715736 1
"116" 9 25 1.78579591836735 8.0989847715736 1
"117" 9 26 1.78579591836735 8.0989847715736 1
"118" 9 27 1.78579591836735 8.0989847715736 1
"119" 9 28 1.78579591836735 8.0989847715736 1
"120" 9 29 1.78579591836735 8.0989847715736 1
"121" 9 30 1.78579591836735 8.0989847715736 1
"122" 9 31 1.78579591836735 8.04314720812183 1
"123" 9 32 1.78579591836735 8.0989847715736 1
"124" 9 38 0 8.0989847715736 0
"125" 9 39 1.86020408163265 8.0989847715736 1
"126" 9 40 1.78579591836735 8.04314720812183 1
"127" 9 41 1.78579591836735 8.0989847715736 1
"128" 9 42 1.78579591836735 8.0989847715736 1
"129" 9 43 1.78579591836735 8.0989847715736 1
"130" 9 44 1.78579591836735 8.0989847715736 1
"131" 9 45 1.78579591836735 8.0989847715736 1
"132" 9 46 1.78579591836735 8.21065989847716 1
"133" 9 48 0 8.15482233502538 1
"134" 10 13 0 8.65736040609137 0
"135" 10 18 1.78579591836735 8.0989847715736 0
"136" 10 19 1.78579591836735 8.0989847715736 0
"137" 10 20 1.78579591836735 8.15482233502538 1
"138" 10 21 1.78579591836735 8.15482233502538 1
"139" 10 22 1.78579591836735 8.15482233502538 1
"140" 10 23 1.78579591836735 8.15482233502538 1
"141" 10 24 1.78579591836735 8.15482233502538 1
"142" 10 25 1.78579591836735 8.15482233502538 1
"143" 10 26 1.78579591836735 8.15482233502538 1
"144" 10 27 1.78579591836735 8.15482233502538 1
"145" 10 28 1.78579591836735 8.15482233502538 1
"146" 10 29 1.78579591836735 8.15482233502538 1
"147" 10 30 1.78579591836735 8.15482233502538 1
"148" 10 31 1.78579591836735 8.15482233502538 1
"149" 10 32 1.78579591836735 8.15482233502538 1
"150" 10 35 1.71138775510204 8.0989847715736 0
"151" 10 36 1.78579591836735 8.0989847715736 1
"152" 10 37 1.78579591836735 8.0989847715736 1
"153" 10 38 1.78579591836735 8.15482233502538 1
"154" 10 39 1.86020408163265 8.15482233502538 1
"155" 10 40 1.78579591836735 8.15482233502538 1
"156" 10 41 1.78579591836735 8.15482233502538 1
"157" 10 42 1.78579591836735 8.15482233502538 1
"158" 10 43 1.78579591836735 8.15482233502538 1
"159" 10 44 1.78579591836735 8.15482233502538 0
"160" 10 45 1.78579591836735 8.15482233502538 1
"161" 10 46 1.78579591836735 7.98730964467005 1
"162" 10 47 1.63697959183673 8.21065989847716 0
"163" 10 48 1.78579591836735 8.04314720812183 1
"164" 10 49 0 8.15482233502538 0
"165" 11 17 0 0.114213197969544 0
"166" 11 18 0 0.281725888324875 1
"167" 11 19 1.78579591836735 8.0989847715736 1
"168" 11 20 1.78579591836735 8.15482233502538 1
"169" 11 21 1.78579591836735 8.15482233502538 1
"170" 11 22 1.78579591836735 8.15482233502538 1
"171" 11 23 1.78579591836735 8.15482233502538 1
"172" 11 24 1.78579591836735 8.15482233502538 1
"173" 11 25 1.78579591836735 8.15482233502538 1
"174" 11 26 1.78579591836735 8.15482233502538 1
"175" 11 27 1.78579591836735 8.15482233502538 1
"176" 11 28 1.78579591836735 8.15482233502538 1
"177" 11 29 1.78579591836735 8.15482233502538 1
"178" 11 30 1.78579591836735 8.15482233502538 1
"179" 11 31 0 8.15482233502538 1
"180" 11 32 1.78579591836735 8.15482233502538 1
"181" 11 33 0 8.0989847715736 0
"182" 11 34 1.78579591836735 8.15482233502538 0
"183" 11 35 1.78579591836735 8.15482233502538 1
```

```
"184" 11 36 1.78579591836735 8.15482233502538 1
"185" 11 37 1.78579591836735 8.15482233502538 1
"186" 11 38 1.78579591836735 8.15482233502538 1
"187" 11 39 1.78579591836735 8.15482233502538 1
"188" 11 40 1.78579591836735 8.15482233502538 1
"189" 11 41 1.78579591836735 8.15482233502538 1
"190" 11 42 1.78579591836735 8.15482233502538 1
"191" 11 43 1.78579591836735 8.15482233502538 1
"192" 11 44 1.78579591836735 8.15482233502538 0
"193" 11 45 1.78579591836735 8.15482233502538 1
"194" 11 46 1.78579591836735 8.15482233502538 1
"195" 11 47 1.78579591836735 8.15482233502538 1
"196" 11 48 1.78579591836735 8.15482233502538 1
"197" 11 49 1.78579591836735 8.15482233502538 1
"198" 11 50 1.78579591836735 8.15482233502538 1
"199" 12 16 1.78579591836735 8.15482233502538 1
"200" 12 17 0.0744081632653071 8.15482233502538 1
"201" 12 19 1.78579591836735 8.15482233502538 1
"202" 12 20 1.78579591836735 8.15482233502538 1
"203" 12 21 1.78579591836735 8.15482233502538 1
"204" 12 22 1.78579591836735 8.15482233502538 1
"205" 12 23 1.78579591836735 8.15482233502538 1
"206" 12 24 1.78579591836735 8.15482233502538 1
"207" 12 25 1.78579591836735 8.15482233502538 1
"208" 12 26 1.78579591836735 8.15482233502538 1
"209" 12 27 1.78579591836735 8.15482233502538 1
"210" 12 28 1.78579591836735 8.15482233502538 1
"211" 12 29 1.78579591836735 8.15482233502538 1
"212" 12 30 1.78579591836735 8.15482233502538 1
"213" 12 31 1.71138775510204 8.15482233502538 1
"214" 12 32 1.71138775510204 8.15482233502538 1
"215" 12 33 1.78579591836735 8.15482233502538 1
"216" 12 34 1.78579591836735 8.15482233502538 0
"217" 12 35 1.78579591836735 8.15482233502538 0
"218" 12 36 1.78579591836735 8.15482233502538 1
"219" 12 37 1.78579591836735 8.15482233502538 1
"220" 12 38 1.78579591836735 8.15482233502538 1
"221" 12 39 1.78579591836735 8.15482233502538 1
"222" 12 40 1.78579591836735 8.15482233502538 1
"223" 12 41 1.78579591836735 8.15482233502538 1
"224" 12 42 1.78579591836735 8.15482233502538 1
"225" 12 43 1.78579591836735 8.15482233502538 1
"226" 12 44 1.78579591836735 8.15482233502538 1
"227" 12 45 1.78579591836735 8.15482233502538 1
"228" 12 46 1.78579591836735 8.15482233502538 1
"229" 12 47 1.78579591836735 8.15482233502538 1
"230" 12 48 1.78579591836735 8.15482233502538 1
"231" 12 49 1.78579591836735 8.15482233502538 1
"232" 12 50 1.71138775510204 8.15482233502538 1
"233" 12 51 1.71138775510204 8.0989847715736 1
"234" 13 17 1.78579591836735 7.98730964467005 1
"235" 13 18 1.78579591836735 8.15482233502538 1
"236" 13 19 1.78579591836735 8.15482233502538 0
"237" 13 20 1.78579591836735 8.0989847715736 1
"238" 13 21 1.78579591836735 8.0989847715736 1
"239" 13 22 1.78579591836735 8.0989847715736 1
"240" 13 23 1.78579591836735 8.0989847715736 1
"241" 13 24 1.78579591836735 8.0989847715736 1
"242" 13 25 1.78579591836735 8.0989847715736 1
"243" 13 26 1.78579591836735 8.0989847715736 1
"244" 13 27 1.78579591836735 8.0989847715736 1
"245" 13 28 1.78579591836735 8.0989847715736 1
"246" 13 29 1.78579591836735 8.0989847715736 1
"247" 13 30 1.78579591836735 8.0989847715736 1
"248" 13 31 1.78579591836735 8.0989847715736 1
"249" 13 32 1.78579591836735 8.0989847715736 1
"250" 13 33 1.78579591836735 8.0989847715736 1
"251" 13 34 1.78579591836735 8.0989847715736 1
"252" 13 35 1.78579591836735 8.0989847715736 1
"253" 13 36 1.78579591836735 8.0989847715736 1
"254" 13 37 1.78579591836735 8.0989847715736 1
"255" 13 38 1.78579591836735 8.0989847715736 1
"256" 13 39 1.78579591836735 8.0989847715736 1
"257" 13 40 1.78579591836735 8.0989847715736 1
"258" 13 41 1.78579591836735 8.0989847715736 1
```

```
"259" 13 42 1.78579591836735 8.0989847715736 1
"260" 13 43 1.78579591836735 8.0989847715736 1
"261" 13 44 1.78579591836735 8.0989847715736 1
"262" 13 45 1.78579591836735 8.0989847715736 1
"263" 13 46 1.78579591836735 8.0989847715736 1
"264" 13 47 1.78579591836735 8.0989847715736 1
"265" 13 48 1.78579591836735 8.0989847715736 1
"266" 13 49 1.78579591836735 8.0989847715736 1
"267" 13 50 1.78579591836735 8.0989847715736 0
"268" 13 51 1.78579591836735 8.15482233502538 0
"269" 14 8 0 8.65736040609137 0
"270" 14 9 1.78579591836735 8.0989847715736 1
"271" 14 10 1.78579591836735 8.15482233502538 1
"272" 14 11 1.78579591836735 8.15482233502538 1
"273" 14 12 1.78579591836735 8.15482233502538 1
"274" 14 13 1.78579591836735 8.15482233502538 1
"275" 14 14 1.78579591836735 8.0989847715736 0
"276" 14 15 0.0744081632653071 0.225888324873099 0
"277" 14 16 0.0744081632653071 7.98730964467005 0
"278" 14 17 1.78579591836735 8.04314720812183 1
"279" 14 18 1.78579591836735 8.04314720812183 1
"280" 14 19 1.78579591836735 8.04314720812183 1
"281" 14 20 1.78579591836735 8.04314720812183 1
"282" 14 21 1.78579591836735 8.04314720812183 1
"283" 14 22 1.78579591836735 8.04314720812183 1
"284" 14 23 1.78579591836735 8.04314720812183 1
"285" 14 24 1.78579591836735 8.04314720812183 1
"286" 14 25 1.78579591836735 8.04314720812183 1
"287" 14 26 1.78579591836735 8.04314720812183 1
"288" 14 27 1.78579591836735 8.04314720812183 1
"289" 14 28 1.78579591836735 8.04314720812183 1
"290" 14 29 1.78579591836735 8.04314720812183 1
"291" 14 30 1.78579591836735 8.04314720812183 1
"292" 14 31 1.78579591836735 8.04314720812183 1
"293" 14 32 1.78579591836735 8.04314720812183 1
"294" 14 33 1.78579591836735 8.04314720812183 1
"295" 14 34 1.78579591836735 8.04314720812183 1
"296" 14 35 1.78579591836735 8.04314720812183 1
"297" 14 36 1.78579591836735 8.04314720812183 1
"298" 14 37 1.78579591836735 8.04314720812183 1
"299" 14 38 1.78579591836735 8.04314720812183 1
"300" 14 39 1.78579591836735 8.04314720812183 1
"301" 14 40 1.78579591836735 8.04314720812183 1
"302" 14 41 1.78579591836735 8.04314720812183 1
"303" 14 42 1.78579591836735 8.04314720812183 1
"304" 14 43 1.78579591836735 8.04314720812183 1
"305" 14 44 1.78579591836735 8.04314720812183 0
"306" 14 45 1.78579591836735 8.04314720812183 1
"307" 14 46 1.78579591836735 8.04314720812183 1
"308" 14 47 1.78579591836735 8.04314720812183 1
"309" 14 48 1.78579591836735 8.04314720812183 1
"310" 14 49 1.78579591836735 8.04314720812183 1
"311" 14 50 1.78579591836735 8.04314720812183 1
"312" 14 51 1.78579591836735 8.15482233502538 1
"313" 15 6 1.78579591836735 8.65736040609137 1
"314" 15 7 1.71138775510204 8.60152284263959 1
"315" 15 8 1.78579591836735 8.15482233502538 1
"316" 15 9 1.78579591836735 8.15482233502538 0
"317" 15 10 1.78579591836735 8.15482233502538 1
"318" 15 11 1.78579591836735 8.15482233502538 1
"319" 15 12 1.78579591836735 8.15482233502538 1
"320" 15 13 1.78579591836735 8.15482233502538 1
"321" 15 14 1.78579591836735 8.15482233502538 0
"322" 15 15 0.223224489795918 8.15482233502538 0
"323" 15 16 0 8.15482233502538 0
"324" 15 17 1.78579591836735 8.15482233502538 1
"325" 15 18 1.78579591836735 8.15482233502538 1
"326" 15 19 1.78579591836735 8.15482233502538 1
"327" 15 20 1.78579591836735 8.15482233502538 1
"328" 15 21 1.78579591836735 8.15482233502538 1
"329" 15 22 1.78579591836735 8.15482233502538 1
"330" 15 23 1.78579591836735 8.15482233502538 1
"331" 15 24 1.78579591836735 8.15482233502538 1
"332" 15 25 1.78579591836735 8.15482233502538 1
"333" 15 26 1.78579591836735 8.15482233502538 1
```

```
"334" 15 27 1.78579591836735 8.15482233502538 1
"335" 15 28 1.78579591836735 8.15482233502538 1
"336" 15 29 1.78579591836735 8.15482233502538 1
"337" 15 30 1.78579591836735 8.15482233502538 1
"338" 15 31 1.78579591836735 8.15482233502538 1
"339" 15 32 1.78579591836735 8.15482233502538 1
"340" 15 33 1.78579591836735 8.15482233502538 1
"341" 15 34 1.78579591836735 8.15482233502538 1
"342" 15 35 1.78579591836735 8.15482233502538 1
"343" 15 36 1.78579591836735 8.15482233502538 1
"344" 15 37 1.78579591836735 8.15482233502538 1
"345" 15 38 1.78579591836735 8.15482233502538 1
"346" 15 39 1.78579591836735 8.15482233502538 1
"347" 15 40 1.78579591836735 8.15482233502538 1
"348" 15 41 1.78579591836735 8.15482233502538 1
"349" 15 42 1.78579591836735 8.15482233502538 1
"350" 15 43 1.78579591836735 8.15482233502538 1
"351" 15 44 1.78579591836735 8.15482233502538 0
"352" 15 45 1.78579591836735 8.15482233502538 1
"353" 15 46 1.78579591836735 8.15482233502538 1
"354" 15 47 1.78579591836735 8.15482233502538 1
"355" 15 48 1.78579591836735 8.15482233502538 1
"356" 15 49 1.78579591836735 8.15482233502538 1
"357" 15 50 1.78579591836735 8.15482233502538 1
"358" 15 51 1.78579591836735 8.0989847715736 1
"359" 16 5 0 8.60152284263959 0
"360" 16 6 1.78579591836735 8.60152284263959 1
"361" 16 7 0 8.15482233502538 0
"362" 16 8 1.78579591836735 8.0989847715736 0
"363" 16 9 1.78579591836735 8.15482233502538 0
"364" 16 10 1.78579591836735 8.15482233502538 1
"365" 16 11 1.78579591836735 8.15482233502538 1
"366" 16 12 1.78579591836735 8.15482233502538 1
"367" 16 13 1.78579591836735 8.15482233502538 1
"368" 16 14 1.78579591836735 8.15482233502538 1
"369" 16 15 1.78579591836735 8.15482233502538 0
"370" 16 16 1.78579591836735 8.15482233502538 0
"371" 16 17 1.78579591836735 8.15482233502538 1
"372" 16 18 1.78579591836735 8.15482233502538 1
"373" 16 19 1.78579591836735 8.15482233502538 1
"374" 16 20 1.78579591836735 8.15482233502538 1
"375" 16 21 1.78579591836735 8.15482233502538 1
"376" 16 22 1.78579591836735 8.15482233502538 1
"377" 16 23 1.78579591836735 8.15482233502538 1
"378" 16 24 1.78579591836735 8.15482233502538 1
"379" 16 25 1.78579591836735 8.60152284263959 1
"380" 16 26 1.78579591836735 8.15482233502538 1
"381" 16 27 1.78579591836735 8.15482233502538 1
"382" 16 28 1.78579591836735 8.15482233502538 1
"383" 16 29 1.78579591836735 8.15482233502538 1
"384" 16 30 1.78579591836735 8.15482233502538 1
"385" 16 31 1.78579591836735 8.15482233502538 1
"386" 16 32 1.78579591836735 8.15482233502538 1
"387" 16 33 1.78579591836735 8.15482233502538 1
"388" 16 34 1.78579591836735 8.15482233502538 1
"389" 16 35 1.78579591836735 8.15482233502538 1
"390" 16 36 1.78579591836735 8.15482233502538 1
"391" 16 37 1.78579591836735 8.15482233502538 1
"392" 16 38 1.78579591836735 8.15482233502538 1
"393" 16 39 1.78579591836735 8.15482233502538 1
"394" 16 40 1.78579591836735 8.15482233502538 1
"395" 16 41 1.78579591836735 8.15482233502538 1
"396" 16 42 1.78579591836735 8.15482233502538 1
"397" 16 43 1.78579591836735 8.15482233502538 1
"398" 16 44 1.78579591836735 8.15482233502538 1
"399" 16 45 1.78579591836735 8.15482233502538 1
"400" 16 46 1.78579591836735 8.15482233502538 1
"401" 16 47 1.78579591836735 8.15482233502538 1
"402" 16 48 1.78579591836735 8.15482233502538 1
"403" 16 49 1.78579591836735 8.15482233502538 1
"404" 16 50 1.78579591836735 8.15482233502538 1
"405" 16 51 1.78579591836735 8.0989847715736 1
"406" 17 8 1.78579591836735 8.0989847715736 0
"407" 17 9 1.78579591836735 8.04314720812183 1
"408" 17 10 1.78579591836735 8.04314720812183 1
```

```
"409" 17 11 1.78579591836735 8.04314720812183 1
"410" 17 12 1.78579591836735 8.04314720812183 1
"411" 17 13 1.78579591836735 8.04314720812183 1
"412" 17 14 1.78579591836735 8.04314720812183 1
"413" 17 15 1.78579591836735 8.04314720812183 1
"414" 17 16 1.78579591836735 8.04314720812183 0
"415" 17 17 1.78579591836735 8.04314720812183 1
"416" 17 18 1.78579591836735 8.04314720812183 1
"417" 17 19 1.78579591836735 8.04314720812183 1
"418" 17 20 1.78579591836735 8.04314720812183 1
"419" 17 21 1.78579591836735 8.04314720812183 1
"420" 17 22 1.78579591836735 8.04314720812183 1
"421" 17 23 1.78579591836735 8.04314720812183 1
"422" 17 24 1.78579591836735 8.04314720812183 1
"423" 17 25 1.78579591836735 8.0989847715736 1
"424" 17 26 1.78579591836735 8.04314720812183 1
"425" 17 27 1.78579591836735 8.04314720812183 1
"426" 17 28 1.78579591836735 8.04314720812183 1
"427" 17 29 1.78579591836735 8.04314720812183 1
"428" 17 30 1.78579591836735 8.04314720812183 1
"429" 17 31 1.78579591836735 8.04314720812183 0
"430" 17 32 1.78579591836735 8.04314720812183 1
"431" 17 33 1.78579591836735 8.04314720812183 1
"432" 17 34 1.78579591836735 8.04314720812183 1
"433" 17 35 1.78579591836735 8.04314720812183 1
"434" 17 36 1.78579591836735 8.04314720812183 1
"435" 17 37 1.78579591836735 8.04314720812183 1
"436" 17 38 1.78579591836735 8.04314720812183 1
"437" 17 39 1.78579591836735 8.04314720812183 1
"438" 17 40 1.78579591836735 8.04314720812183 1
"439" 17 41 1.78579591836735 8.04314720812183 1
"440" 17 42 1.78579591836735 8.04314720812183 1
"441" 17 43 1.78579591836735 8.04314720812183 1
"442" 17 44 1.78579591836735 8.04314720812183 1
"443" 17 45 1.78579591836735 8.04314720812183 1
"444" 17 46 1.78579591836735 8.04314720812183 1
"445" 17 47 1.78579591836735 8.04314720812183 1
"446" 17 48 1.78579591836735 8.04314720812183 0
"447" 17 49 1.78579591836735 8.04314720812183 0
"448" 17 50 1.78579591836735 8.04314720812183 1
"449" 17 51 1.78579591836735 7.98730964467005 1
"450" 17 52 1.78579591836735 8.21065989847716 0
"451" 18 8 1.86020408163265 8.15482233502538 0
"452" 18 9 1.78579591836735 8.15482233502538 1
"453" 18 10 1.78579591836735 8.0989847715736 1
"454" 18 11 1.78579591836735 8.0989847715736 1
"455" 18 12 1.78579591836735 8.0989847715736 1
"456" 18 13 1.78579591836735 8.0989847715736 1
"457" 18 14 1.78579591836735 8.0989847715736 1
"458" 18 15 1.78579591836735 8.0989847715736 1
"459" 18 16 1.78579591836735 8.0989847715736 1
"460" 18 17 1.78579591836735 8.0989847715736 1
"461" 18 18 1.78579591836735 8.0989847715736 1
"462" 18 19 1.78579591836735 8.0989847715736 1
"463" 18 20 1.78579591836735 8.0989847715736 1
"464" 18 21 1.78579591836735 8.0989847715736 1
"465" 18 22 1.78579591836735 8.0989847715736 1
"466" 18 23 1.78579591836735 8.0989847715736 1
"467" 18 24 1.78579591836735 8.0989847715736 1
"468" 18 25 1.78579591836735 8.0989847715736 1
"469" 18 26 1.78579591836735 8.0989847715736 1
"470" 18 27 1.78579591836735 8.0989847715736 1
"471" 18 28 1.78579591836735 8.0989847715736 1
"472" 18 29 1.78579591836735 8.0989847715736 1
"473" 18 30 1.78579591836735 8.0989847715736 1
"474" 18 31 1.78579591836735 8.0989847715736 1
"475" 18 32 1.78579591836735 8.0989847715736 1
"476" 18 33 1.78579591836735 8.0989847715736 1
"477" 18 34 1.78579591836735 8.0989847715736 1
"478" 18 35 1.78579591836735 8.0989847715736 1
"479" 18 36 1.78579591836735 8.0989847715736 1
"480" 18 37 1.78579591836735 8.0989847715736 1
"481" 18 38 1.78579591836735 8.0989847715736 1
"482" 18 39 1.78579591836735 8.0989847715736 1
"483" 18 40 1.78579591836735 8.0989847715736 1
```

```
"484" 18 41 1.78579591836735 8.0989847715736 1
"485" 18 42 1.78579591836735 8.0989847715736 1
"486" 18 43 1.78579591836735 8.0989847715736 1
"487" 18 44 1.78579591836735 8.0989847715736 1
"488" 18 45 1.78579591836735 8.0989847715736 1
"489" 18 46 1.78579591836735 8.0989847715736 1
"490" 18 47 1.78579591836735 8.0989847715736 1
"491" 18 48 1.78579591836735 8.0989847715736 0
"492" 18 49 1.78579591836735 8.0989847715736 1
"493" 18 50 1.78579591836735 8.0989847715736 1
"494" 18 51 1.78579591836735 8.0989847715736 1
"495" 18 52 1.78579591836735 8.15482233502538 1
"496" 19 9 1.78579591836735 8.15482233502538 1
"497" 19 10 1.78579591836735 8.15482233502538 1
"498" 19 11 1.78579591836735 8.15482233502538 1
"499" 19 12 1.78579591836735 8.15482233502538 1
"500" 19 13 1.78579591836735 8.15482233502538 1
"501" 19 14 1.78579591836735 8.15482233502538 1
"502" 19 15 1.78579591836735 8.15482233502538 1
"503" 19 16 1.78579591836735 8.15482233502538 1
"504" 19 17 1.78579591836735 8.15482233502538 1
"505" 19 18 1.78579591836735 8.15482233502538 1
"506" 19 19 1.78579591836735 8.15482233502538 1
"507" 19 20 1.78579591836735 8.15482233502538 1
"508" 19 21 1.78579591836735 8.15482233502538 1
"509" 19 22 1.78579591836735 8.15482233502538 1
"510" 19 23 1.78579591836735 8.15482233502538 1
"511" 19 24 1.78579591836735 8.15482233502538 1
"512" 19 25 1.78579591836735 8.15482233502538 1
"513" 19 26 1.78579591836735 8.15482233502538 1
"514" 19 27 1.78579591836735 8.15482233502538 1
"515" 19 28 1.78579591836735 8.15482233502538 1
"516" 19 29 1.78579591836735 8.15482233502538 1
"517" 19 30 1.78579591836735 8.15482233502538 1
"518" 19 31 1.78579591836735 8.15482233502538 1
"519" 19 32 1.78579591836735 8.15482233502538 1
"520" 19 33 1.78579591836735 8.15482233502538 1
"521" 19 34 1.78579591836735 8.15482233502538 1
"522" 19 35 1.78579591836735 8.15482233502538 1
"523" 19 36 1.78579591836735 8.15482233502538 1
"524" 19 37 1.78579591836735 8.15482233502538 1
"525" 19 38 1.78579591836735 8.15482233502538 0
"526" 19 39 1.78579591836735 8.15482233502538 0
"527" 19 40 1.78579591836735 8.15482233502538 1
"528" 19 41 1.78579591836735 8.15482233502538 1
"529" 19 42 1.78579591836735 8.15482233502538 1
"530" 19 43 1.78579591836735 8.15482233502538 1
"531" 19 44 1.78579591836735 8.15482233502538 1
"532" 19 45 1.78579591836735 8.15482233502538 1
"533" 19 46 1.78579591836735 8.15482233502538 1
"534" 19 47 1.78579591836735 8.15482233502538 1
"535" 19 48 1.78579591836735 8.15482233502538 1
"536" 19 49 1.78579591836735 8.15482233502538 1
"537" 19 50 1.78579591836735 8.15482233502538 0
"538" 19 51 1.78579591836735 8.15482233502538 1
"539" 19 52 1.78579591836735 8.15482233502538 1
"540" 20 9 1.71138775510204 8.15482233502538 1
"541" 20 10 1.78579591836735 8.21065989847716 1
"542" 20 11 1.78579591836735 8.21065989847716 1
"543" 20 12 1.78579591836735 8.21065989847716 1
"544" 20 13 1.78579591836735 8.21065989847716 1
"545" 20 14 1.78579591836735 8.21065989847716 1
"546" 20 15 1.78579591836735 8.21065989847716 1
"547" 20 16 1.78579591836735 8.21065989847716 1
"548" 20 17 1.78579591836735 8.21065989847716 1
"549" 20 18 1.78579591836735 8.21065989847716 1
"550" 20 19 1.78579591836735 8.21065989847716 1
"551" 20 20 1.78579591836735 8.21065989847716 1
"552" 20 21 1.78579591836735 8.21065989847716 1
"553" 20 22 1.78579591836735 8.21065989847716 1
"554" 20 23 1.78579591836735 8.21065989847716 1
"555" 20 24 1.78579591836735 8.21065989847716 1
"556" 20 25 1.78579591836735 8.21065989847716 1
"557" 20 26 1.78579591836735 8.21065989847716 1
"558" 20 27 1.78579591836735 8.21065989847716 1
```

```
"559" 20 28 1.78579591836735 8.21065989847716 1
"560" 20 29 1.78579591836735 8.21065989847716 1
"561" 20 30 1.78579591836735 8.21065989847716 1
"562" 20 31 1.78579591836735 8.21065989847716 1
"563" 20 32 1.78579591836735 8.21065989847716 1
"564" 20 33 1.78579591836735 8.21065989847716 1
"565" 20 34 1.78579591836735 8.21065989847716 1
"566" 20 35 1.78579591836735 8.21065989847716 1
"567" 20 36 1.78579591836735 8.21065989847716 1
"568" 20 37 1.78579591836735 8.21065989847716 1
"569" 20 38 1.78579591836735 8.21065989847716 0
"570" 20 39 1.78579591836735 8.21065989847716 0
"571" 20 40 1.78579591836735 8.21065989847716 0
"572" 20 41 1.78579591836735 8.21065989847716 1
"573" 20 42 1.78579591836735 8.21065989847716 1
"574" 20 43 1.78579591836735 8.21065989847716 1
"575" 20 44 1.78579591836735 8.21065989847716 1
"576" 20 45 1.78579591836735 8.21065989847716 1
"577" 20 46 1.78579591836735 8.21065989847716 1
"578" 20 47 1.78579591836735 8.21065989847716 1
"579" 20 48 1.78579591836735 8.21065989847716 1
"580" 20 49 1.78579591836735 8.21065989847716 1
"581" 20 50 1.78579591836735 8.21065989847716 1
"582" 20 51 1.78579591836735 8.21065989847716 1
"583" 20 52 1.78579591836735 8.15482233502538 1
"584" 21 9 1.71138775510204 8.15482233502538 1
"585" 21 10 1.78579591836735 8.0989847715736 1
"586" 21 11 1.78579591836735 8.0989847715736 1
"587" 21 12 1.78579591836735 8.0989847715736 1
"588" 21 13 1.78579591836735 8.0989847715736 1
"589" 21 14 1.78579591836735 8.0989847715736 1
"590" 21 15 1.78579591836735 8.0989847715736 1
"591" 21 16 1.78579591836735 8.0989847715736 1
"592" 21 17 1.78579591836735 8.0989847715736 1
"593" 21 18 1.78579591836735 8.0989847715736 1
"594" 21 19 1.78579591836735 8.0989847715736 1
"595" 21 20 1.78579591836735 8.0989847715736 1
"596" 21 21 1.78579591836735 8.0989847715736 1
"597" 21 22 1.78579591836735 8.0989847715736 1
"598" 21 23 1.78579591836735 8.0989847715736 1
"599" 21 24 1.78579591836735 8.0989847715736 1
"600" 21 25 1.78579591836735 8.0989847715736 1
"601" 21 26 1.78579591836735 8.0989847715736 1
"602" 21 27 1.78579591836735 8.0989847715736 1
"603" 21 28 1.78579591836735 8.0989847715736 1
"604" 21 29 1.78579591836735 8.0989847715736 1
"605" 21 30 1.78579591836735 8.0989847715736 1
"606" 21 31 1.78579591836735 8.0989847715736 1
"607" 21 32 1.78579591836735 8.0989847715736 1
"608" 21 33 1.78579591836735 8.0989847715736 1
"609" 21 34 1.78579591836735 8.0989847715736 1
"610" 21 35 1.78579591836735 8.0989847715736 1
"611" 21 36 1.78579591836735 8.0989847715736 0
"612" 21 37 1.78579591836735 8.0989847715736 1
"613" 21 38 1.78579591836735 8.0989847715736 0
"614" 21 39 1.78579591836735 8.0989847715736 1
"615" 21 40 1.78579591836735 8.0989847715736 1
"616" 21 41 1.78579591836735 8.0989847715736 1
"617" 21 42 1.78579591836735 8.0989847715736 1
"618" 21 43 1.78579591836735 8.0989847715736 1
"619" 21 44 1.78579591836735 8.0989847715736 1
"620" 21 45 1.78579591836735 8.0989847715736 1
"621" 21 46 1.78579591836735 8.0989847715736 1
"622" 21 47 1.78579591836735 8.0989847715736 1
"623" 21 48 1.78579591836735 8.0989847715736 1
"624" 21 49 1.78579591836735 8.0989847715736 1
"625" 21 50 1.78579591836735 8.0989847715736 1
"626" 21 51 1.78579591836735 8.0989847715736 1
"627" 21 52 1.78579591836735 8.0989847715736 0
"628" 22 8 1.78579591836735 8.0989847715736 1
"629" 22 9 1.78579591836735 8.0989847715736 1
"630" 22 10 1.78579591836735 8.0989847715736 1
"631" 22 11 1.78579591836735 8.0989847715736 1
"632" 22 12 1.78579591836735 8.0989847715736 1
"633" 22 13 1.78579591836735 8.0989847715736 1
```

```
"634" 22 14 1.78579591836735 8.0989847715736 1
"635" 22 15 1.78579591836735 8.0989847715736 1
"636" 22 16 1.78579591836735 8.0989847715736 1
"637" 22 17 1.78579591836735 8.0989847715736 1
"638" 22 18 1.78579591836735 8.0989847715736 1
"639" 22 19 1.78579591836735 8.0989847715736 1
"640" 22 20 1.78579591836735 8.0989847715736 1
"641" 22 21 1.78579591836735 8.0989847715736 1
"642" 22 22 1.78579591836735 8.0989847715736 1
"643" 22 23 1.78579591836735 8.0989847715736 1
"644" 22 24 1.78579591836735 8.0989847715736 1
"645" 22 25 1.78579591836735 8.0989847715736 1
"646" 22 26 1.78579591836735 8.0989847715736 1
"647" 22 27 1.78579591836735 8.0989847715736 1
"648" 22 28 1.78579591836735 8.0989847715736 1
"649" 22 29 1.78579591836735 8.0989847715736 1
"650" 22 30 1.78579591836735 8.0989847715736 1
"651" 22 31 1.78579591836735 8.0989847715736 1
"652" 22 32 1.78579591836735 8.0989847715736 1
"653" 22 33 1.78579591836735 8.0989847715736 1
"654" 22 34 1.78579591836735 8.0989847715736 1
"655" 22 35 1.78579591836735 8.0989847715736 1
"656" 22 36 1.78579591836735 8.0989847715736 1
"657" 22 37 1.78579591836735 8.0989847715736 1
"658" 22 38 1.78579591836735 8.0989847715736 1
"659" 22 39 1.78579591836735 8.0989847715736 0
"660" 22 40 1.78579591836735 8.0989847715736 1
"661" 22 41 1.78579591836735 8.0989847715736 1
"662" 22 42 1.78579591836735 8.0989847715736 1
"663" 22 43 1.78579591836735 8.0989847715736 1
"664" 22 44 1.78579591836735 8.0989847715736 1
"665" 22 45 1.78579591836735 8.0989847715736 1
"666" 22 46 1.78579591836735 8.0989847715736 1
"667" 22 47 1.78579591836735 8.0989847715736 1
"668" 22 48 1.78579591836735 8.0989847715736 1
"669" 22 49 1.78579591836735 8.0989847715736 1
"670" 22 50 1.78579591836735 8.0989847715736 1
"671" 22 51 1.78579591836735 8.0989847715736 1
"672" 23 8 1.78579591836735 8.54568527918782 1
"673" 23 9 1.78579591836735 8.15482233502538 1
"674" 23 10 1.78579591836735 8.15482233502538 1
"675" 23 11 1.78579591836735 8.0989847715736 1
"676" 23 12 1.78579591836735 8.15482233502538 1
"677" 23 13 1.78579591836735 8.21065989847716 1
"678" 23 14 1.78579591836735 8.21065989847716 1
"679" 23 15 1.78579591836735 8.21065989847716 1
"680" 23 16 1.78579591836735 8.21065989847716 1
"681" 23 17 1.78579591836735 8.21065989847716 1
"682" 23 18 1.78579591836735 8.21065989847716 1
"683" 23 19 1.78579591836735 8.21065989847716 1
"684" 23 20 1.78579591836735 8.21065989847716 1
"685" 23 21 1.78579591836735 8.21065989847716 1
"686" 23 22 1.78579591836735 8.21065989847716 1
"687" 23 23 1.78579591836735 8.21065989847716 1
"688" 23 24 1.78579591836735 8.21065989847716 1
"689" 23 25 1.78579591836735 8.21065989847716 1
"690" 23 26 1.78579591836735 8.21065989847716 1
"691" 23 27 1.78579591836735 8.21065989847716 1
"692" 23 28 1.78579591836735 8.21065989847716 1
"693" 23 29 1.78579591836735 8.21065989847716 1
"694" 23 30 1.78579591836735 8.21065989847716 1
"695" 23 31 1.78579591836735 8.21065989847716 1
"696" 23 32 1.78579591836735 8.21065989847716 1
"697" 23 33 1.78579591836735 8.21065989847716 1
"698" 23 34 1.78579591836735 8.21065989847716 1
"699" 23 35 1.78579591836735 8.21065989847716 1
"700" 23 36 1.78579591836735 8.21065989847716 1
"701" 23 37 1.78579591836735 8.21065989847716 0
"702" 23 38 1.78579591836735 8.21065989847716 1
"703" 23 39 1.78579591836735 8.21065989847716 1
"704" 23 40 1.78579591836735 8.21065989847716 0
"705" 23 41 1.78579591836735 8.21065989847716 1
"706" 23 42 1.78579591836735 8.21065989847716 1
"707" 23 43 1.78579591836735 8.21065989847716 1
"708" 23 44 1.78579591836735 8.21065989847716 1
```

```
"709" 23 45 1.78579591836735 8.21065989847716 1
"710" 23 46 1.78579591836735 8.21065989847716 1
"711" 23 47 1.78579591836735 8.21065989847716 1
"712" 23 48 1.78579591836735 8.21065989847716 1
"713" 23 49 1.78579591836735 8.21065989847716 1
"714" 23 50 1.78579591836735 8.21065989847716 1
"715" 23 51 1.78579591836735 8.15482233502538 0
"716" 24 8 1.78579591836735 8.60152284263959 1
"717" 24 9 1.78579591836735 8.60152284263959 1
"718" 24 10 1.78579591836735 8.60152284263959 1
"719" 24 11 1.78579591836735 8.15482233502538 1
"720" 24 12 1.78579591836735 8.60152284263959 1
"721" 24 13 1.78579591836735 8.15482233502538 1
"722" 24 14 1.78579591836735 8.15482233502538 1
"723" 24 15 1.78579591836735 8.15482233502538 1
"724" 24 16 1.78579591836735 8.15482233502538 1
"725" 24 17 1.78579591836735 8.15482233502538 1
"726" 24 18 1.78579591836735 8.15482233502538 1
"727" 24 19 1.78579591836735 8.15482233502538 1
"728" 24 20 1.78579591836735 8.15482233502538 1
"729" 24 21 1.78579591836735 8.15482233502538 1
"730" 24 22 1.78579591836735 8.15482233502538 1
"731" 24 23 1.78579591836735 8.15482233502538 1
"732" 24 24 1.78579591836735 8.15482233502538 1
"733" 24 25 1.78579591836735 8.15482233502538 1
"734" 24 26 1.78579591836735 8.15482233502538 1
"735" 24 27 1.78579591836735 8.15482233502538 1
"736" 24 28 1.78579591836735 8.15482233502538 1
"737" 24 29 1.78579591836735 8.15482233502538 1
"738" 24 30 1.78579591836735 8.15482233502538 1
"739" 24 31 1.78579591836735 8.15482233502538 1
"740" 24 32 1.78579591836735 8.15482233502538 1
"741" 24 33 1.78579591836735 8.15482233502538 1
"742" 24 34 1.78579591836735 8.15482233502538 1
"743" 24 35 1.78579591836735 8.15482233502538 1
"744" 24 36 1.78579591836735 8.15482233502538 1
"745" 24 37 1.78579591836735 8.15482233502538 1
"746" 24 38 1.78579591836735 8.15482233502538 1
"747" 24 39 1.78579591836735 8.15482233502538 1
"748" 24 40 1.78579591836735 8.15482233502538 1
"749" 24 41 1.78579591836735 8.15482233502538 1
"750" 24 42 1.78579591836735 8.15482233502538 1
"751" 24 43 1.78579591836735 8.15482233502538 1
"752" 24 44 1.78579591836735 8.15482233502538 1
"753" 24 45 1.78579591836735 8.15482233502538 1
"754" 24 46 1.78579591836735 8.15482233502538 1
"755" 24 47 1.78579591836735 8.15482233502538 1
"756" 24 48 1.78579591836735 8.15482233502538 1
"757" 24 49 1.78579591836735 8.15482233502538 1
"758" 24 50 1.78579591836735 8.15482233502538 1
"759" 24 51 1.78579591836735 8.21065989847716 1
"760" 24 52 1.86020408163265 8.15482233502538 0
"761" 25 6 1.48816326530612 8.65736040609137 1
"762" 25 7 1.78579591836735 8.60152284263959 1
"763" 25 8 1.78579591836735 8.60152284263959 1
"764" 25 9 1.78579591836735 8.60152284263959 1
"765" 25 10 1.78579591836735 8.60152284263959 1
"766" 25 11 1.78579591836735 8.60152284263959 1
"767" 25 12 1.78579591836735 8.60152284263959 1
"768" 25 13 1.78579591836735 8.54568527918782 1
"769" 25 14 1.78579591836735 8.0989847715736 1
"770" 25 15 1.78579591836735 8.0989847715736 1
"771" 25 16 1.78579591836735 8.0989847715736 1
"772" 25 17 1.78579591836735 8.0989847715736 1
"773" 25 18 1.78579591836735 8.0989847715736 1
"774" 25 19 1.78579591836735 8.0989847715736 1
"775" 25 20 1.78579591836735 8.0989847715736 1
"776" 25 21 1.78579591836735 8.0989847715736 1
"777" 25 22 1.78579591836735 8.0989847715736 1
"778" 25 23 1.78579591836735 8.0989847715736 1
"779" 25 24 1.78579591836735 8.0989847715736 1
"780" 25 25 1.78579591836735 8.0989847715736 1
"781" 25 26 1.78579591836735 8.0989847715736 1
"782" 25 27 1.78579591836735 8.0989847715736 1
"783" 25 28 1.78579591836735 8.0989847715736 1
```

```
"784" 25 29 1.78579591836735 8.0989847715736 1
"785" 25 30 1.78579591836735 8.0989847715736 1
"786" 25 31 1.78579591836735 8.0989847715736 1
"787" 25 32 1.78579591836735 8.0989847715736 1
"788" 25 33 1.78579591836735 8.0989847715736 1
"789" 25 34 1.78579591836735 8.0989847715736 0
"790" 25 35 1.78579591836735 8.0989847715736 0
"791" 25 36 1.78579591836735 8.0989847715736 1
"792" 25 37 1.78579591836735 8.0989847715736 1
"793" 25 38 1.78579591836735 8.0989847715736 1
"794" 25 39 1.78579591836735 8.0989847715736 1
"795" 25 40 1.78579591836735 8.0989847715736 1
"796" 25 41 1.78579591836735 8.0989847715736 1
"797" 25 42 1.78579591836735 8.0989847715736 1
"798" 25 43 1.78579591836735 8.0989847715736 1
"799" 25 44 1.78579591836735 8.0989847715736 1
"800" 25 45 1.78579591836735 8.0989847715736 1
"801" 25 46 1.78579591836735 8.0989847715736 1
"802" 25 47 1.78579591836735 8.0989847715736 1
"803" 25 48 1.78579591836735 8.0989847715736 1
"804" 25 49 1.78579591836735 8.0989847715736 1
"805" 25 50 1.78579591836735 8.0989847715736 1
"806" 25 51 1.78579591836735 8.0989847715736 1
"807" 25 52 1.78579591836735 7.98730964467005 0
"808" 25 53 1.71138775510204 8.21065989847716 1
"809" 26 6 1.78579591836735 8.60152284263959 1
"810" 26 7 1.78579591836735 8.60152284263959 1
"811" 26 8 1.78579591836735 8.60152284263959 1
"812" 26 9 1.78579591836735 8.60152284263959 1
"813" 26 10 1.78579591836735 8.60152284263959 1
"814" 26 11 1.78579591836735 8.60152284263959 1
"815" 26 12 1.78579591836735 8.15482233502538 1
"816" 26 13 1.78579591836735 8.15482233502538 1
"817" 26 14 1.78579591836735 8.0989847715736 1
"818" 26 15 1.78579591836735 8.0989847715736 1
"819" 26 16 1.78579591836735 8.15482233502538 1
"820" 26 17 1.78579591836735 8.15482233502538 1
"821" 26 18 1.78579591836735 8.15482233502538 1
"822" 26 19 1.78579591836735 8.15482233502538 1
"823" 26 20 1.78579591836735 8.60152284263959 1
"824" 26 21 1.78579591836735 8.15482233502538 1
"825" 26 22 1.78579591836735 8.15482233502538 1
"826" 26 23 1.78579591836735 8.15482233502538 1
"827" 26 24 1.78579591836735 8.15482233502538 1
"828" 26 25 1.78579591836735 8.15482233502538 1
"829" 26 26 1.78579591836735 8.15482233502538 1
"830" 26 27 1.78579591836735 8.15482233502538 1
"831" 26 28 1.78579591836735 8.15482233502538 1
"832" 26 29 1.78579591836735 8.15482233502538 1
"833" 26 30 1.78579591836735 8.15482233502538 1
"834" 26 31 1.78579591836735 8.15482233502538 1
"835" 26 32 1.78579591836735 8.15482233502538 1
"836" 26 33 1.78579591836735 8.15482233502538 1
"837" 26 34 1.78579591836735 8.15482233502538 1
"838" 26 35 1.78579591836735 8.15482233502538 1
"839" 26 36 1.78579591836735 8.15482233502538 1
"840" 26 37 1.78579591836735 8.15482233502538 1
"841" 26 38 1.78579591836735 8.15482233502538 1
"842" 26 39 1.78579591836735 8.15482233502538 1
"843" 26 40 1.78579591836735 8.15482233502538 1
"844" 26 41 1.78579591836735 8.15482233502538 1
"845" 26 42 1.78579591836735 8.15482233502538 1
"846" 26 43 1.78579591836735 8.15482233502538 1
"847" 26 44 1.78579591836735 8.65736040609137 1
"848" 26 45 1.78579591836735 8.60152284263959 1
"849" 26 46 1.78579591836735 8.60152284263959 1
"850" 26 47 1.78579591836735 8.65736040609137 0
"851" 26 48 1.78579591836735 8.15482233502538 1
"852" 26 49 1.78579591836735 8.15482233502538 1
"853" 26 50 1.78579591836735 8.15482233502538 1
"854" 26 51 1.78579591836735 8.15482233502538 0
"855" 26 52 1.78579591836735 8.15482233502538 1
"856" 26 53 1.78579591836735 8.15482233502538 0
"857" 27 6 1.71138775510204 8.65736040609137 1
"858" 27 7 1.71138775510204 8.65736040609137 1
```

```
"859" 27 8 1.78579591836735 8.54568527918782 1
"860" 27 9 1.78579591836735 8.60152284263959 1
"861" 27 10 1.78579591836735 8.60152284263959 1
"862" 27 11 1.78579591836735 8.60152284263959 1
"863" 27 12 1.78579591836735 8.60152284263959 1
"864" 27 13 1.78579591836735 8.15482233502538 1
"865" 27 14 1.78579591836735 8.15482233502538 1
"866" 27 15 1.78579591836735 8.60152284263959 1
"867" 27 16 1.78579591836735 8.21065989847716 1
"868" 27 17 1.78579591836735 8.0989847715736 1
"869" 27 18 1.78579591836735 8.15482233502538 1
"870" 27 19 1.78579591836735 8.60152284263959 0
"871" 27 20 1.78579591836735 8.60152284263959 1
"872" 27 21 1.78579591836735 8.60152284263959 1
"873" 27 22 1.78579591836735 8.15482233502538 1
"874" 27 23 1.78579591836735 8.15482233502538 1
"875" 27 24 1.78579591836735 8.15482233502538 1
"876" 27 25 1.78579591836735 8.15482233502538 1
"877" 27 26 1.78579591836735 8.15482233502538 1
"878" 27 27 1.78579591836735 8.15482233502538 1
"879" 27 28 1.78579591836735 8.15482233502538 1
"880" 27 29 1.78579591836735 8.65736040609137 1
"881" 27 30 1.78579591836735 8.65736040609137 1
"882" 27 31 1.78579591836735 8.21065989847716 1
"883" 27 32 1.78579591836735 8.15482233502538 1
"884" 27 33 1.78579591836735 8.15482233502538 1
"885" 27 34 1.78579591836735 8.15482233502538 1
"886" 27 35 1.78579591836735 8.15482233502538 1
"887" 27 36 1.78579591836735 8.15482233502538 1
"888" 27 37 1.78579591836735 8.15482233502538 1
"889" 27 38 1.78579591836735 8.15482233502538 1
"890" 27 39 1.78579591836735 8.15482233502538 1
"891" 27 40 1.78579591836735 8.15482233502538 1
"892" 27 41 1.78579591836735 8.15482233502538 1
"893" 27 42 1.78579591836735 8.15482233502538 1
"894" 27 43 1.78579591836735 8.15482233502538 1
"895" 27 44 1.78579591836735 8.15482233502538 1
"896" 27 45 1.78579591836735 8.60152284263959 1
"897" 27 46 1.78579591836735 8.65736040609137 0
"898" 27 47 1.78579591836735 8.60152284263959 1
"899" 27 48 1.78579591836735 8.15482233502538 1
"900" 27 49 1.78579591836735 8.15482233502538 1
"901" 27 50 1.78579591836735 8.15482233502538 1
"902" 27 51 1.78579591836735 8.15482233502538 0
"903" 27 52 1.78579591836735 8.15482233502538 0
"904" 27 53 1.78579591836735 8.15482233502538 0
"905" 28 8 1.78579591836735 8.60152284263959 1
"906" 28 9 1.78579591836735 8.60152284263959 1
"907" 28 10 1.78579591836735 8.60152284263959 1
"908" 28 11 1.78579591836735 8.60152284263959 1
"909" 28 12 1.78579591836735 8.54568527918782 1
"910" 28 13 1.78579591836735 8.60152284263959 1
"911" 28 14 1.78579591836735 8.21065989847716 1
"912" 28 15 1.78579591836735 8.60152284263959 1
"913" 28 16 1.78579591836735 8.60152284263959 1
"914" 28 17 1.78579591836735 8.0989847715736 1
"915" 28 18 1.78579591836735 8.60152284263959 1
"916" 28 19 1.78579591836735 8.60152284263959 1
"917" 28 20 1.71138775510204 8.60152284263959 1
"918" 28 21 1.78579591836735 8.60152284263959 1
"919" 28 22 1.78579591836735 8.0989847715736 1
"920" 28 23 1.78579591836735 8.0989847715736 0
"921" 28 24 1.78579591836735 8.0989847715736 0
"922" 28 25 1.78579591836735 8.0989847715736 1
"923" 28 26 1.78579591836735 8.0989847715736 1
"924" 28 27 1.78579591836735 8.0989847715736 1
"925" 28 28 1.78579591836735 8.0989847715736 1
"926" 28 29 1.78579591836735 8.04314720812183 1
"927" 28 30 1.78579591836735 8.15482233502538 1
"928" 28 31 1.71138775510204 8.0989847715736 1
"929" 28 32 1.78579591836735 8.0989847715736 0
"930" 28 33 1.78579591836735 8.0989847715736 1
"931" 28 34 1.78579591836735 8.0989847715736 1
"932" 28 35 1.78579591836735 8.0989847715736 1
"933" 28 36 1.78579591836735 8.0989847715736 1
```

```
"934" 28 37 1.78579591836735 8.0989847715736 0
"935" 28 38 1.78579591836735 8.0989847715736 1
"936" 28 39 1.78579591836735 8.0989847715736 1
"937" 28 40 1.78579591836735 8.0989847715736 1
"938" 28 41 1.78579591836735 8.0989847715736 1
"939" 28 42 1.78579591836735 8.0989847715736 1
"940" 28 43 1.78579591836735 8.0989847715736 1
"941" 28 44 1.78579591836735 8.0989847715736 1
"942" 28 45 1.78579591836735 8.60152284263959 1
"943" 28 46 1.78579591836735 8.04314720812183 1
"944" 28 47 1.78579591836735 8.15482233502538 1
"945" 28 48 1.78579591836735 8.0989847715736 1
"946" 28 49 1.78579591836735 8.0989847715736 1
"947" 28 50 1.78579591836735 8.0989847715736 1
"948" 28 51 1.78579591836735 8.0989847715736 1
"949" 28 52 1.78579591836735 8.0989847715736 1
"950" 28 53 1.71138775510204 8.0989847715736 1
"951" 29 7 1.78579591836735 8.54568527918782 0
"952" 29 8 1.78579591836735 8.54568527918782 1
"953" 29 9 1.78579591836735 8.60152284263959 1
"954" 29 10 1.78579591836735 8.60152284263959 1
"955" 29 11 1.78579591836735 8.60152284263959 1
"956" 29 12 1.78579591836735 8.60152284263959 1
"957" 29 13 1.78579591836735 8.65736040609137 1
"958" 29 14 1.78579591836735 8.60152284263959 0
"959" 29 15 1.78579591836735 8.15482233502538 1
"960" 29 16 1.78579591836735 8.0989847715736 1
"961" 29 17 1.78579591836735 8.0989847715736 1
"962" 29 18 1.78579591836735 8.60152284263959 1
"963" 29 19 1.78579591836735 8.60152284263959 1
"964" 29 20 1.78579591836735 8.04314720812183 0
"965" 29 21 3.57159183673469 8.0989847715736 0
"966" 29 22 3.57159183673469 8.0989847715736 0
"967" 29 23 3.57159183673469 8.0989847715736 0
"968" 29 24 1.78579591836735 8.0989847715736 0
"969" 29 25 1.78579591836735 8.0989847715736 0
"970" 29 26 1.78579591836735 8.0989847715736 0
"971" 29 27 1.78579591836735 8.0989847715736 1
"972" 29 28 1.78579591836735 8.0989847715736 0
"973" 29 29 1.78579591836735 8.0989847715736 1
"974" 29 30 3.49718367346939 8.0989847715736 1
"975" 29 31 1.78579591836735 8.0989847715736 1
"976" 29 32 3.57159183673469 8.0989847715736 1
"977" 29 33 3.57159183673469 8.0989847715736 0
"978" 29 34 1.78579591836735 8.0989847715736 0
"979" 29 35 1.78579591836735 8.0989847715736 1
"980" 29 36 1.78579591836735 8.0989847715736 1
"981" 29 37 1.78579591836735 8.0989847715736 1
"982" 29 38 1.78579591836735 8.0989847715736 1
"983" 29 39 1.78579591836735 8.0989847715736 1
"984" 29 40 1.78579591836735 8.0989847715736 1
"985" 29 41 1.78579591836735 8.0989847715736 1
"986" 29 42 1.78579591836735 8.0989847715736 1
"987" 29 43 1.78579591836735 8.0989847715736 1
"988" 29 44 1.78579591836735 8.0989847715736 1
"989" 29 45 1.78579591836735 8.60152284263959 1
"990" 29 46 1.78579591836735 8.60152284263959 1
"991" 29 47 1.78579591836735 8.0989847715736 1
"992" 29 48 1.78579591836735 8.0989847715736 1
"993" 29 49 1.78579591836735 8.0989847715736 1
"994" 29 50 1.78579591836735 8.0989847715736 1
"995" 29 51 1.78579591836735 8.0989847715736 1
"996" 29 52 1.78579591836735 8.0989847715736 1
"997" 29 53 1.78579591836735 8.0989847715736 1
"998" 29 54 1.86020408163265 8.26649746192893 1
"999" 30 6 1.78579591836735 8.54568527918782 1
"1000" 30 7 1.78579591836735 8.65736040609137 1
"1001" 30 8 1.78579591836735 8.60152284263959 1
"1002" 30 9 1.78579591836735 8.60152284263959 1
"1003" 30 10 1.78579591836735 8.60152284263959 1
"1004" 30 11 1.78579591836735 8.60152284263959 1
"1005" 30 12 1.78579591836735 8.60152284263959 1
"1006" 30 13 1.78579591836735 8.65736040609137 1
"1007" 30 14 3.57159183673469 8.15482233502538 1
"1008" 30 15 3.49718367346939 8.15482233502538 1
```

```
"1009" 30 16 1.78579591836735 8.15482233502538 1
"1010" 30 17 1.78579591836735 8.60152284263959 1
"1011" 30 18 3.49718367346939 8.60152284263959 1
"1012" 30 19 3.57159183673469 8.0989847715736 1
"1013" 30 20 3.57159183673469 8.15482233502538 0
"1014" 30 21 1.78579591836735 8.15482233502538 1
"1015" 30 22 3.7204081632653 8.15482233502538 1
"1016" 30 23 3.57159183673469 8.15482233502538 1
"1017" 30 24 1.78579591836735 8.15482233502538 0
"1018" 30 25 3.57159183673469 8.15482233502538 1
"1019" 30 26 1.86020408163265 8.15482233502538 0
"1020" 30 27 1.78579591836735 8.15482233502538 0
"1021" 30 28 1.78579591836735 8.15482233502538 0
"1022" 30 29 1.78579591836735 8.15482233502538 0
"1023" 30 30 3.57159183673469 8.15482233502538 0
"1024" 30 31 3.57159183673469 8.15482233502538 0
"1025" 30 32 3.57159183673469 8.15482233502538 0
"1026" 30 33 3.57159183673469 8.15482233502538 1
"1027" 30 34 1.78579591836735 8.15482233502538 0
"1028" 30 35 1.78579591836735 8.15482233502538 0
"1029" 30 36 1.78579591836735 8.15482233502538 1
"1030" 30 37 1.78579591836735 8.15482233502538 0
"1031" 30 38 1.78579591836735 8.15482233502538 0
"1032" 30 39 1.78579591836735 8.15482233502538 0
"1033" 30 40 3.57159183673469 8.15482233502538 1
"1034" 30 41 1.78579591836735 8.15482233502538 1
"1035" 30 42 1.78579591836735 8.15482233502538 1
"1036" 30 43 1.78579591836735 8.15482233502538 1
"1037" 30 44 1.78579591836735 8.15482233502538 1
"1038" 30 45 1.78579591836735 8.15482233502538 1
"1039" 30 46 1.78579591836735 8.15482233502538 1
"1040" 30 47 1.78579591836735 8.15482233502538 1
"1041" 30 48 1.78579591836735 8.15482233502538 1
"1042" 30 49 1.78579591836735 8.15482233502538 1
"1043" 30 50 1.78579591836735 8.15482233502538 1
"1044" 30 51 1.78579591836735 8.15482233502538 1
"1045" 30 52 1.78579591836735 8.15482233502538 1
"1046" 30 53 1.78579591836735 8.15482233502538 1
"1047" 30 54 1.78579591836735 8.0989847715736 0
"1048" 31 6 1.56257142857143 8.71319796954315 1
"1049" 31 7 1.78579591836735 8.60152284263959 1
"1050" 31 8 1.78579591836735 8.60152284263959 1
"1051" 31 9 1.78579591836735 8.60152284263959 0
"1052" 31 10 1.78579591836735 8.60152284263959 0
"1053" 31 11 1.78579591836735 8.60152284263959 1
"1054" 31 12 1.78579591836735 8.60152284263959 1
"1055" 31 13 1.78579591836735 8.60152284263959 1
"1056" 31 14 1.78579591836735 8.60152284263959 1
"1057" 31 15 1.86020408163265 8.60152284263959 1
"1058" 31 16 3.57159183673469 8.21065989847716 1
"1059" 31 17 1.93461224489796 8.60152284263959 1
"1060" 31 18 1.71138775510204 8.60152284263959 1
"1061" 31 19 3.57159183673469 8.21065989847716 1
"1062" 31 20 3.57159183673469 8.15482233502538 0
"1063" 31 21 3.57159183673469 8.15482233502538 1
"1064" 31 22 1.71138775510204 8.15482233502538 0
"1065" 31 23 3.57159183673469 8.15482233502538 0
"1066" 31 24 3.57159183673469 8.15482233502538 0
"1067" 31 25 3.57159183673469 8.15482233502538 0
"1068" 31 26 3.57159183673469 8.15482233502538 0
"1069" 31 27 3.57159183673469 8.15482233502538 0
"1070" 31 28 3.646 8.15482233502538 0
"1071" 31 29 1.78579591836735 8.15482233502538 0
"1072" 31 30 3.57159183673469 8.15482233502538 1
"1073" 31 31 1.78579591836735 8.15482233502538 0
"1074" 31 32 1.78579591836735 8.15482233502538 0
"1075" 31 33 3.57159183673469 8.15482233502538 0
"1076" 31 34 1.86020408163265 8.15482233502538 1
"1077" 31 35 1.78579591836735 8.0989847715736 0
"1078" 31 36 1.78579591836735 8.65736040609137 1
"1079" 31 37 1.78579591836735 8.15482233502538 1
"1080" 31 38 1.78579591836735 8.15482233502538 1
"1081" 31 39 1.78579591836735 8.21065989847716 1
"1082" 31 40 3.57159183673469 8.15482233502538 1
"1083" 31 41 3.57159183673469 8.15482233502538 1
```

```
"1084" 31 42 1.78579591836735 8.15482233502538 1
"1085" 31 43 1.78579591836735 8.15482233502538 1
"1086" 31 44 1.78579591836735 8.15482233502538 1
"1087" 31 45 1.78579591836735 8.15482233502538 1
"1088" 31 46 1.78579591836735 8.15482233502538 1
"1089" 31 47 1.78579591836735 8.15482233502538 1
"1090" 31 48 1.78579591836735 8.15482233502538 1
"1091" 31 49 1.78579591836735 8.15482233502538 1
"1092" 31 50 1.78579591836735 8.15482233502538 1
"1093" 31 51 1.78579591836735 8.15482233502538 1
"1094" 31 52 1.78579591836735 8.15482233502538 1
"1095" 31 53 1.78579591836735 8.15482233502538 1
"1096" 31 54 1.78579591836735 8.0989847715736 1
"1097" 32 2 1.78579591836735 8.76903553299493 0
"1098" 32 3 1.71138775510204 0.281725888324875 0
"1099" 32 5 1.78579591836735 8.60152284263959 1
"1100" 32 6 1.78579591836735 8.60152284263959 1
"1101" 32 7 1.78579591836735 8.60152284263959 1
"1102" 32 8 1.78579591836735 8.60152284263959 1
"1103" 32 9 1.78579591836735 8.60152284263959 1
"1104" 32 10 1.78579591836735 8.60152284263959 0
"1105" 32 11 1.78579591836735 8.60152284263959 1
"1106" 32 12 1.78579591836735 8.60152284263959 1
"1107" 32 13 1.78579591836735 8.60152284263959 1
"1108" 32 14 1.78579591836735 8.60152284263959 1
"1109" 32 15 1.78579591836735 8.60152284263959 1
"1110" 32 16 1.86020408163265 8.60152284263959 1
"1111" 32 17 3.646 8.04314720812183 1
"1112" 32 18 3.57159183673469 8.0989847715736 1
"1113" 32 19 3.57159183673469 8.0989847715736 0
"1114" 32 20 3.57159183673469 8.0989847715736 0
"1115" 32 21 3.57159183673469 8.0989847715736 0
"1116" 32 22 3.57159183673469 8.0989847715736 0
"1117" 32 23 3.57159183673469 8.0989847715736 0
"1118" 32 24 3.57159183673469 8.0989847715736 1
"1119" 32 25 3.57159183673469 8.0989847715736 0
"1120" 32 26 3.57159183673469 8.0989847715736 0
"1121" 32 27 5.28297959183674 8.0989847715736 0
"1122" 32 28 3.57159183673469 8.04314720812183 0
"1123" 32 29 3.49718367346939 8.04314720812183 0
"1124" 32 30 3.57159183673469 8.0989847715736 0
"1125" 32 31 3.646 8.0989847715736 0
"1126" 32 32 3.42277551020408 8.0989847715736 0
"1127" 32 33 1.78579591836735 8.0989847715736 0
"1128" 32 34 1.71138775510204 8.0989847715736 1
"1129" 32 35 1.78579591836735 8.0989847715736 0
"1130" 32 36 1.78579591836735 8.65736040609137 0
"1131" 32 37 1.78579591836735 8.60152284263959 1
"1132" 32 38 1.78579591836735 8.60152284263959 1
"1133" 32 39 1.78579591836735 8.65736040609137 1
"1134" 32 40 1.71138775510204 8.15482233502538 1
"1135" 32 41 3.57159183673469 8.0989847715736 1
"1136" 32 42 1.78579591836735 8.0989847715736 1
"1137" 32 43 1.78579591836735 8.0989847715736 1
"1138" 32 44 1.78579591836735 8.0989847715736 1
"1139" 32 45 1.78579591836735 8.0989847715736 1
"1140" 32 46 1.78579591836735 8.0989847715736 1
"1141" 32 47 1.78579591836735 8.0989847715736 1
"1142" 32 48 1.78579591836735 8.0989847715736 1
"1143" 32 49 1.78579591836735 8.0989847715736 1
"1144" 32 50 1.78579591836735 8.0989847715736 1
"1145" 32 51 1.78579591836735 8.0989847715736 1
"1146" 32 52 1.78579591836735 8.0989847715736 1
"1147" 32 53 1.78579591836735 8.0989847715736 1
"1148" 32 54 1.78579591836735 8.04314720812183 0
"1149" 33 1 1.78579591836735 8.60152284263959 1
"1150" 33 2 1.86020408163265 8.60152284263959 1
"1151" 33 3 1.78579591836735 8.60152284263959 1
"1152" 33 4 1.78579591836735 8.71319796954315 1
"1153" 33 5 1.78579591836735 8.60152284263959 1
"1154" 33 6 1.78579591836735 8.60152284263959 1
"1155" 33 7 1.78579591836735 8.60152284263959 1
"1156" 33 8 1.78579591836735 8.60152284263959 1
"1157" 33 9 1.78579591836735 8.60152284263959 1
"1158" 33 10 1.78579591836735 8.60152284263959 1
```

```
"1159" 33 11 1.78579591836735 8.60152284263959 1
"1160" 33 12 1.78579591836735 8.60152284263959 1
"1161" 33 13 1.78579591836735 8.60152284263959 0
"1162" 33 14 1.78579591836735 8.60152284263959 1
"1163" 33 15 1.78579591836735 8.60152284263959 1
"1164" 33 16 1.78579591836735 8.60152284263959 1
"1165" 33 17 1.78579591836735 8.60152284263959 1
"1166" 33 18 3.57159183673469 8.0989847715736 1
"1167" 33 19 5.28297959183674 8.0989847715736 1
"1168" 33 20 3.57159183673469 8.04314720812183 0
"1169" 33 21 3.57159183673469 8.04314720812183 1
"1170" 33 22 3.57159183673469 8.15482233502538 0
"1171" 33 23 5.28297959183674 8.04314720812183 0
"1172" 33 24 3.57159183673469 8.0989847715736 0
"1173" 33 25 3.57159183673469 8.04314720812183 0
"1174" 33 26 3.57159183673469 8.0989847715736 0
"1175" 33 27 6.54791836734694 7.48477157360406 1
"1176" 33 28 6.77114285714286 6.25634517766497 0
"1177" 33 29 8.25930612244898 6.25634517766497 0
"1178" 33 30 5.13416326530612 7.48477157360406 0
"1179" 33 31 3.646 8.15482233502538 1
"1180" 33 32 3.57159183673469 8.0989847715736 1
"1181" 33 33 3.57159183673469 8.04314720812183 0
"1182" 33 34 3.57159183673469 8.15482233502538 0
"1183" 33 35 1.86020408163265 8.49984771573604 1
"1184" 33 36 1.78579591836735 8.60152284263959 0
"1185" 33 37 1.78579591836735 8.60152284263959 1
"1186" 33 38 1.78579591836735 8.60152284263959 1
"1187" 33 39 1.78579591836735 8.65736040609137 1
"1188" 33 40 1.78579591836735 8.54568527918782 1
"1189" 33 41 1.86020408163265 8.60152284263959 1
"1190" 33 42 1.78579591836735 8.04314720812183 1
"1191" 33 43 1.78579591836735 8.04314720812183 1
"1192" 33 44 1.78579591836735 8.04314720812183 1
"1193" 33 45 1.78579591836735 8.04314720812183 1
"1194" 33 46 1.78579591836735 8.04314720812183 1
"1195" 33 47 1.78579591836735 8.04314720812183 1
"1196" 33 48 1.78579591836735 8.04314720812183 1
"1197" 33 49 1.78579591836735 8.04314720812183 1
"1198" 33 50 1.78579591836735 8.60152284263959 1
"1199" 33 51 1.78579591836735 8.04314720812183 1
"1200" 33 52 1.78579591836735 8.0989847715736 1
"1201" 33 53 1.63697959183673 8.0989847715736 1
"1202" 34 1 1.86020408163265 8.60152284263959 0
"1203" 34 2 1.78579591836735 8.60152284263959 1
"1204" 34 3 1.78579591836735 8.60152284263959 1
"1205" 34 4 1.78579591836735 8.65736040609137 1
"1206" 34 5 1.78579591836735 8.60152284263959 1
"1207" 34 6 1.78579591836735 8.60152284263959 1
"1208" 34 7 1.78579591836735 8.60152284263959 1
"1209" 34 8 1.78579591836735 8.60152284263959 1
"1210" 34 9 1.78579591836735 8.60152284263959 1
"1211" 34 10 1.78579591836735 8.60152284263959 1
"1212" 34 11 1.78579591836735 8.60152284263959 0
"1213" 34 12 1.78579591836735 8.60152284263959 0
"1214" 34 13 1.78579591836735 8.60152284263959 0
"1215" 34 14 1.78579591836735 8.60152284263959 1
"1216" 34 15 1.78579591836735 8.60152284263959 1
"1217" 34 16 1.93461224489796 8.60152284263959 1
"1218" 34 17 1.86020408163265 8.60152284263959 1
"1219" 34 18 1.78579591836735 8.60152284263959 1
"1220" 34 19 5.13416326530612 8.15482233502538 1
"1221" 34 20 3.57159183673469 8.15482233502538 1
"1222" 34 21 3.57159183673469 8.15482233502538 0
"1223" 34 22 3.57159183673469 7.98730964467005 0
"1224" 34 23 6.77114285714286 7.54060913705584 0
"1225" 34 24 5.28297959183674 8.0989847715736 0
"1226" 34 25 3.57159183673469 8.15482233502538 0
"1227" 34 26 3.57159183673469 8.15482233502538 0
"1228" 34 27 5.13416326530612 7.48477157360406 0
"1229" 34 28 8.18489795918367 6.25634517766497 0
"1230" 34 29 9.6730612244898 5.47461928934011 0
"1231" 34 30 8.18489795918367 6.25634517766497 0
"1232" 34 31 6.84555102040816 7.54060913705584 0
"1233" 34 32 5.20857142857143 8.15482233502538 0
```

```
"1234" 34 33 3.72040816326531 8.15482233502538 0
"1235" 34 34 3.57159183673469 8.15482233502538 0
"1236" 34 35 3.57159183673469 8.0989847715736 0
"1237" 34 36 1.71138775510204 8.54568527918782 0
"1238" 34 37 1.78579591836735 8.60152284263959 0
"1239" 34 38 1.78579591836735 8.60152284263959 1
"1240" 34 39 1.78579591836735 8.60152284263959 1
"1241" 34 40 1.78579591836735 8.60152284263959 1
"1242" 34 41 1.78579591836735 8.60152284263959 1
"1243" 34 42 1.78579591836735 8.65736040609137 1
"1244" 34 43 1.78579591836735 8.15482233502538 1
"1245" 34 44 1.78579591836735 8.15482233502538 1
"1246" 34 45 1.78579591836735 8.15482233502538 1
"1247" 34 46 1.78579591836735 8.15482233502538 1
"1248" 34 47 1.78579591836735 8.15482233502538 1
"1249" 34 48 1.78579591836735 8.15482233502538 1
"1250" 34 49 1.78579591836735 8.15482233502538 1
"1251" 34 50 1.78579591836735 8.15482233502538 1
"1252" 34 51 1.78579591836735 8.15482233502538 1
"1253" 34 52 1.78579591836735 8.15482233502538 1
"1254" 34 53 1.78579591836735 8.15482233502538 1
"1255" 34 54 1.78579591836735 8.54568527918782 0
"1256" 35 2 1.78579591836735 8.71319796954315 1
"1257" 35 3 1.78579591836735 8.60152284263959 1
"1258" 35 4 1.78579591836735 8.60152284263959 1
"1259" 35 5 1.78579591836735 9.27157360406091 1
"1260" 35 6 1.78579591836735 9.27157360406091 1
"1261" 35 7 1.78579591836735 9.27157360406091 1
"1262" 35 8 1.78579591836735 8.65736040609137 1
"1263" 35 9 1.78579591836735 8.60152284263959 1
"1264" 35 10 1.78579591836735 8.60152284263959 0
"1265" 35 11 1.78579591836735 8.60152284263959 1
"1266" 35 12 1.78579591836735 8.60152284263959 0
"1267" 35 13 1.78579591836735 8.60152284263959 0
"1268" 35 14 1.78579591836735 8.60152284263959 0
"1269" 35 15 1.78579591836735 8.60152284263959 1
"1270" 35 16 1.71138775510204 8.60152284263959 1
"1271" 35 17 3.57159183673469 8.15482233502538 1
"1272" 35 18 3.57159183673469 8.15482233502538 0
"1273" 35 19 5.20857142857143 8.15482233502538 1
"1274" 35 20 3.57159183673469 8.15482233502538 0
"1275" 35 21 3.57159183673469 8.15482233502538 0
"1276" 35 22 3.57159183673469 8.0989847715736 0
"1277" 35 23 3.646 8.0989847715736 0
"1278" 35 24 5.13416326530612 8.15482233502538 0
"1279" 35 25 3.57159183673469 8.15482233502538 0
"1280" 35 26 3.57159183673469 8.15482233502538 0
"1281" 35 27 3.646 8.0989847715736 0
"1282" 35 28 5.28297959183674 7.48477157360406 0
"1283" 35 29 6.69673469387755 6.92639593908629 0
"1284" 35 30 6.77114285714286 6.92639593908629 0
"1285" 35 31 6.69673469387755 6.92639593908629 0
"1286" 35 32 6.77114285714286 7.48477157360406 0
"1287" 35 33 5.13416326530612 8.15482233502538 0
"1288" 35 34 3.646 8.15482233502538 0
"1289" 35 35 3.57159183673469 8.0989847715736 0
"1290" 35 36 1.71138775510204 8.21065989847716 0
"1291" 35 37 1.78579591836735 8.54568527918782 0
"1292" 35 38 1.78579591836735 8.60152284263959 1
"1293" 35 39 1.78579591836735 8.60152284263959 1
"1294" 35 40 1.78579591836735 8.60152284263959 1
"1295" 35 41 3.57159183673469 8.60152284263959 1
"1296" 35 42 1.78579591836735 8.15482233502538 1
"1297" 35 43 1.78579591836735 8.15482233502538 1
"1298" 35 44 1.78579591836735 8.15482233502538 1
"1299" 35 45 1.78579591836735 8.15482233502538 1
"1300" 35 46 1.78579591836735 8.15482233502538 1
"1301" 35 47 1.78579591836735 8.15482233502538 1
"1302" 35 48 1.78579591836735 8.15482233502538 1
"1303" 35 49 1.78579591836735 8.15482233502538 1
"1304" 35 50 1.78579591836735 8.15482233502538 1
"1305" 35 51 1.78579591836735 8.15482233502538 1
"1306" 35 52 1.78579591836735 8.15482233502538 1
"1307" 35 53 1.78579591836735 8.15482233502538 1
"1308" 35 54 1.78579591836735 8.15482233502538 1
```

```
"1309" 36 2 1.63697959183673 9.27157360406091 1
"1310" 36 3 1.78579591836735 9.27157360406091 1
"1311" 36 4 1.78579591836735 9.21573604060914 1
"1312" 36 5 1.78579591836735 10.2208121827411 0
"1313" 36 6 1.78579591836735 10.2208121827411 1
"1314" 36 7 1.78579591836735 9.27157360406091 1
"1315" 36 8 1.78579591836735 9.21573604060914 0
"1316" 36 9 1.78579591836735 8.60152284263959 1
"1317" 36 10 1.78579591836735 8.60152284263959 1
"1318" 36 11 1.78579591836735 8.60152284263959 0
"1319" 36 12 1.86020408163265 8.60152284263959 0
"1320" 36 13 3.646 8.60152284263959 0
"1321" 36 14 3.57159183673469 8.60152284263959 0
"1322" 36 15 3.646 8.0989847715736 0
"1323" 36 16 5.13416326530612 8.15482233502538 1
"1324" 36 17 5.20857142857143 8.15482233502538 0
"1325" 36 18 5.13416326530612 8.15482233502538 0
"1326" 36 19 5.13416326530612 8.15482233502538 0
"1327" 36 20 3.57159183673469 8.15482233502538 0
"1328" 36 21 3.57159183673469 8.15482233502538 0
"1329" 36 22 3.57159183673469 8.15482233502538 0
"1330" 36 23 5.05975510204082 8.15482233502538 0
"1331" 36 24 5.13416326530612 8.15482233502538 0
"1332" 36 25 3.646 8.15482233502538 0
"1333" 36 26 5.28297959183674 8.15482233502538 0
"1334" 36 27 3.57159183673469 8.15482233502538 0
"1335" 36 28 5.28297959183674 8.0989847715736 0
"1336" 36 29 3.57159183673469 8.0989847715736 1
"1337" 36 30 3.57159183673469 8.0989847715736 0
"1338" 36 31 5.20857142857143 7.48477157360406 0
"1339" 36 32 5.13416326530612 7.48477157360406 0
"1340" 36 33 5.13416326530612 8.0989847715736 0
"1341" 36 34 5.13416326530612 8.15482233502538 0
"1342" 36 35 3.646 8.15482233502538 0
"1343" 36 36 3.57159183673469 8.0989847715736 0
"1344" 36 37 1.78579591836735 8.54568527918782 0
"1345" 36 38 1.78579591836735 8.60152284263959 0
"1346" 36 39 1.78579591836735 8.60152284263959 0
"1347" 36 40 1.71138775510204 8.60152284263959 1
"1348" 36 41 1.86020408163265 8.60152284263959 1
"1349" 36 42 1.71138775510204 8.60152284263959 1
"1350" 36 43 1.71138775510204 8.15482233502538 1
"1351" 36 44 1.78579591836735 8.15482233502538 1
"1352" 36 45 1.78579591836735 8.0989847715736 1
"1353" 36 46 1.78579591836735 8.15482233502538 1
"1354" 36 47 1.78579591836735 8.15482233502538 1
"1355" 36 48 1.78579591836735 8.15482233502538 1
"1356" 36 49 1.78579591836735 8.15482233502538 1
"1357" 36 50 1.78579591836735 8.0989847715736 1
"1358" 36 51 1.78579591836735 8.60152284263959 1
"1359" 36 52 1.78579591836735 8.65736040609137 1
"1360" 36 53 1.78579591836735 8.15482233502538 1
"1361" 36 54 1.71138775510204 8.04314720812183 1
"1362" 37 3 1.78579591836735 9.43908629441624 1
"1363" 37 4 1.78579591836735 9.32741116751269 1
"1364" 37 5 1.78579591836735 10.2766497461929 1
"1365" 37 6 1.78579591836735 10.2208121827411 1
"1366" 37 7 1.71138775510204 9.27157360406091 1
"1367" 37 8 1.78579591836735 9.27157360406091 0
"1368" 37 9 1.78579591836735 8.60152284263959 0
"1369" 37 10 3.57159183673469 8.60152284263959 1
"1370" 37 11 3.49718367346939 8.54568527918782 1
"1371" 37 12 3.57159183673469 8.48984771573604 1
"1372" 37 13 3.57159183673469 8.0989847715736 1
"1373" 37 14 3.57159183673469 8.15482233502538 1
"1374" 37 15 4.98534693877551 8.04314720812183 1
"1375" 37 16 5.13416326530612 8.0989847715736 0
"1376" 37 17 5.13416326530612 8.0989847715736 1
"1377" 37 18 6.69673469387755 8.04314720812183 0
"1378" 37 19 5.05975510204082 8.15482233502538 0
"1379" 37 20 3.57159183673469 8.0989847715736 0
"1380" 37 21 3.57159183673469 8.0989847715736 0
"1381" 37 22 3.57159183673469 8.0989847715736 0
"1382" 37 23 5.05975510204082 8.0989847715736 0
"1383" 37 24 5.13416326530612 8.04314720812183 0
```

```
"1384" 37 25 3.57159183673469 8.09898477715736 0
"1385" 37 26 5.13416326530612 8.15482233502538 0
"1386" 37 27 5.13416326530612 8.09898477715736 0
"1387" 37 28 5.05975510204082 7.48477157360406 0
"1388" 37 29 5.13416326530612 8.09898477715736 0
"1389" 37 30 3.57159183673469 8.09898477715736 0
"1390" 37 31 3.57159183673469 8.09898477715736 0
"1391" 37 32 3.57159183673469 8.09898477715736 0
"1392" 37 33 3.57159183673469 8.09898477715736 0
"1393" 37 34 3.57159183673469 8.09898477715736 0
"1394" 37 35 3.57159183673469 8.15482233502538 0
"1395" 37 36 3.57159183673469 8.49984771573604 0
"1396" 37 37 1.78579591836735 8.60152284263959 0
"1397" 37 38 1.78579591836735 8.60152284263959 0
"1398" 37 39 1.78579591836735 8.60152284263959 0
"1399" 37 40 1.78579591836735 8.60152284263959 0
"1400" 37 41 1.78579591836735 8.60152284263959 1
"1401" 37 42 1.78579591836735 8.60152284263959 1
"1402" 37 43 3.57159183673469 8.09898477715736 1
"1403" 37 44 3.57159183673469 8.09898477715736 1
"1404" 37 45 1.78579591836735 8.60152284263959 1
"1405" 37 46 1.78579591836735 8.09898477715736 1
"1406" 37 47 1.78579591836735 8.09898477715736 1
"1407" 37 48 1.78579591836735 8.09898477715736 1
"1408" 37 49 1.78579591836735 8.09898477715736 1
"1409" 37 50 1.86020408163265 8.15482233502538 1
"1410" 37 51 1.86020408163265 8.54568527918782 1
"1411" 37 52 1.78579591836735 8.54568527918782 1
"1412" 37 53 1.78579591836735 8.15482233502538 1
"1413" 37 54 1.78579591836735 8.09898477715736 1
"1414" 37 55 1.78579591836735 8.21065989847716 1
"1415" 38 2 1.78579591836735 9.27157360406091 0
"1416" 38 3 1.78579591836735 9.27157360406091 1
"1417" 38 4 1.78579591836735 9.27157360406091 1
"1418" 38 5 1.78579591836735 9.27157360406091 1
"1419" 38 6 1.78579591836735 10.2766497461929 1
"1420" 38 7 1.71138775510204 9.27157360406091 1
"1421" 38 8 3.57159183673469 8.65736040609137 1
"1422" 38 9 3.57159183673469 8.60152284263959 1
"1423" 38 10 3.57159183673469 8.54568527918782 1
"1424" 38 11 3.57159183673469 8.09898477715736 1
"1425" 38 12 5.13416326530612 8.09898477715736 1
"1426" 38 13 5.13416326530612 8.09898477715736 1
"1427" 38 14 5.13416326530612 8.15482233502538 0
"1428" 38 15 5.13416326530612 7.54060913705584 1
"1429" 38 16 6.69673469387755 7.48477157360406 0
"1430" 38 17 8.18489795918367 6.81472081218274 0
"1431" 38 18 6.77114285714286 7.48477157360406 0
"1432" 38 19 5.05975510204082 8.04314720812183 0
"1433" 38 20 5.13416326530612 8.09898477715736 0
"1434" 38 21 5.13416326530612 8.09898477715736 0
"1435" 38 22 5.13416326530612 8.09898477715736 0
"1436" 38 23 3.49718367346939 8.09898477715736 0
"1437" 38 24 5.13416326530612 7.54060913705584 1
"1438" 38 25 3.57159183673469 8.04314720812183 0
"1439" 38 26 5.13416326530612 8.09898477715736 0
"1440" 38 27 6.77114285714286 7.48477157360406 0
"1441" 38 28 5.05975510204082 7.42893401015228 0
"1442" 38 29 6.77114285714286 7.42893401015228 0
"1443" 38 30 5.05975510204082 8.09898477715736 0
"1444" 38 31 5.13416326530612 8.09898477715736 0
"1445" 38 32 3.72040816326531 8.09898477715736 0
"1446" 38 33 3.57159183673469 8.09898477715736 0
"1447" 38 34 3.57159183673469 8.09898477715736 0
"1448" 38 35 3.57159183673469 8.09898477715736 0
"1449" 38 36 3.57159183673469 8.15482233502538 0
"1450" 38 37 1.78579591836735 8.54568527918782 0
"1451" 38 38 1.78579591836735 8.60152284263959 0
"1452" 38 39 1.78579591836735 8.60152284263959 0
"1453" 38 40 1.78579591836735 8.60152284263959 0
"1454" 38 41 3.57159183673469 8.60152284263959 1
"1455" 38 42 3.72040816326531 8.09898477715736 1
"1456" 38 43 3.57159183673469 8.09898477715736 1
"1457" 38 44 3.57159183673469 8.09898477715736 1
"1458" 38 45 1.78579591836735 8.60152284263959 0
```

```
"1459" 38 46 1.78579591836735 8.65736040609137 1
"1460" 38 47 1.78579591836735 8.0989847715736 1
"1461" 38 48 3.646 8.0989847715736 1
"1462" 38 49 1.71138775510204 8.0989847715736 1
"1463" 38 50 3.57159183673469 8.0989847715736 1
"1464" 38 51 1.86020408163265 8.15482233502538 1
"1465" 38 52 1.78579591836735 8.0989847715736 1
"1466" 38 53 1.86020408163265 8.0989847715736 1
"1467" 38 54 3.57159183673469 8.0989847715736 1
"1468" 38 55 3.57159183673469 8.15482233502538 1
"1469" 38 56 3.57159183673469 0.337563451776651 0
"1470" 39 2 1.78579591836735 9.27157360406091 1
"1471" 39 3 1.78579591836735 9.32741116751269 1
"1472" 39 4 1.78579591836735 9.32741116751269 1
"1473" 39 5 1.78579591836735 10.2766497461929 0
"1474" 39 6 1.78579591836735 10.3324873096447 0
"1475" 39 7 1.78579591836735 8.65736040609137 0
"1476" 39 8 3.57159183673469 8.54568527918782 1
"1477" 39 9 5.05975510204082 8.15482233502538 0
"1478" 39 10 5.13416326530612 8.15482233502538 1
"1479" 39 11 5.05975510204082 8.0989847715736 0
"1480" 39 12 5.13416326530612 8.21065989847716 0
"1481" 39 13 5.13416326530612 8.15482233502538 0
"1482" 39 14 6.69673469387755 7.48477157360406 0
"1483" 39 15 6.77114285714286 7.48477157360406 0
"1484" 39 16 8.25930612244898 6.31218274111675 0
"1485" 39 17 9.6730612244898 6.25634517766497 1
"1486" 39 18 6.69673469387755 7.48477157360406 0
"1487" 39 19 5.13416326530612 8.21065989847716 0
"1488" 39 20 5.13416326530612 8.21065989847716 0
"1489" 39 21 5.05975510204082 8.21065989847716 0
"1490" 39 22 3.57159183673469 8.21065989847716 0
"1491" 39 23 5.13416326530612 8.21065989847716 0
"1492" 39 24 6.77114285714286 7.42893401015228 0
"1493" 39 25 5.13416326530612 8.15482233502538 0
"1494" 39 26 5.05975510204082 8.0989847715736 0
"1495" 39 27 6.69673469387755 7.54060913705584 0
"1496" 39 28 5.28297959183674 8.15482233502538 0
"1497" 39 29 5.05975510204082 8.21065989847716 0
"1498" 39 30 5.05975510204082 8.15482233502538 0
"1499" 39 31 3.57159183673469 8.15482233502538 0
"1500" 39 32 3.57159183673469 8.21065989847716 0
"1501" 39 33 3.57159183673469 8.21065989847716 0
"1502" 39 34 3.57159183673469 8.21065989847716 0
"1503" 39 35 3.57159183673469 8.21065989847716 0
"1504" 39 36 3.646 8.21065989847716 0
"1505" 39 37 3.57159183673469 8.60152284263959 0
"1506" 39 38 1.78579591836735 8.60152284263959 0
"1507" 39 39 1.78579591836735 8.60152284263959 0
"1508" 39 40 1.86020408163265 8.60152284263959 0
"1509" 39 41 3.57159183673469 8.60152284263959 1
"1510" 39 42 3.57159183673469 8.15482233502538 1
"1511" 39 43 3.57159183673469 8.21065989847716 1
"1512" 39 44 3.57159183673469 8.21065989847716 0
"1513" 39 45 3.57159183673469 8.15482233502538 0
"1514" 39 46 3.57159183673469 8.15482233502538 1
"1515" 39 47 3.57159183673469 8.15482233502538 1
"1516" 39 48 3.57159183673469 8.49984771573604 1
"1517" 39 49 3.57159183673469 8.15482233502538 1
"1518" 39 50 3.57159183673469 8.21065989847716 1
"1519" 39 51 3.57159183673469 8.21065989847716 1
"1520" 39 52 3.49718367346939 8.21065989847716 1
"1521" 39 53 3.57159183673469 8.21065989847716 1
"1522" 39 54 3.57159183673469 8.21065989847716 1
"1523" 39 55 3.57159183673469 8.21065989847716 1
"1524" 39 56 3.57159183673469 8.0989847715736 0
"1525" 40 2 1.78579591836735 9.32741116751269 1
"1526" 40 3 1.78579591836735 8.65736040609137 1
"1527" 40 4 1.78579591836735 8.54568527918782 0
"1528" 40 5 1.78579591836735 9.27157360406091 0
"1529" 40 6 1.78579591836735 10.2766497461929 0
"1530" 40 7 1.86020408163265 9.32741116751269 1
"1531" 40 8 3.57159183673469 8.60152284263959 0
"1532" 40 9 5.05975510204082 8.15482233502538 1
"1533" 40 10 5.13416326530612 8.15482233502538 0
```

```
"1534" 40 11 5.28297959183674 8.15482233502538 0
"1535" 40 12 6.77114285714286 8.15482233502538 1
"1536" 40 13 5.13416326530612 8.15482233502538 0
"1537" 40 14 5.20857142857143 7.48477157360406 0
"1538" 40 15 6.77114285714286 7.48477157360406 0
"1539" 40 16 8.33371428571429 6.2005076142132 1
"1540" 40 17 8.11048979591837 6.92639593908629 0
"1541" 40 18 5.13416326530612 8.15482233502538 0
"1542" 40 19 5.13416326530612 8.0989847715736 0
"1543" 40 20 5.20857142857143 8.15482233502538 0
"1544" 40 21 3.49718367346939 8.15482233502538 0
"1545" 40 22 3.49718367346939 8.15482233502538 0
"1546" 40 23 5.13416326530612 8.15482233502538 0
"1547" 40 24 5.13416326530612 7.48477157360406 0
"1548" 40 25 5.13416326530612 8.0989847715736 0
"1549" 40 26 5.13416326530612 8.15482233502538 0
"1550" 40 27 5.13416326530612 8.15482233502538 0
"1551" 40 28 5.20857142857143 8.15482233502538 0
"1552" 40 29 3.57159183673469 8.15482233502538 0
"1553" 40 30 3.57159183673469 8.15482233502538 0
"1554" 40 31 3.57159183673469 8.15482233502538 0
"1555" 40 32 3.57159183673469 8.15482233502538 0
"1556" 40 33 3.57159183673469 8.15482233502538 0
"1557" 40 34 3.57159183673469 8.15482233502538 0
"1558" 40 35 3.57159183673469 8.15482233502538 0
"1559" 40 36 3.57159183673469 8.15482233502538 1
"1560" 40 37 3.57159183673469 8.15482233502538 1
"1561" 40 38 3.57159183673469 8.15482233502538 0
"1562" 40 39 3.57159183673469 8.60152284263959 0
"1563" 40 40 3.57159183673469 8.60152284263959 0
"1564" 40 41 3.57159183673469 8.15482233502538 1
"1565" 40 42 3.57159183673469 8.15482233502538 1
"1566" 40 43 3.57159183673469 8.15482233502538 1
"1567" 40 44 3.57159183673469 8.15482233502538 0
"1568" 40 45 3.57159183673469 8.15482233502538 0
"1569" 40 46 3.57159183673469 8.15482233502538 1
"1570" 40 47 3.57159183673469 8.65736040609137 1
"1571" 40 48 3.57159183673469 8.04314720812183 1
"1572" 40 49 3.57159183673469 8.15482233502538 1
"1573" 40 50 5.13416326530612 8.15482233502538 1
"1574" 40 51 5.13416326530612 8.15482233502538 1
"1575" 40 52 3.57159183673469 8.15482233502538 1
"1576" 40 53 5.05975510204082 8.15482233502538 1
"1577" 40 54 3.57159183673469 8.15482233502538 1
"1578" 40 55 3.57159183673469 8.21065989847716 0
"1579" 40 56 3.57159183673469 8.15482233502538 0
"1580" 41 1 1.78579591836735 8.60152284263959 1
"1581" 41 2 1.78579591836735 8.49984771573604 0
"1582" 41 3 1.78579591836735 8.60152284263959 1
"1583" 41 4 1.78579591836735 8.60152284263959 0
"1584" 41 5 1.78579591836735 8.60152284263959 0
"1585" 41 6 1.78579591836735 9.27157360406091 0
"1586" 41 7 1.78579591836735 10.2766497461929 1
"1587" 41 8 1.71138775510204 9.27157360406091 1
"1588" 41 9 3.57159183673469 8.0989847715736 1
"1589" 41 10 5.13416326530612 8.0989847715736 0
"1590" 41 11 5.13416326530612 8.0989847715736 0
"1591" 41 12 5.13416326530612 8.0989847715736 0
"1592" 41 13 6.84555102040816 8.15482233502538 0
"1593" 41 14 6.77114285714286 7.48477157360406 0
"1594" 41 15 8.11048979591837 6.92639593908629 0
"1595" 41 16 8.18489795918367 6.87055837563452 0
"1596" 41 17 6.84555102040816 7.48477157360406 0
"1597" 41 18 5.05975510204082 8.04314720812183 0
"1598" 41 19 5.05975510204082 7.48477157360406 0
"1599" 41 20 6.77114285714286 7.54060913705584 1
"1600" 41 21 3.57159183673469 8.0989847715736 0
"1601" 41 22 5.28297959183674 8.0989847715736 0
"1602" 41 23 5.13416326530612 8.0989847715736 0
"1603" 41 24 5.13416326530612 8.0989847715736 0
"1604" 41 25 5.13416326530612 7.48477157360406 0
"1605" 41 26 5.13416326530612 8.0989847715736 0
"1606" 41 27 5.13416326530612 8.0989847715736 0
"1607" 41 28 5.13416326530612 8.0989847715736 0
"1608" 41 29 5.13416326530612 8.0989847715736 0
```

```
"1609" 41 30 5.13416326530612 8.0989847715736 0
"1610" 41 31 3.646 8.0989847715736 0
"1611" 41 32 3.57159183673469 8.0989847715736 0
"1612" 41 33 3.57159183673469 8.0989847715736 0
"1613" 41 34 3.57159183673469 8.0989847715736 0
"1614" 41 35 3.57159183673469 8.0989847715736 0
"1615" 41 36 3.57159183673469 8.0989847715736 0
"1616" 41 37 3.57159183673469 8.0989847715736 0
"1617" 41 38 3.57159183673469 8.0989847715736 0
"1618" 41 39 3.57159183673469 8.0989847715736 0
"1619" 41 40 3.57159183673469 8.0989847715736 0
"1620" 41 41 3.57159183673469 8.04314720812183 0
"1621" 41 42 3.57159183673469 8.0989847715736 1
"1622" 41 43 3.57159183673469 8.0989847715736 0
"1623" 41 44 3.49718367346939 8.0989847715736 0
"1624" 41 45 5.05975510204082 8.0989847715736 1
"1625" 41 46 3.57159183673469 8.0989847715736 0
"1626" 41 47 3.57159183673469 8.04314720812183 0
"1627" 41 48 3.57159183673469 8.60152284263959 1
"1628" 41 49 3.57159183673469 8.15482233502538 1
"1629" 41 50 5.13416326530612 8.0989847715736 1
"1630" 41 51 5.13416326530612 8.0989847715736 1
"1631" 41 52 5.05975510204082 8.0989847715736 1
"1632" 41 53 5.20857142857143 8.0989847715736 0
"1633" 41 54 5.05975510204082 8.0989847715736 0
"1634" 41 55 3.57159183673469 8.0989847715736 0
"1635" 42 1 1.71138775510204 8.60152284263959 1
"1636" 42 2 1.78579591836735 8.60152284263959 1
"1637" 42 3 1.78579591836735 8.60152284263959 0
"1638" 42 4 1.78579591836735 8.60152284263959 0
"1639" 42 5 1.78579591836735 8.60152284263959 0
"1640" 42 6 1.78579591836735 9.27157360406091 0
"1641" 42 7 1.78579591836735 10.2208121827411 0
"1642" 42 8 1.78579591836735 9.27157360406091 1
"1643" 42 9 3.57159183673469 8.54568527918782 0
"1644" 42 10 3.57159183673469 8.15482233502538 0
"1645" 42 11 3.57159183673469 8.15482233502538 0
"1646" 42 12 5.13416326530612 8.15482233502538 0
"1647" 42 13 5.05975510204082 8.0989847715736 0
"1648" 42 14 5.28297959183674 8.15482233502538 0
"1649" 42 15 8.25930612244898 6.87055837563452 0
"1650" 42 16 6.77114285714286 7.48477157360406 0
"1651" 42 17 5.13416326530612 7.48477157360406 0
"1652" 42 18 5.13416326530612 8.0989847715736 1
"1653" 42 19 5.28297959183674 8.0989847715736 1
"1654" 42 20 5.13416326530612 8.0989847715736 0
"1655" 42 21 3.57159183673469 8.04314720812183 0
"1656" 42 22 6.77114285714286 8.15482233502538 0
"1657" 42 23 6.84555102040816 8.15482233502538 0
"1658" 42 24 5.13416326530612 8.04314720812183 0
"1659" 42 25 5.05975510204082 8.04314720812183 0
"1660" 42 26 3.57159183673469 8.15482233502538 1
"1661" 42 27 5.13416326530612 8.15482233502538 0
"1662" 42 28 5.05975510204082 8.15482233502538 0
"1663" 42 29 5.05975510204082 8.0989847715736 0
"1664" 42 30 5.13416326530612 8.15482233502538 1
"1665" 42 31 3.49718367346939 8.15482233502538 0
"1666" 42 32 5.13416326530612 8.15482233502538 0
"1667" 42 33 5.13416326530612 8.15482233502538 0
"1668" 42 34 5.13416326530612 8.15482233502538 0
"1669" 42 35 5.13416326530612 8.15482233502538 0
"1670" 42 36 5.13416326530612 8.15482233502538 1
"1671" 42 37 5.13416326530612 8.15482233502538 1
"1672" 42 38 3.646 8.15482233502538 0
"1673" 42 39 3.49718367346939 8.15482233502538 0
"1674" 42 40 3.57159183673469 8.15482233502538 0
"1675" 42 41 3.49718367346939 8.15482233502538 0
"1676" 42 42 3.646 8.15482233502538 1
"1677" 42 43 5.13416326530612 8.15482233502538 0
"1678" 42 44 5.13416326530612 8.15482233502538 0
"1679" 42 45 5.13416326530612 8.15482233502538 0
"1680" 42 46 5.13416326530612 8.04314720812183 0
"1681" 42 47 5.13416326530612 8.15482233502538 0
"1682" 42 48 5.20857142857143 8.0989847715736 0
"1683" 42 49 3.49718367346939 8.0989847715736 0
```

```
"1684" 42 50 3.57159183673469 8.04314720812183 0
"1685" 42 51 5.13416326530612 8.15482233502538 0
"1686" 42 52 6.77114285714286 8.04314720812183 0
"1687" 42 53 6.77114285714286 8.0989847715736 0
"1688" 42 54 5.13416326530612 8.0989847715736 0
"1689" 42 55 5.28297959183674 8.32233502538071 0
"1690" 43 1 3.57159183673469 8.60152284263959 0
"1691" 43 2 3.57159183673469 8.60152284263959 0
"1692" 43 3 3.57159183673469 8.60152284263959 0
"1693" 43 4 1.78579591836735 8.60152284263959 0
"1694" 43 5 1.78579591836735 8.60152284263959 0
"1695" 43 6 1.71138775510204 8.60152284263959 0
"1696" 43 7 1.78579591836735 9.27157360406091 0
"1697" 43 8 1.86020408163265 9.27157360406091 1
"1698" 43 9 3.57159183673469 8.60152284263959 1
"1699" 43 10 3.57159183673469 8.0989847715736 1
"1700" 43 11 3.57159183673469 8.0989847715736 0
"1701" 43 12 5.28297959183674 8.15482233502538 0
"1702" 43 13 5.28297959183674 8.21065989847716 0
"1703" 43 14 5.28297959183674 8.0989847715736 0
"1704" 43 15 6.69673469387755 7.48477157360406 0
"1705" 43 16 6.77114285714286 7.48477157360406 0
"1706" 43 17 5.13416326530612 8.15482233502538 0
"1707" 43 18 3.49718367346939 8.15482233502538 0
"1708" 43 19 3.646 8.15482233502538 0
"1709" 43 20 5.05975510204082 8.15482233502538 0
"1710" 43 21 3.57159183673469 8.15482233502538 0
"1711" 43 22 6.77114285714286 8.15482233502538 0
"1712" 43 23 5.28297959183674 8.15482233502538 0
"1713" 43 24 5.13416326530612 8.15482233502538 0
"1714" 43 25 5.13416326530612 8.15482233502538 0
"1715" 43 26 4.98534693877551 8.15482233502538 0
"1716" 43 27 5.13416326530612 8.15482233502538 0
"1717" 43 28 8.25930612244898 6.25634517766497 0
"1718" 43 29 6.84555102040816 6.92639593908629 0
"1719" 43 30 5.05975510204082 8.21065989847716 0
"1720" 43 31 5.13416326530612 8.15482233502538 0
"1721" 43 32 5.13416326530612 8.15482233502538 0
"1722" 43 33 5.13416326530612 8.15482233502538 0
"1723" 43 34 5.13416326530612 8.15482233502538 0
"1724" 43 35 5.13416326530612 8.15482233502538 0
"1725" 43 36 5.13416326530612 8.15482233502538 0
"1726" 43 37 5.13416326530612 8.15482233502538 1
"1727" 43 38 5.13416326530612 8.15482233502538 0
"1728" 43 39 5.13416326530612 8.15482233502538 1
"1729" 43 40 3.646 8.15482233502538 1
"1730" 43 41 5.20857142857143 8.15482233502538 0
"1731" 43 42 5.13416326530612 8.15482233502538 0
"1732" 43 43 5.13416326530612 8.21065989847716 0
"1733" 43 44 5.28297959183674 8.21065989847716 1
"1734" 43 45 6.69673469387755 8.15482233502538 1
"1735" 43 46 6.77114285714286 7.37309644670051 1
"1736" 43 47 8.25930612244898 6.92639593908629 0
"1737" 43 48 6.84555102040816 6.92639593908629 0
"1738" 43 49 6.77114285714286 7.48477157360406 0
"1739" 43 50 6.69673469387755 8.26649746192893 0
"1740" 43 51 5.20857142857143 8.15482233502538 0
"1741" 43 52 5.05975510204082 8.15482233502538 0
"1742" 43 53 6.77114285714286 7.48477157360406 0
"1743" 43 54 5.13416326530612 7.42893401015228 0
"1744" 43 55 5.13416326530612 0.337563451776651 0
"1745" 44 1 3.57159183673469 8.71319796954315 0
"1746" 44 2 3.646 8.60152284263959 1
"1747" 44 3 3.646 8.15482233502538 0
"1748" 44 4 3.57159183673469 8.60152284263959 0
"1749" 44 5 3.57159183673469 8.54568527918782 0
"1750" 44 6 1.86020408163265 8.54568527918782 0
"1751" 44 7 3.49718367346939 8.65736040609137 0
"1752" 44 8 1.78579591836735 9.27157360406091 1
"1753" 44 9 1.78579591836735 8.60152284263959 0
"1754" 44 10 3.49718367346939 8.60152284263959 0
"1755" 44 11 3.49718367346939 8.15482233502538 0
"1756" 44 12 3.57159183673469 8.15482233502538 0
"1757" 44 13 5.13416326530612 8.15482233502538 0
"1758" 44 14 6.77114285714286 7.54060913705584 0
```

```
"1759" 44 15 5.05975510204082 8.15482233502538 0
"1760" 44 16 6.77114285714286 7.54060913705584 0
"1761" 44 17 5.13416326530612 8.15482233502538 0
"1762" 44 18 3.57159183673469 8.15482233502538 0
"1763" 44 19 5.13416326530612 8.15482233502538 0
"1764" 44 20 5.13416326530612 8.15482233502538 0
"1765" 44 21 5.13416326530612 8.0989847715736 0
"1766" 44 22 5.13416326530612 8.15482233502538 0
"1767" 44 23 5.13416326530612 8.15482233502538 0
"1768" 44 24 5.05975510204082 8.15482233502538 0
"1769" 44 25 5.13416326530612 8.15482233502538 0
"1770" 44 26 6.77114285714286 7.42893401015228 0
"1771" 44 27 5.28297959183674 7.48477157360406 0
"1772" 44 28 5.28297959183674 7.48477157360406 0
"1773" 44 29 8.18489795918367 6.31218274111675 0
"1774" 44 30 8.25930612244898 6.92639593908629 0
"1775" 44 31 6.84555102040816 8.15482233502538 0
"1776" 44 32 6.69673469387755 8.15482233502538 0
"1777" 44 33 5.28297959183674 8.15482233502538 0
"1778" 44 34 5.13416326530612 8.15482233502538 0
"1779" 44 35 5.05975510204082 8.15482233502538 0
"1780" 44 36 5.13416326530612 8.15482233502538 1
"1781" 44 37 5.28297959183674 8.15482233502538 1
"1782" 44 38 5.13416326530612 8.15482233502538 0
"1783" 44 39 5.13416326530612 8.15482233502538 0
"1784" 44 40 5.05975510204082 8.15482233502538 0
"1785" 44 41 5.20857142857143 8.15482233502538 0
"1786" 44 42 6.77114285714286 8.15482233502538 1
"1787" 44 43 6.77114285714286 8.0989847715736 0
"1788" 44 44 6.84555102040816 7.48477157360406 0
"1789" 44 45 6.77114285714286 7.54060913705584 0
"1790" 44 46 8.18489795918367 6.31218274111675 0
"1791" 44 47 9.6730612244898 6.2005076142132 0
"1792" 44 48 9.82187755102041 5.53045685279188 0
"1793" 44 49 8.25930612244898 5.58629441624366 0
"1794" 45 2 5.13416326530612 8.21065989847716 0
"1795" 45 3 6.54791836734694 8.15482233502538 1
"1796" 45 4 5.13416326530612 8.04314720812183 0
"1797" 45 5 5.13416326530612 8.0989847715736 0
"1798" 45 6 3.57159183673469 8.60152284263959 0
"1799" 45 7 3.646 8.60152284263959 0
"1800" 45 8 3.57159183673469 8.65736040609137 0
"1801" 45 9 3.57159183673469 8.60152284263959 0
"1802" 45 10 3.57159183673469 8.0989847715736 0
"1803" 45 11 3.57159183673469 8.0989847715736 0
"1804" 45 12 5.13416326530612 8.0989847715736 0
"1805" 45 13 6.84555102040816 8.0989847715736 0
"1806" 45 14 8.25930612244898 6.92639593908629 0
"1807" 45 15 5.28297959183674 7.42893401015228 0
"1808" 45 16 5.35738775510204 8.0989847715736 0
"1809" 45 17 5.13416326530612 8.0989847715736 0
"1810" 45 18 3.49718367346939 8.0989847715736 0
"1811" 45 19 5.13416326530612 8.0989847715736 0
"1812" 45 20 5.13416326530612 8.0989847715736 0
"1813" 45 21 6.69673469387755 7.48477157360406 0
"1814" 45 22 5.05975510204082 8.0989847715736 0
"1815" 45 23 5.13416326530612 8.0989847715736 0
"1816" 45 24 5.13416326530612 8.04314720812183 0
"1817" 45 25 6.77114285714286 7.48477157360406 0
"1818" 45 26 8.25930612244898 6.92639593908629 0
"1819" 45 27 6.77114285714286 7.48477157360406 0
"1820" 45 28 6.69673469387755 7.48477157360406 0
"1821" 45 29 6.69673469387755 6.87055837563452 0
"1822" 45 30 9.82187755102041 5.53045685279188 0
"1823" 45 31 9.6730612244898 6.31218274111675 0
"1824" 45 32 6.84555102040816 7.48477157360406 0
"1825" 45 33 6.69673469387755 7.48477157360406 0
"1826" 45 34 5.13416326530612 8.0989847715736 0
"1827" 45 35 5.05975510204082 8.0989847715736 0
"1828" 45 36 6.77114285714286 7.48477157360406 1
"1829" 45 37 6.77114285714286 8.15482233502538 1
"1830" 45 38 5.13416326530612 8.0989847715736 0
"1831" 45 39 5.13416326530612 8.0989847715736 0
"1832" 45 40 5.13416326530612 8.0989847715736 1
"1833" 45 41 6.77114285714286 8.0989847715736 1
```

```
"1834" 45 42 6.84555102040816 7.48477157360406 1
"1835" 45 43 8.18489795918367 6.87055837563452 0
"1836" 45 44 8.25930612244898 6.81472081218274 0
"1837" 45 45 8.25930612244898 6.25634517766497 0
"1838" 45 46 9.82187755102041 6.25634517766497 0
"1839" 45 47 9.82187755102041 5.53045685279188 0
"1840" 46 3 8.18489795918367 7.42893401015228 1
"1841" 46 4 6.77114285714286 8.0989847715736 0
"1842" 46 5 6.77114285714286 8.15482233502538 0
"1843" 46 6 5.13416326530612 8.15482233502538 0
"1844" 46 7 5.28297959183674 8.15482233502538 0
"1845" 46 8 3.49718367346939 8.60152284263959 0
"1846" 46 9 3.57159183673469 8.54568527918782 0
"1847" 46 10 5.05975510204082 8.0989847715736 0
"1848" 46 11 5.13416326530612 8.0989847715736 0
"1849" 46 12 5.13416326530612 8.0989847715736 0
"1850" 46 13 5.13416326530612 8.0989847715736 0
"1851" 46 14 5.13416326530612 8.21065989847716 0
"1852" 46 15 5.13416326530612 8.15482233502538 0
"1853" 46 16 3.57159183673469 8.15482233502538 0
"1854" 46 17 3.57159183673469 8.15482233502538 0
"1855" 46 18 3.57159183673469 8.0989847715736 0
"1856" 46 19 3.57159183673469 8.15482233502538 0
"1857" 46 20 5.13416326530612 8.0989847715736 0
"1858" 46 21 8.18489795918367 6.98223350253807 0
"1859" 46 22 6.77114285714286 8.15482233502538 0
"1860" 46 23 5.13416326530612 8.15482233502538 0
"1861" 46 24 5.13416326530612 8.15482233502538 0
"1862" 46 25 8.18489795918367 6.87055837563452 0
"1863" 46 26 8.18489795918367 6.31218274111675 0
"1864" 46 27 6.77114285714286 7.48477157360406 0
"1865" 46 28 6.77114285714286 7.42893401015228 0
"1866" 46 29 6.77114285714286 7.48477157360406 0
"1867" 46 30 8.11048979591837 6.92639593908629 0
"1868" 46 31 9.7474693877551 6.31218274111675 0
"1869" 46 32 9.7474693877551 6.25634517766497 0
"1870" 46 33 8.18489795918367 6.31218274111675 0
"1871" 46 34 6.84555102040816 6.92639593908629 0
"1872" 46 35 8.11048979591837 6.87055837563452 0
"1873" 46 36 6.77114285714286 7.48477157360406 0
"1874" 46 37 6.84555102040816 6.87055837563452 0
"1875" 46 38 6.77114285714286 7.54060913705584 0
"1876" 46 39 6.77114285714286 8.15482233502538 1
"1877" 46 40 6.77114285714286 7.48477157360406 1
"1878" 46 41 8.25930612244898 6.92639593908629 1
"1879" 46 42 8.18489795918367 6.92639593908629 0
"1880" 46 43 9.6730612244898 6.25634517766497 1
"1881" 46 44 9.82187755102041 6.25634517766497 0
"1882" 46 45 9.7474693877551 5.58629441624366 0
"1883" 47 3 8.18489795918367 6.92639593908629 0
"1884" 47 4 8.25930612244898 6.81472081218274 0
"1885" 47 5 8.18489795918367 7.48477157360406 0
"1886" 47 6 6.77114285714286 8.15482233502538 0
"1887" 47 7 5.20857142857143 8.15482233502538 0
"1888" 47 8 6.77114285714286 8.15482233502538 0
"1889" 47 9 5.13416326530612 8.15482233502538 1
"1890" 47 10 3.57159183673469 8.60152284263959 0
"1891" 47 11 3.57159183673469 8.60152284263959 1
"1892" 47 12 5.28297959183674 8.0989847715736 0
"1893" 47 13 5.28297959183674 8.0989847715736 0
"1894" 47 14 3.49718367346939 8.60152284263959 0
"1895" 47 15 3.646 8.15482233502538 0
"1896" 47 16 5.13416326530612 8.15482233502538 0
"1897" 47 17 5.13416326530612 8.0989847715736 0
"1898" 47 18 3.57159183673469 8.15482233502538 0
"1899" 47 19 3.57159183673469 8.54568527918782 0
"1900" 47 20 3.57159183673469 8.0989847715736 0
"1901" 47 21 6.77114285714286 7.54060913705584 0
"1902" 47 22 6.77114285714286 8.15482233502538 0
"1903" 47 23 5.28297959183674 8.15482233502538 0
"1904" 47 24 6.84555102040816 7.42893401015228 0
"1905" 47 25 9.7474693877551 6.31218274111675 0
"1906" 47 26 8.18489795918367 6.87055837563452 0
"1907" 47 27 5.35738775510204 8.0989847715736 0
"1908" 47 28 5.13416326530612 8.15482233502538 0
```

```
"1909" 47 29 6.77114285714286 8.21065989847716 0
"1910" 47 30 6.77114285714286 7.48477157360406 0
"1911" 47 31 8.25930612244898 6.92639593908629 0
"1912" 47 32 9.59865306122449 6.25634517766497 0
"1913" 47 33 9.7474693877551 6.25634517766497 0
"1914" 47 34 9.6730612244898 6.31218274111675 0
"1915" 47 35 8.18489795918367 6.31218274111675 0
"1916" 47 36 6.84555102040816 6.87055837563452 0
"1917" 47 37 8.18489795918367 6.25634517766497 1
"1918" 47 38 6.99436734693878 8.15482233502538 0
"1919" 47 39 8.18489795918367 7.48477157360406 0
"1920" 47 40 9.7474693877551 6.31218274111675 1
"1921" 47 41 10.8635918367347 5.58629441624366 0
"1922" 47 42 10.938 5.58629441624366 0
"1923" 47 43 10.938 4.97208121827411 0
"1924" 47 44 9.7474693877551 5.47461928934011 0
"1925" 48 3 8.18489795918367 8.04314720812183 0
"1926" 48 4 8.25930612244898 6.92639593908629 0
"1927" 48 5 6.77114285714286 7.48477157360406 0
"1928" 48 6 6.77114285714286 8.15482233502538 0
"1929" 48 7 6.77114285714286 8.0989847715736 0
"1930" 48 8 6.77114285714286 8.0989847715736 0
"1931" 48 9 5.13416326530612 8.15482233502538 0
"1932" 48 10 3.57159183673469 8.60152284263959 0
"1933" 48 11 3.57159183673469 9.27157360406091 0
"1934" 48 12 5.13416326530612 8.15482233502538 0
"1935" 48 13 3.57159183673469 8.0989847715736 0
"1936" 48 14 3.57159183673469 8.60152284263959 0
"1937" 48 15 5.13416326530612 8.15482233502538 0
"1938" 48 16 5.13416326530612 8.0989847715736 0
"1939" 48 17 5.13416326530612 8.0989847715736 0
"1940" 48 18 3.57159183673469 8.60152284263959 0
"1941" 48 19 3.57159183673469 8.60152284263959 0
"1942" 48 20 3.57159183673469 8.0989847715736 0
"1943" 48 21 5.13416326530612 8.04314720812183 0
"1944" 48 22 5.13416326530612 8.0989847715736 0
"1945" 48 23 5.05975510204082 8.0989847715736 0
"1946" 48 24 6.77114285714286 7.42893401015228 0
"1947" 48 25 8.18489795918367 6.87055837563452 0
"1948" 48 26 6.77114285714286 8.0989847715736 0
"1949" 48 27 5.13416326530612 8.0989847715736 0
"1950" 48 28 5.05975510204082 8.0989847715736 0
"1951" 48 29 5.13416326530612 8.0989847715736 0
"1952" 48 30 5.13416326530612 8.15482233502538 0
"1953" 48 31 6.77114285714286 8.0989847715736 0
"1954" 48 32 6.77114285714286 8.0989847715736 1
"1955" 48 33 6.84555102040816 7.48477157360406 0
"1956" 48 34 8.18489795918367 6.87055837563452 0
"1957" 48 35 8.25930612244898 6.31218274111675 0
"1958" 48 36 8.18489795918367 6.31218274111675 0
"1959" 48 37 8.18489795918367 6.14467005076142 0
"1960" 48 38 8.18489795918367 6.87055837563452 1
"1961" 48 39 8.25930612244898 6.87055837563452 1
"1962" 48 40 9.7474693877551 6.31218274111675 0
"1963" 48 41 10.938 4.91624365482233 0
"1964" 49 3 6.99436734693878 8.15482233502538 0
"1965" 49 4 8.03608163265306 7.48477157360406 0
"1966" 49 5 6.77114285714286 7.54060913705584 0
"1967" 49 6 6.77114285714286 7.48477157360406 0
"1968" 49 7 6.84555102040816 8.15482233502538 0
"1969" 49 8 6.77114285714286 8.04314720812183 0
"1970" 49 9 5.28297959183674 8.15482233502538 0
"1971" 49 10 3.7204081632653 8.60152284263959 0
"1972" 49 11 3.79481632653061 8.60152284263959 0
"1973" 49 12 5.13416326530612 8.15482233502538 0
"1974" 49 13 5.05975510204082 8.0989847715736 0
"1975" 49 14 5.28297959183674 8.15482233502538 0
"1976" 49 15 5.13416326530612 8.0989847715736 0
"1977" 49 16 6.77114285714286 7.48477157360406 0
"1978" 49 17 5.13416326530612 8.04314720812183 0
"1979" 49 18 3.57159183673469 8.65736040609137 0
"1980" 49 19 3.57159183673469 8.54568527918782 0
"1981" 49 20 3.646 8.60152284263959 0
"1982" 49 21 3.57159183673469 8.15482233502538 0
"1983" 49 22 5.05975510204082 8.04314720812183 0
```

```
"1984" 49 23 5.13416326530612 8.15482233502538 0
"1985" 49 24 5.28297959183674 8.04314720812183 0
"1986" 49 25 6.77114285714286 8.04314720812183 0
"1987" 49 26 5.13416326530612 8.04314720812183 0
"1988" 49 27 5.13416326530612 8.04314720812183 0
"1989" 49 28 5.13416326530612 8.04314720812183 0
"1990" 49 29 5.13416326530612 8.04314720812183 0
"1991" 49 30 5.13416326530612 8.04314720812183 0
"1992" 49 31 5.35738775510204 8.04314720812183 0
"1993" 49 32 5.05975510204082 8.04314720812183 0
"1994" 49 33 5.05975510204082 8.0989847715736 0
"1995" 49 34 6.69673469387755 7.48477157360406 0
"1996" 49 35 8.03608163265306 6.31218274111675 0
"1997" 49 36 8.11048979591837 6.25634517766497 0
"1998" 49 37 8.18489795918367 6.25634517766497 1
"1999" 49 38 8.18489795918367 6.25634517766497 1
"2000" 49 39 8.18489795918367 6.25634517766497 1
"2001" 49 40 8.25930612244898 6.25634517766497 0
"2002" 50 2 6.77114285714286 8.0989847715736 0
"2003" 50 3 6.77114285714286 8.15482233502538 0
"2004" 50 4 6.77114285714286 8.15482233502538 0
"2005" 50 5 6.69673469387755 7.48477157360406 0
"2006" 50 6 6.77114285714286 7.48477157360406 0
"2007" 50 7 6.77114285714286 8.15482233502538 1
"2008" 50 8 5.13416326530612 8.15482233502538 0
"2009" 50 9 3.79481632653061 8.15482233502538 0
"2010" 50 10 5.13416326530612 8.15482233502538 0
"2011" 50 11 6.84555102040816 8.15482233502538 0
"2012" 50 12 5.13416326530612 8.15482233502538 0
"2013" 50 13 5.13416326530612 8.15482233502538 0
"2014" 50 14 6.69673469387755 8.15482233502538 0
"2015" 50 15 5.05975510204082 8.15482233502538 0
"2016" 50 16 5.28297959183674 8.15482233502538 0
"2017" 50 17 3.57159183673469 9.32741116751269 0
"2018" 50 18 3.57159183673469 9.27157360406091 1
"2019" 50 19 1.78579591836735 8.60152284263959 1
"2020" 50 20 3.646 8.60152284263959 1
"2021" 50 21 5.13416326530612 8.0989847715736 1
"2022" 50 22 6.84555102040816 7.48477157360406 0
"2023" 50 23 5.13416326530612 8.21065989847716 0
"2024" 50 24 5.13416326530612 8.15482233502538 0
"2025" 50 25 5.13416326530612 8.15482233502538 0
"2026" 50 26 5.13416326530612 8.15482233502538 0
"2027" 50 27 5.20857142857143 8.15482233502538 0
"2028" 50 28 5.13416326530612 8.15482233502538 0
"2029" 50 29 5.13416326530612 8.15482233502538 0
"2030" 50 30 5.13416326530612 8.15482233502538 0
"2031" 50 31 5.13416326530612 8.15482233502538 0
"2032" 50 32 5.28297959183674 8.26649746192893 0
"2033" 50 33 5.28297959183674 8.15482233502538 0
"2034" 50 34 6.69673469387755 8.15482233502538 0
"2035" 50 35 6.77114285714286 6.92639593908629 0
"2036" 50 36 8.25930612244898 6.2005076142132 0
"2037" 50 37 8.11048979591837 6.25634517766497 1
"2038" 50 38 8.11048979591837 6.2005076142132 0
"2039" 50 39 8.18489795918367 6.25634517766497 0
"2040" 50 40 6.77114285714286 0.225888324873099 0
"2041" 51 2 6.77114285714286 8.0989847715736 0
"2042" 51 3 6.77114285714286 8.15482233502538 0
"2043" 51 4 6.84555102040816 8.0989847715736 0
"2044" 51 5 5.05975510204082 8.15482233502538 0
"2045" 51 6 5.05975510204082 8.15482233502538 0
"2046" 51 7 5.05975510204082 8.0989847715736 0
"2047" 51 8 3.49718367346939 8.15482233502538 0
"2048" 51 9 5.13416326530612 8.15482233502538 0
"2049" 51 10 6.84555102040816 8.15482233502538 0
"2050" 51 11 6.77114285714286 8.15482233502538 0
"2051" 51 12 5.13416326530612 8.15482233502538 0
"2052" 51 13 5.13416326530612 8.15482233502538 0
"2053" 51 14 3.646 8.60152284263959 1
"2054" 51 15 3.57159183673469 8.60152284263959 0
"2055" 51 16 3.49718367346939 8.60152284263959 0
"2056" 51 17 1.71138775510204 8.60152284263959 1
"2057" 51 18 1.71138775510204 8.60152284263959 1
"2058" 51 19 1.78579591836735 8.545685279187821 1
```

```
"2059" 51 20 3.646 8.60152284263959 1
"2060" 51 21 5.28297959183674 8.15482233502538 0
"2061" 51 22 5.05975510204082 8.21065989847716 0
"2062" 51 23 5.13416326530612 7.42893401015228 0
"2063" 51 24 5.13416326530612 8.15482233502538 0
"2064" 51 25 5.13416326530612 8.15482233502538 0
"2065" 51 26 5.13416326530612 8.15482233502538 0
"2066" 51 27 3.57159183673469 8.15482233502538 0
"2067" 51 28 5.13416326530612 8.15482233502538 0
"2068" 51 29 5.13416326530612 8.15482233502538 0
"2069" 51 30 5.13416326530612 8.15482233502538 0
"2070" 51 31 5.13416326530612 8.15482233502538 0
"2071" 51 32 6.69673469387755 7.42893401015228 0
"2072" 51 33 6.84555102040816 6.92639593908629 0
"2073" 51 34 5.13416326530612 8.15482233502538 0
"2074" 51 35 6.84555102040816 7.48477157360406 0
"2075" 51 36 9.7474693877551 6.31218274111675 0
"2076" 51 37 9.89628571428572 6.25634517766497 0
"2077" 51 38 8.18489795918367 6.25634517766497 0
"2078" 51 39 8.18489795918367 6.92639593908629 0
"2079" 51 40 6.77114285714286 6.98223350253807 0
"2080" 52 2 6.77114285714286 8.21065989847716 1
"2081" 52 3 5.13416326530612 7.98730964467005 0
"2082" 52 4 5.13416326530612 8.04314720812183 0
"2083" 52 5 5.13416326530612 8.0989847715736 0
"2084" 52 6 5.05975510204082 8.0989847715736 0
"2085" 52 7 3.57159183673469 8.54568527918782 0
"2086" 52 8 3.57159183673469 8.60152284263959 0
"2087" 52 9 6.69673469387755 8.0989847715736 0
"2088" 52 10 5.20857142857143 8.0989847715736 0
"2089" 52 11 6.84555102040816 7.48477157360406 0
"2090" 52 12 4.98534693877551 8.0989847715736 1
"2091" 52 13 3.57159183673469 8.60152284263959 0
"2092" 52 14 3.49718367346939 8.60152284263959 0
"2093" 52 15 3.49718367346939 8.60152284263959 0
"2094" 52 16 1.78579591836735 9.27157360406091 1
"2095" 52 17 1.71138775510204 8.65736040609137 1
"2096" 52 18 3.57159183673469 8.60152284263959 0
"2097" 52 19 1.86020408163265 8.60152284263959 1
"2098" 52 20 1.78579591836735 8.60152284263959 1
"2099" 52 21 5.13416326530612 8.15482233502538 0
"2100" 52 22 6.84555102040816 8.15482233502538 0
"2101" 52 23 5.13416326530612 8.0989847715736 0
"2102" 52 24 3.57159183673469 8.15482233502538 0
"2103" 52 25 5.05975510204082 8.04314720812183 0
"2104" 52 26 3.49718367346939 8.04314720812183 0
"2105" 52 27 3.49718367346939 8.04314720812183 0
"2106" 52 28 5.13416326530612 8.04314720812183 0
"2107" 52 29 5.13416326530612 8.04314720812183 0
"2108" 52 30 5.13416326530612 8.04314720812183 0
"2109" 52 31 5.28297959183674 7.98730964467005 0
"2110" 52 32 6.69673469387755 7.48477157360406 0
"2111" 52 33 6.84555102040816 7.98730964467005 0
"2112" 52 34 6.77114285714286 7.48477157360406 0
"2113" 52 35 5.13416326530612 8.15482233502538 0
"2114" 52 36 8.25930612244898 6.36802030456853 1
"2115" 52 37 7.96167346938775 6.25634517766497 1
"2116" 52 38 9.7474693877551 6.25634517766497 0
"2117" 52 39 8.18489795918367 6.87055837563452 0
"2118" 52 40 8.18489795918367 6.2005076142132 0
"2119" 52 41 9.6730612244898 6.2005076142132 0
"2120" 53 3 5.28297959183674 8.0989847715736 0
"2121" 53 4 5.13416326530612 8.0989847715736 0
"2122" 53 5 5.13416326530612 8.0989847715736 0
"2123" 53 6 3.57159183673469 8.60152284263959 0
"2124" 53 7 3.57159183673469 8.60152284263959 0
"2125" 53 8 6.69673469387755 8.0989847715736 0
"2126" 53 9 8.03608163265306 7.48477157360406 1
"2127" 53 10 5.13416326530612 8.04314720812183 0
"2128" 53 11 5.13416326530612 8.0989847715736 0
"2129" 53 12 3.57159183673469 8.60152284263959 0
"2130" 53 13 3.49718367346939 8.54568527918782 0
"2131" 53 14 3.646 8.60152284263959 0
"2132" 53 15 3.57159183673469 8.60152284263959 0
"2133" 53 16 1.86020408163265 9.27157360406091 1
```

```
"2134" 53 17 1.71138775510204 9.32741116751269 0
"2135" 53 18 3.57159183673469 8.60152284263959 0
"2136" 53 19 3.57159183673469 8.60152284263959 0
"2137" 53 20 3.646 8.15482233502538 0
"2138" 53 21 5.13416326530612 8.0989847715736 0
"2139" 53 22 5.13416326530612 8.0989847715736 0
"2140" 53 23 3.57159183673469 8.0989847715736 0
"2141" 53 24 5.13416326530612 8.0989847715736 0
"2142" 53 25 5.13416326530612 8.0989847715736 0
"2143" 53 26 3.57159183673469 8.0989847715736 0
"2144" 53 27 3.57159183673469 8.0989847715736 0
"2145" 53 28 5.13416326530612 8.0989847715736 0
"2146" 53 29 5.13416326530612 8.0989847715736 0
"2147" 53 30 5.13416326530612 8.0989847715736 0
"2148" 53 31 3.57159183673469 8.0989847715736 0
"2149" 53 32 6.77114285714286 8.0989847715736 0
"2150" 53 33 6.77114285714286 8.0989847715736 0
"2151" 53 34 6.77114285714286 8.15482233502538 0
"2152" 53 35 6.77114285714286 7.48477157360406 0
"2153" 53 36 6.77114285714286 7.48477157360406 0
"2154" 53 37 8.18489795918367 6.87055837563452 1
"2155" 53 38 8.18489795918367 6.87055837563452 0
"2156" 53 39 6.69673469387755 7.48477157360406 0
"2157" 53 40 8.03608163265306 6.26634517766497 0
"2158" 53 41 8.25930612244898 6.31218274111675 0
"2159" 54 4 5.13416326530612 8.65736040609137 0
"2160" 54 5 3.646 8.54568527918782 0
"2161" 54 6 5.05975510204082 8.0989847715736 0
"2162" 54 7 6.62232653061224 8.15482233502538 0
"2163" 54 8 8.25930612244898 7.54060913705584 0
"2164" 54 9 6.84555102040816 8.0989847715736 0
"2165" 54 10 5.20857142857143 8.15482233502538 0
"2166" 54 11 5.13416326530612 8.15482233502538 0
"2167" 54 12 3.646 8.15482233502538 0
"2168" 54 13 5.05975510204082 8.15482233502538 0
"2169" 54 14 3.57159183673469 8.60152284263959 0
"2170" 54 15 3.57159183673469 8.60152284263959 0
"2171" 54 16 1.71138775510204 10.2766497461929 1
"2172" 54 17 3.57159183673469 9.27157360406091 0
"2173" 54 18 5.13416326530612 8.15482233502538 0
"2174" 54 19 5.13416326530612 8.15482233502538 0
"2175" 54 20 5.28297959183674 8.15482233502538 0
"2176" 54 21 3.49718367346939 8.15482233502538 0
"2177" 54 22 5.28297959183674 8.15482233502538 0
"2178" 54 23 3.57159183673469 8.15482233502538 0
"2179" 54 24 5.13416326530612 8.15482233502538 0
"2180" 54 25 5.13416326530612 8.15482233502538 0
"2181" 54 26 3.57159183673469 8.15482233502538 0
"2182" 54 27 3.57159183673469 8.15482233502538 1
"2183" 54 28 5.13416326530612 8.15482233502538 0
"2184" 54 29 5.13416326530612 8.15482233502538 0
"2185" 54 30 5.13416326530612 8.15482233502538 0
"2186" 54 31 5.20857142857143 8.15482233502538 1
"2187" 54 32 5.28297959183674 8.15482233502538 0
"2188" 54 33 6.77114285714286 8.15482233502538 0
"2189" 54 34 6.77114285714286 8.15482233502538 0
"2190" 54 35 6.77114285714286 7.48477157360406 0
"2191" 54 36 6.77114285714286 7.48477157360406 1
"2192" 54 37 6.77114285714286 7.48477157360406 0
"2193" 54 38 6.77114285714286 7.48477157360406 0
"2194" 54 39 6.77114285714286 7.48477157360406 0
"2195" 54 40 9.89628571428572 6.26634517766497 0
"2196" 55 4 3.57159183673469 8.21065989847716 0
"2197" 55 5 5.20857142857143 8.60152284263959 0
"2198" 55 6 6.77114285714286 8.15482233502538 0
"2199" 55 7 6.77114285714286 8.21065989847716 0
"2200" 55 8 8.18489795918367 7.48477157360406 1
"2201" 55 9 6.77114285714286 8.21065989847716 0
"2202" 55 10 6.84555102040816 8.21065989847716 0
"2203" 55 11 5.28297959183674 8.21065989847716 0
"2204" 55 12 5.05975510204082 8.21065989847716 0
"2205" 55 13 5.13416326530612 8.0989847715736 0
"2206" 55 14 3.57159183673469 8.60152284263959 0
"2207" 55 15 3.57159183673469 9.27157360406091 0
"2208" 55 16 1.78579591836735 10.2766497461929 1
```

```
"2209" 55 17 1.93461224489796 9.27157360406091 1
"2210" 55 18 5.13416326530612 8.60152284263959 0
"2211" 55 19 5.13416326530612 8.04314720812183 1
"2212" 55 20 6.69673469387755 8.21065989847716 1
"2213" 55 21 5.28297959183674 8.21065989847716 0
"2214" 55 22 5.13416326530612 8.21065989847716 0
"2215" 55 23 5.05975510204082 8.21065989847716 0
"2216" 55 24 5.13416326530612 8.21065989847716 0
"2217" 55 25 5.05975510204082 8.21065989847716 0
"2218" 55 26 3.57159183673469 8.21065989847716 0
"2219" 55 27 5.05975510204082 8.21065989847716 0
"2220" 55 28 5.13416326530612 8.21065989847716 0
"2221" 55 29 5.13416326530612 8.21065989847716 0
"2222" 55 30 5.13416326530612 8.21065989847716 1
"2223" 55 31 5.13416326530612 8.21065989847716 1
"2224" 55 32 5.13416326530612 8.21065989847716 0
"2225" 55 33 6.77114285714286 8.21065989847716 0
"2226" 55 34 6.77114285714286 7.48477157360406 0
"2227" 55 35 6.77114285714286 7.48477157360406 0
"2228" 55 36 6.77114285714286 8.15482233502538 0
"2229" 55 37 6.69673469387755 8.15482233502538 1
"2230" 55 38 5.35738775510204 8.0989847715736 0
"2231" 55 39 6.69673469387755 7.54060913705584 0
"2232" 55 40 8.18489795918367 6.92639593908629 0
"2233" 55 41 8.18489795918367 7.54060913705584 0
"2234" 56 4 5.13416326530612 8.65736040609137 0
"2235" 56 5 5.13416326530612 8.15482233502538 1
"2236" 56 6 6.69673469387755 8.15482233502538 1
"2237" 56 7 5.05975510204082 8.0989847715736 1
"2238" 56 8 6.77114285714286 8.0989847715736 0
"2239" 56 9 6.77114285714286 8.0989847715736 0
"2240" 56 10 5.05975510204082 8.0989847715736 0
"2241" 56 11 5.28297959183674 8.0989847715736 0
"2242" 56 12 5.13416326530612 8.15482233502538 1
"2243" 56 13 5.13416326530612 8.15482233502538 0
"2244" 56 14 3.57159183673469 8.60152284263959 0
"2245" 56 15 3.34836734693877 9.27157360406091 0
"2246" 56 16 1.78579591836735 10.2208121827411 0
"2247" 56 17 1.78579591836735 9.27157360406091 0
"2248" 56 18 5.13416326530612 8.0989847715736 0
"2249" 56 19 5.13416326530612 8.0989847715736 1
"2250" 56 20 6.77114285714286 8.0989847715736 0
"2251" 56 21 6.77114285714286 8.0989847715736 1
"2252" 56 22 5.05975510204082 8.0989847715736 0
"2253" 56 23 5.13416326530612 8.0989847715736 0
"2254" 56 24 5.13416326530612 8.0989847715736 0
"2255" 56 25 5.13416326530612 8.0989847715736 0
"2256" 56 26 5.13416326530612 8.0989847715736 0
"2257" 56 27 5.28297959183674 8.0989847715736 0
"2258" 56 28 5.13416326530612 8.0989847715736 0
"2259" 56 29 5.28297959183674 8.0989847715736 0
"2260" 56 30 5.13416326530612 8.0989847715736 0
"2261" 56 31 5.13416326530612 8.0989847715736 0
"2262" 56 32 5.13416326530612 8.0989847715736 0
"2263" 56 33 5.13416326530612 8.0989847715736 0
"2264" 56 34 6.77114285714286 7.48477157360406 0
"2265" 56 35 6.77114285714286 6.92639593908629 0
"2266" 56 36 6.77114285714286 8.04314720812183 0
"2267" 56 37 6.77114285714286 7.48477157360406 0
"2268" 56 38 6.77114285714286 8.15482233502538 0
"2269" 56 39 6.77114285714286 7.48477157360406 0
"2270" 56 40 8.18489795918367 6.92639593908629 0
"2271" 56 41 9.7474693877551 6.2005076142132 0
"2272" 57 4 5.13416326530612 8.0989847715736 0
"2273" 57 5 5.13416326530612 8.60152284263959 0
"2274" 57 6 5.13416326530612 8.21065989847716 1
"2275" 57 7 5.13416326530612 8.0989847715736 0
"2276" 57 8 5.13416326530612 8.0989847715736 0
"2277" 57 9 5.13416326530612 8.0989847715736 0
"2278" 57 10 5.13416326530612 8.0989847715736 1
"2279" 57 11 3.57159183673469 8.0989847715736 1
"2280" 57 12 5.13416326530612 8.54568527918782 1
"2281" 57 13 5.13416326530612 8.04314720812183 1
"2282" 57 14 5.13416326530612 8.04314720812183 1
"2283" 57 15 1.86020408163265 9.215736040609140
```

```
"2284" 57 16 1.78579591836735 10.2766497461929 0
"2285" 57 17 1.78579591836735 10.3324873096447 0
"2286" 57 18 5.13416326530612 8.60152284263959 0
"2287" 57 19 5.13416326530612 8.15482233502538 0
"2288" 57 20 6.77114285714286 8.0989847715736 1
"2289" 57 21 5.05975510204082 8.0989847715736 1
"2290" 57 22 5.13416326530612 8.0989847715736 0
"2291" 57 23 5.13416326530612 8.0989847715736 0
"2292" 57 24 5.13416326530612 8.0989847715736 0
"2293" 57 25 5.13416326530612 8.0989847715736 0
"2294" 57 26 5.13416326530612 8.0989847715736 0
"2295" 57 27 6.84555102040816 8.0989847715736 0
"2296" 57 28 6.77114285714286 8.0989847715736 0
"2297" 57 29 6.69673469387755 8.0989847715736 0
"2298" 57 30 5.13416326530612 8.0989847715736 0
"2299" 57 31 5.13416326530612 8.0989847715736 1
"2300" 57 32 5.13416326530612 8.0989847715736 0
"2301" 57 33 5.13416326530612 8.0989847715736 0
"2302" 57 34 6.77114285714286 7.48477157360406 0
"2303" 57 35 8.18489795918367 6.92639593908629 0
"2304" 57 36 6.84555102040816 7.48477157360406 0
"2305" 57 37 6.77114285714286 8.15482233502538 0
"2306" 57 38 6.77114285714286 8.15482233502538 0
"2307" 57 39 6.77114285714286 8.15482233502538 0
"2308" 57 40 8.11048979591837 7.48477157360406 0
"2309" 57 41 8.18489795918367 6.92639593908629 0
"2310" 57 42 8.18489795918367 6.98223350253807 0
"2311" 58 5 3.49718367346939 8.54568527918782 0
"2312" 58 6 3.57159183673469 8.60152284263959 0
"2313" 58 7 5.05975510204082 8.60152284263959 0
"2314" 58 8 5.13416326530612 8.15482233502538 1
"2315" 58 9 5.13416326530612 8.15482233502538 1
"2316" 58 10 5.13416326530612 8.15482233502538 1
"2317" 58 11 5.05975510204082 8.15482233502538 1
"2318" 58 12 5.13416326530612 8.15482233502538 1
"2319" 58 13 5.13416326530612 8.04314720812183 0
"2320" 58 14 5.13416326530612 8.60152284263959 0
"2321" 58 15 1.78579591836735 10.2208121827411 1
"2322" 58 16 1.78579591836735 10.2766497461929 1
"2323" 58 17 1.78579591836735 10.2766497461929 0
"2324" 58 18 3.57159183673469 8.60152284263959 0
"2325" 58 19 3.57159183673469 8.60152284263959 0
"2326" 58 20 3.57159183673469 8.60152284263959 0
"2327" 58 21 5.13416326530612 8.0989847715736 0
"2328" 58 22 5.13416326530612 8.15482233502538 0
"2329" 58 23 5.13416326530612 8.04314720812183 0
"2330" 58 24 5.13416326530612 8.15482233502538 0
"2331" 58 25 6.77114285714286 8.15482233502538 0
"2332" 58 26 6.77114285714286 8.15482233502538 0
"2333" 58 27 6.77114285714286 8.15482233502538 0
"2334" 58 28 6.77114285714286 8.15482233502538 0
"2335" 58 29 6.77114285714286 8.15482233502538 0
"2336" 58 30 5.13416326530612 8.15482233502538 0
"2337" 58 31 5.13416326530612 8.15482233502538 0
"2338" 58 32 5.13416326530612 8.15482233502538 0
"2339" 58 33 5.13416326530612 8.15482233502538 0
"2340" 58 34 6.77114285714286 7.48477157360406 0
"2341" 58 35 8.18489795918367 6.87055837563452 0
"2342" 58 36 6.84555102040816 8.0989847715736 0
"2343" 58 37 6.77114285714286 8.15482233502538 0
"2344" 58 38 5.13416326530612 8.15482233502538 1
"2345" 58 39 6.69673469387755 8.15482233502538 1
"2346" 58 40 6.77114285714286 8.15482233502538 1
"2347" 58 41 8.18489795918367 7.54060913705584 0
"2348" 58 42 8.11048979591837 6.75888324873097 0
"2349" 58 43 6.69673469387755 8.21065989847716 0
"2350" 58 44 8.11048979591837 0.393401015228429 0
"2351" 59 5 4.98534693877551 8.60152284263959 0
"2352" 59 6 3.57159183673469 8.65736040609137 0
"2353" 59 7 3.79481632653061 9.27157360406091 0
"2354" 59 8 5.20857142857143 8.60152284263959 0
"2355" 59 9 5.05975510204082 8.60152284263959 1
"2356" 59 10 5.13416326530612 8.15482233502538 0
"2357" 59 11 5.13416326530612 8.15482233502538 1
"2358" 59 12 5.13416326530612 8.15482233502538 0
```

```
"2359" 59 13 5.13416326530612 8.15482233502538 0
"2360" 59 14 5.13416326530612 8.60152284263959 1
"2361" 59 15 1.78579591836735 9.21573604060914 0
"2362" 59 16 1.78579591836735 10.2766497461929 1
"2363" 59 17 1.86020408163265 10.2766497461929 1
"2364" 59 18 3.646 8.60152284263959 0
"2365" 59 19 3.57159183673469 8.60152284263959 0
"2366" 59 20 3.57159183673469 8.60152284263959 0
"2367" 59 21 3.57159183673469 8.60152284263959 0
"2368" 59 22 3.646 8.60152284263959 0
"2369" 59 23 5.13416326530612 8.15482233502538 0
"2370" 59 24 5.13416326530612 8.15482233502538 0
"2371" 59 25 6.77114285714286 8.15482233502538 0
"2372" 59 26 6.77114285714286 8.15482233502538 0
"2373" 59 27 6.77114285714286 8.15482233502538 0
"2374" 59 28 6.77114285714286 8.15482233502538 0
"2375" 59 29 6.77114285714286 8.15482233502538 0
"2376" 59 30 5.28297959183674 8.15482233502538 0
"2377" 59 31 5.13416326530612 8.15482233502538 0
"2378" 59 32 5.13416326530612 8.15482233502538 0
"2379" 59 33 5.13416326530612 8.15482233502538 1
"2380" 59 34 6.69673469387755 8.15482233502538 0
"2381" 59 35 8.11048979591837 8.15482233502538 0
"2382" 59 36 6.84555102040816 8.15482233502538 0
"2383" 59 37 6.77114285714286 8.15482233502538 0
"2384" 59 38 6.69673469387755 8.15482233502538 1
"2385" 59 39 5.13416326530612 8.15482233502538 1
"2386" 59 40 6.77114285714286 8.15482233502538 0
"2387" 59 41 6.77114285714286 8.15482233502538 0
"2388" 59 42 6.77114285714286 8.15482233502538 0
"2389" 59 43 8.25930612244898 6.87055837563452 0
"2390" 59 44 9.82187755102041 5.58629441624366 0
"2391" 60 6 3.49718367346939 8.71319796954315 0
"2392" 60 7 5.28297959183674 9.27157360406091 0
"2393" 60 8 3.646 8.54568527918782 0
"2394" 60 9 5.28297959183674 8.15482233502538 0
"2395" 60 10 5.13416326530612 8.15482233502538 1
"2396" 60 11 5.20857142857143 8.0989847715736 1
"2397" 60 12 5.13416326530612 8.15482233502538 1
"2398" 60 13 5.13416326530612 8.54568527918782 1
"2399" 60 14 3.646 8.60152284263959 0
"2400" 60 15 1.78579591836735 9.21573604060914 1
"2401" 60 16 1.71138775510204 10.2766497461929 1
"2402" 60 17 3.57159183673469 9.32741116751269 0
"2403" 60 18 3.646 8.60152284263959 0
"2404" 60 19 3.57159183673469 8.60152284263959 0
"2405" 60 20 3.57159183673469 8.60152284263959 0
"2406" 60 21 3.72040816326531 8.60152284263959 0
"2407" 60 22 5.28297959183674 8.60152284263959 0
"2408" 60 23 5.13416326530612 8.15482233502538 0
"2409" 60 24 5.13416326530612 8.0989847715736 0
"2410" 60 25 6.77114285714286 8.0989847715736 0
"2411" 60 26 6.77114285714286 8.0989847715736 0
"2412" 60 27 6.77114285714286 8.0989847715736 0
"2413" 60 28 6.77114285714286 8.0989847715736 0
"2414" 60 29 6.77114285714286 8.0989847715736 0
"2415" 60 30 6.77114285714286 8.0989847715736 0
"2416" 60 31 6.77114285714286 8.0989847715736 0
"2417" 60 32 6.84555102040816 8.0989847715736 0
"2418" 60 33 6.77114285714286 8.0989847715736 0
"2419" 60 34 6.77114285714286 8.0989847715736 0
"2420" 60 35 6.77114285714286 8.0989847715736 0
"2421" 60 36 6.77114285714286 8.0989847715736 0
"2422" 60 37 6.77114285714286 8.0989847715736 0
"2423" 60 38 5.28297959183674 8.0989847715736 0
"2424" 60 39 6.77114285714286 8.15482233502538 0
"2425" 60 40 6.69673469387755 7.48477157360406 0
"2426" 60 41 6.77114285714286 7.48477157360406 0
"2427" 60 42 6.77114285714286 7.48477157360406 0
"2428" 60 43 8.25930612244898 6.36802030456853 0
"2429" 60 44 11.0124081632653 5.58629441624366 0
"2430" 60 45 12.2773469387755 5.69796954314721 0
"2431" 61 11 5.05975510204082 8.15482233502538 1
"2432" 61 12 5.05975510204082 8.15482233502538 1
"2433" 61 13 5.05975510204082 8.60152284263959 1
```

```
"2434" 61 14 3.646 9.32741116751269 1
"2435" 61 15 1.86020408163265 9.27157360406091 0
"2436" 61 16 1.71138775510204 10.3324873096447 0
"2437" 61 17 3.57159183673469 8.60152284263959 0
"2438" 61 18 3.49718367346939 8.60152284263959 0
"2439" 61 19 5.05975510204082 8.60152284263959 0
"2440" 61 20 3.57159183673469 8.60152284263959 0
"2441" 61 21 3.646 8.54568527918782 0
"2442" 61 22 6.84555102040816 8.0989847715736 0
"2443" 61 23 6.77114285714286 8.04314720812183 0
"2444" 61 24 6.77114285714286 8.0989847715736 0
"2445" 61 25 6.77114285714286 8.0989847715736 0
"2446" 61 26 6.77114285714286 8.0989847715736 0
"2447" 61 27 6.77114285714286 8.0989847715736 1
"2448" 61 28 6.77114285714286 8.0989847715736 0
"2449" 61 29 6.77114285714286 8.0989847715736 0
"2450" 61 30 6.69673469387755 8.0989847715736 0
"2451" 61 31 6.77114285714286 8.0989847715736 0
"2452" 61 32 8.11048979591837 8.0989847715736 0
"2453" 61 33 6.77114285714286 8.0989847715736 0
"2454" 61 34 6.77114285714286 8.0989847715736 0
"2455" 61 35 6.77114285714286 8.0989847715736 0
"2456" 61 36 6.77114285714286 8.0989847715736 0
"2457" 61 37 6.77114285714286 8.0989847715736 0
"2458" 61 38 5.13416326530612 8.0989847715736 0
"2459" 61 39 6.77114285714286 8.15482233502538 0
"2460" 61 40 6.84555102040816 7.48477157360406 0
"2461" 61 41 8.18489795918367 7.48477157360406 0
"2462" 61 42 8.25930612244898 6.31218274111675 0
"2463" 61 43 8.18489795918367 6.92639593908629 0
"2464" 61 44 9.6730612244898 6.25634517766497 0
"2465" 61 45 9.6730612244898 6.25634517766497 0
"2466" 61 46 11.9797142857143 4.86040609137056 0
"2467" 62 14 3.57159183673469 9.27157360406091 0
"2468" 62 15 1.86020408163265 9.27157360406091 0
"2469" 62 16 3.57159183673469 8.60152284263959 0
"2470" 62 17 3.57159183673469 8.65736040609137 0
"2471" 62 18 5.13416326530612 8.15482233502538 0
"2472" 62 19 3.49718367346939 8.54568527918782 0
"2473" 62 20 3.646 8.60152284263959 0
"2474" 62 21 3.57159183673469 8.60152284263959 0
"2475" 62 22 5.13416326530612 8.0989847715736 0
"2476" 62 23 6.77114285714286 8.0989847715736 0
"2477" 62 24 6.77114285714286 8.0989847715736 0
"2478" 62 25 6.84555102040816 8.0989847715736 0
"2479" 62 26 6.77114285714286 8.0989847715736 0
"2480" 62 27 8.18489795918367 8.0989847715736 0
"2481" 62 28 6.77114285714286 8.0989847715736 0
"2482" 62 29 6.77114285714286 8.0989847715736 0
"2483" 62 30 8.11048979591837 8.0989847715736 0
"2484" 62 31 6.77114285714286 8.0989847715736 0
"2485" 62 32 8.11048979591837 8.0989847715736 0
"2486" 62 33 6.77114285714286 8.0989847715736 0
"2487" 62 34 6.77114285714286 8.0989847715736 0
"2488" 62 35 6.77114285714286 8.0989847715736 0
"2489" 62 36 6.84555102040816 8.0989847715736 0
"2490" 62 37 6.84555102040816 8.0989847715736 0
"2491" 62 38 5.13416326530612 8.0989847715736 0
"2492" 62 39 5.13416326530612 8.0989847715736 0
"2493" 62 40 5.13416326530612 8.0989847715736 0
"2494" 62 41 5.13416326530612 8.15482233502538 0
"2495" 62 42 6.77114285714286 8.0989847715736 0
"2496" 62 43 9.7474693877551 6.2005076142132 0
"2497" 62 44 9.7474693877551 6.31218274111675 0
"2498" 62 45 9.7474693877551 6.25634517766497 0
"2499" 62 46 12.2029387755102 5.58629441624366 0
"2500" 62 47 13.3934693877551 5.58629441624366 0
"2501" 63 14 1.78579591836735 10.2766497461929 0
"2502" 63 15 3.7204081632653 9.21573604060914 0
"2503" 63 16 5.13416326530612 8.15482233502538 0
"2504" 63 17 6.77114285714286 8.0989847715736 0
"2505" 63 18 6.69673469387755 8.15482233502538 0
"2506" 63 19 6.77114285714286 8.15482233502538 0
"2507" 63 20 5.13416326530612 8.60152284263959 0
"2508" 63 21 3.646 8.60152284263959 0
```

```
"2509" 63 22 5.28297959183674 8.15482233502538 0
"2510" 63 23 5.13416326530612 8.15482233502538 0
"2511" 63 24 6.69673469387755 8.15482233502538 0
"2512" 63 25 6.77114285714286 8.15482233502538 0
"2513" 63 26 8.18489795918367 7.42893401015228 0
"2514" 63 27 9.59865306122449 7.42893401015228 0
"2515" 63 28 6.84555102040816 8.0989847715736 0
"2516" 63 29 5.28297959183674 8.15482233502538 0
"2517" 63 30 6.77114285714286 8.15482233502538 0
"2518" 63 31 6.77114285714286 8.15482233502538 0
"2519" 63 32 6.84555102040816 8.15482233502538 0
"2520" 63 33 6.77114285714286 8.15482233502538 0
"2521" 63 34 6.77114285714286 8.15482233502538 0
"2522" 63 35 6.77114285714286 8.15482233502538 0
"2523" 63 36 5.13416326530612 8.15482233502538 0
"2524" 63 37 5.13416326530612 8.15482233502538 0
"2525" 63 38 5.13416326530612 8.15482233502538 0
"2526" 63 39 5.13416326530612 8.15482233502538 0
"2527" 63 40 5.28297959183674 8.15482233502538 0
"2528" 63 41 5.13416326530612 8.15482233502538 0
"2529" 63 42 5.28297959183674 8.15482233502538 0
"2530" 63 43 6.69673469387755 8.0989847715736 0
"2531" 63 44 8.18489795918367 6.87055837563452 0
"2532" 63 45 9.7474693877551 6.31218274111675 0
"2533" 63 46 9.82187755102041 6.92639593908629 0
"2534" 63 47 11.0124081632653 6.2005076142132 0
"2535" 63 48 10.938 5.58629441624366 0
"2536" 64 13 1.78579591836735 9.21573604060914 0
"2537" 64 14 1.86020408163265 10.2766497461929 1
"2538" 64 15 5.05975510204082 8.60152284263959 0
"2539" 64 16 8.18489795918367 7.54060913705584 0
"2540" 64 17 9.7474693877551 7.48477157360406 0
"2541" 64 18 8.18489795918367 8.0989847715736 0
"2542" 64 19 6.84555102040816 8.0989847715736 0
"2543" 64 20 6.77114285714286 8.15482233502538 0
"2544" 64 21 5.13416326530612 8.60152284263959 0
"2545" 64 22 5.13416326530612 8.0989847715736 0
"2546" 64 23 5.13416326530612 8.0989847715736 0
"2547" 64 24 6.84555102040816 8.0989847715736 0
"2548" 64 25 8.11048979591837 7.48477157360406 0
"2549" 64 26 9.7474693877551 6.87055837563452 0
"2550" 64 27 8.25930612244898 7.42893401015228 0
"2551" 64 28 8.11048979591837 8.0989847715736 0
"2552" 64 29 6.84555102040816 8.0989847715736 0
"2553" 64 30 6.77114285714286 8.0989847715736 0
"2554" 64 31 6.77114285714286 8.0989847715736 0
"2555" 64 32 6.77114285714286 8.0989847715736 0
"2556" 64 33 5.13416326530612 8.0989847715736 0
"2557" 64 34 5.13416326530612 8.0989847715736 0
"2558" 64 35 5.13416326530612 8.0989847715736 0
"2559" 64 36 5.20857142857143 8.0989847715736 0
"2560" 64 37 6.77114285714286 8.0989847715736 0
"2561" 64 38 6.77114285714286 8.0989847715736 0
"2562" 64 39 6.77114285714286 8.0989847715736 0
"2563" 64 40 6.77114285714286 8.0989847715736 0
"2564" 64 41 5.13416326530612 8.0989847715736 0
"2565" 64 42 5.05975510204082 8.0989847715736 0
"2566" 64 43 5.13416326530612 8.15482233502538 0
"2567" 64 44 5.13416326530612 8.15482233502538 0
"2568" 64 45 6.77114285714286 7.48477157360406 0
"2569" 64 46 6.77114285714286 7.48477157360406 0
"2570" 64 47 8.25930612244898 7.42893401015228 0
"2571" 64 48 11.0124081632653 6.25634517766497 0
"2572" 64 49 11.0124081632653 6.2005076142132 0
"2573" 65 12 1.63697959183673 10.2766497461929 0
"2574" 65 13 1.78579591836735 9.27157360406091 1
"2575" 65 14 3.646 9.27157360406091 0
"2576" 65 15 9.82187755102041 6.81472081218274 0
"2577" 65 16 9.7474693877551 6.87055837563452 0
"2578" 65 17 9.59865306122449 6.92639593908629 0
"2579" 65 18 6.84555102040816 8.04314720812183 0
"2580" 65 19 6.77114285714286 8.04314720812183 0
"2581" 65 20 6.69673469387755 8.04314720812183 0
"2582" 65 21 5.13416326530612 8.04314720812183 0
"2583" 65 22 5.13416326530612 8.15482233502538 0
```

```
"2584" 65 23 6.84555102040816 7.98730964467005 0
"2585" 65 24 6.77114285714286 8.0989847715736 0
"2586" 65 25 9.7474693877551 7.48477157360406 0
"2587" 65 26 8.18489795918367 6.87055837563452 0
"2588" 65 27 8.11048979591837 7.37309644670051 0
"2589" 65 28 6.77114285714286 8.0989847715736 0
"2590" 65 29 6.69673469387755 8.04314720812183 0
"2591" 65 30 5.28297959183674 8.04314720812183 0
"2592" 65 31 5.13416326530612 8.04314720812183 0
"2593" 65 32 6.69673469387755 8.04314720812183 0
"2594" 65 33 5.13416326530612 8.04314720812183 0
"2595" 65 34 5.13416326530612 8.04314720812183 0
"2596" 65 35 5.13416326530612 8.04314720812183 0
"2597" 65 36 6.77114285714286 8.04314720812183 0
"2598" 65 37 6.77114285714286 8.04314720812183 0
"2599" 65 38 6.77114285714286 8.04314720812183 0
"2600" 65 39 6.77114285714286 8.04314720812183 0
"2601" 65 40 6.77114285714286 8.04314720812183 0
"2602" 65 41 6.69673469387755 8.04314720812183 0
"2603" 65 42 5.35738775510204 8.04314720812183 0
"2604" 65 43 5.13416326530612 8.04314720812183 0
"2605" 65 44 5.13416326530612 8.04314720812183 0
"2606" 65 45 5.13416326530612 8.15482233502538 0
"2607" 65 46 5.13416326530612 8.0989847715736 0
"2608" 65 47 6.84555102040816 7.48477157360406 0
"2609" 65 48 9.7474693877551 6.25634517766497 0
"2610" 65 49 9.82187755102041 0.337563451776651 0
"2611" 66 12 1.78579591836735 10.2766497461929 1
"2612" 66 13 1.78579591836735 9.32741116751269 0
"2613" 66 14 5.13416326530612 8.60152284263959 0
"2614" 66 15 9.7474693877551 6.92639593908629 0
"2615" 66 16 9.6730612244898 6.98223350253807 0
"2616" 66 17 8.18489795918367 7.48477157360406 0
"2617" 66 18 6.77114285714286 8.15482233502538 0
"2618" 66 19 6.77114285714286 8.15482233502538 0
"2619" 66 20 6.77114285714286 8.0989847715736 0
"2620" 66 21 5.13416326530612 8.0989847715736 0
"2621" 66 22 8.25930612244898 7.48477157360406 0
"2622" 66 23 9.7474693877551 6.87055837563452 0
"2623" 66 24 9.7474693877551 6.98223350253807 0
"2624" 66 25 9.7474693877551 8.15482233502538 0
"2625" 66 26 8.25930612244898 8.15482233502538 0
"2626" 66 27 6.69673469387755 8.15482233502538 0
"2627" 66 28 5.13416326530612 8.15482233502538 0
"2628" 66 29 6.77114285714286 8.15482233502538 0
"2629" 66 30 6.77114285714286 8.15482233502538 0
"2630" 66 31 6.77114285714286 8.15482233502538 0
"2631" 66 32 6.77114285714286 8.15482233502538 0
"2632" 66 33 6.77114285714286 8.15482233502538 0
"2633" 66 34 6.69673469387755 8.15482233502538 0
"2634" 66 35 6.69673469387755 8.15482233502538 0
"2635" 66 36 6.77114285714286 8.15482233502538 0
"2636" 66 37 6.77114285714286 8.15482233502538 0
"2637" 66 38 6.77114285714286 8.15482233502538 0
"2638" 66 39 6.77114285714286 8.15482233502538 0
"2639" 66 40 6.77114285714286 8.15482233502538 0
"2640" 66 41 6.77114285714286 8.15482233502538 0
"2641" 66 42 6.77114285714286 8.15482233502538 0
"2642" 66 43 5.20857142857143 8.15482233502538 0
"2643" 66 44 5.05975510204082 8.15482233502538 0
"2644" 66 45 5.13416326530612 8.15482233502538 0
"2645" 66 46 5.05975510204082 8.15482233502538 0
"2646" 66 47 6.77114285714286 8.15482233502538 0
"2647" 66 48 8.25930612244898 7.48477157360406 0
"2648" 67 12 1.71138775510204 10.2766497461929 1
"2649" 67 13 3.49718367346939 9.27157360406091 0
"2650" 67 14 6.84555102040816 8.0989847715736 0
"2651" 67 15 9.7474693877551 6.814720812182740
"2652" 67 16 9.6730612244898 7.42893401015228 0
"2653" 67 17 8.18489795918367 7.48477157360406 0
"2654" 67 18 6.77114285714286 8.15482233502538 0
"2655" 67 19 6.69673469387755 8.15482233502538 0
"2656" 67 20 6.77114285714286 8.21065989847716 0
"2657" 67 21 8.18489795918367 6.98223350253807 0
"2658" 67 22 9.52424489795918 6.92639593908629 0
```

```
"2659" 67 23 9.7474693877551 6.25634517766497 0
"2660" 67 24 9.6730612244898 7.48477157360406 0
"2661" 67 25 8.11048979591837 8.21065989847716 0
"2662" 67 26 8.18489795918367 8.15482233502538 0
"2663" 67 27 6.77114285714286 8.15482233502538 0
"2664" 67 28 6.77114285714286 8.15482233502538 0
"2665" 67 29 6.69673469387755 8.15482233502538 0
"2666" 67 30 6.69673469387755 8.15482233502538 0
"2667" 67 31 6.69673469387755 8.15482233502538 0
"2668" 67 32 6.69673469387755 8.15482233502538 0
"2669" 67 33 6.69673469387755 8.15482233502538 0
"2670" 67 34 8.25930612244898 8.15482233502538 0
"2671" 67 35 6.77114285714286 8.15482233502538 0
"2672" 67 36 6.77114285714286 8.15482233502538 0
"2673" 67 37 6.77114285714286 8.15482233502538 0
"2674" 67 38 6.77114285714286 8.15482233502538 0
"2675" 67 39 6.77114285714286 8.15482233502538 0
"2676" 67 40 6.77114285714286 8.15482233502538 0
"2677" 67 41 6.77114285714286 8.15482233502538 0
"2678" 67 42 6.77114285714286 8.15482233502538 0
"2679" 67 43 6.84555102040816 8.15482233502538 0
"2680" 67 44 6.69673469387755 8.15482233502538 0
"2681" 67 45 5.13416326530612 8.15482233502538 0
"2682" 67 46 5.28297959183674 8.26649746192893 0
"2683" 68 12 3.57159183673469 10.5 1
"2684" 68 13 5.13416326530612 8.60152284263959 0
"2685" 68 14 6.77114285714286 8.15482233502538 0
"2686" 68 15 8.25930612244898 7.48477157360406 0
"2687" 68 16 8.25930612244898 7.48477157360406 0
"2688" 68 17 8.18489795918367 7.48477157360406 0
"2689" 68 18 8.18489795918367 8.0989847715736 0
"2690" 68 19 8.33371428571429 8.0989847715736 0
"2691" 68 20 9.7474693877551 6.98223350253807 0
"2692" 68 21 9.82187755102041 6.25634517766497 0
"2693" 68 22 9.7474693877551 6.2005076142132 0
"2694" 68 23 9.7474693877551 6.92639593908629 0
"2695" 68 24 8.25930612244898 8.04314720812183 0
"2696" 68 25 8.18489795918367 8.04314720812183 0
"2697" 68 26 8.18489795918367 8.04314720812183 0
"2698" 68 27 8.11048979591837 8.04314720812183 0
"2699" 68 28 8.18489795918367 8.04314720812183 0
"2700" 68 29 8.18489795918367 8.04314720812183 1
"2701" 68 30 8.18489795918367 8.04314720812183 0
"2702" 68 31 8.18489795918367 8.04314720812183 0
"2703" 68 32 8.18489795918367 8.04314720812183 0
"2704" 68 33 8.18489795918367 8.04314720812183 0
"2705" 68 34 8.25930612244898 8.04314720812183 0
"2706" 68 35 6.77114285714286 8.04314720812183 0
"2707" 68 36 6.69673469387755 8.04314720812183 0
"2708" 68 37 6.84555102040816 8.04314720812183 0
"2709" 68 38 6.77114285714286 8.04314720812183 0
"2710" 68 39 6.77114285714286 8.04314720812183 0
"2711" 68 40 6.77114285714286 8.04314720812183 0
"2712" 68 41 6.77114285714286 8.04314720812183 0
"2713" 68 42 6.77114285714286 8.04314720812183 0
"2714" 68 43 6.77114285714286 8.04314720812183 0
"2715" 68 44 6.69673469387755 8.04314720812183 0
"2716" 68 45 5.05975510204082 8.04314720812183 0
"2717" 68 46 5.13416326530612 8.04314720812183 0
"2718" 69 11 3.57159183673469 10.3324873096447 0
"2719" 69 12 3.57159183673469 9.27157360406091 1
"2720" 69 13 5.13416326530612 8.60152284263959 0
"2721" 69 14 6.99436734693878 8.0989847715736 0
"2722" 69 15 8.11048979591837 7.42893401015228 0
"2723" 69 16 9.7474693877551 7.54060913705584 0
"2724" 69 17 9.7474693877551 7.48477157360406 0
"2725" 69 18 9.7474693877551 7.48477157360406 0
"2726" 69 19 9.7474693877551 6.92639593908629 0
"2727" 69 20 10.938 6.25634517766497 0
"2728" 69 21 9.7474693877551 6.14467005076142 0
"2729" 69 22 9.82187755102041 6.87055837563452 0
"2730" 69 23 8.18489795918367 8.0989847715736 0
"2731" 69 24 8.18489795918367 8.0989847715736 0
"2732" 69 25 8.18489795918367 8.0989847715736 0
"2733" 69 26 8.18489795918367 8.0989847715736 0
```

```
"2734" 69 27 8.18489795918367 8.0989847715736 0
"2735" 69 28 8.18489795918367 8.0989847715736 0
"2736" 69 29 8.18489795918367 8.0989847715736 0
"2737" 69 30 8.18489795918367 8.0989847715736 0
"2738" 69 31 8.18489795918367 8.0989847715736 0
"2739" 69 32 8.18489795918367 8.0989847715736 0
"2740" 69 33 8.18489795918367 8.0989847715736 0
"2741" 69 34 8.18489795918367 8.0989847715736 0
"2742" 69 35 8.18489795918367 8.0989847715736 0
"2743" 69 36 8.18489795918367 8.0989847715736 0
"2744" 69 37 8.25930612244898 8.0989847715736 0
"2745" 69 38 6.69673469387755 8.0989847715736 0
"2746" 69 39 6.69673469387755 8.0989847715736 0
"2747" 69 40 6.77114285714286 8.0989847715736 0
"2748" 69 41 5.05975510204082 8.0989847715736 0
"2749" 69 42 6.77114285714286 8.0989847715736 1
"2750" 69 43 6.77114285714286 7.98730964467005 0
"2751" 69 44 5.28297959183674 8.26649746192893 0
"2752" 69 45 4.98534693877551 8.04314720812183 0
"2753" 70 11 3.57159183673469 9.32741116751269 0
"2754" 70 12 3.57159183673469 9.21573604060914 1
"2755" 70 13 5.13416326530612 8.54568527918782 0
"2756" 70 14 9.7474693877551 7.42893401015228 0
"2757" 70 15 10.7891836734694 6.31218274111675 0
"2758" 70 16 9.7474693877551 6.92639593908629 0
"2759" 70 17 9.7474693877551 7.48477157360406 0
"2760" 70 18 10.938 6.92639593908629 0
"2761" 70 19 10.8635918367347 6.31218274111675 0
"2762" 70 20 9.7474693877551 6.256345177766497 0
"2763" 70 21 9.7474693877551 6.92639593908629 0
"2764" 70 22 8.18489795918367 7.42893401015228 0
"2765" 70 23 8.18489795918367 8.21065989847716 0
"2766" 70 24 8.18489795918367 8.15482233502538 0
"2767" 70 25 8.18489795918367 8.15482233502538 0
"2768" 70 26 8.18489795918367 8.15482233502538 1
"2769" 70 27 8.18489795918367 8.15482233502538 0
"2770" 70 28 8.18489795918367 8.15482233502538 0
"2771" 70 29 8.18489795918367 8.15482233502538 0
"2772" 70 30 8.18489795918367 8.15482233502538 0
"2773" 70 31 8.18489795918367 8.15482233502538 0
"2774" 70 32 8.18489795918367 8.15482233502538 0
"2775" 70 33 8.18489795918367 8.15482233502538 0
"2776" 70 34 8.18489795918367 8.15482233502538 0
"2777" 70 35 8.18489795918367 8.15482233502538 0
"2778" 70 36 8.18489795918367 8.15482233502538 0
"2779" 70 37 8.18489795918367 8.15482233502538 0
"2780" 70 38 8.18489795918367 8.15482233502538 1
"2781" 70 39 6.69673469387755 8.15482233502538 0
"2782" 70 40 6.77114285714286 8.15482233502538 0
"2783" 70 41 6.69673469387755 8.15482233502538 0
"2784" 70 42 6.77114285714286 8.15482233502538 0
"2785" 70 43 6.77114285714286 8.21065989847716 0
"2786" 71 11 3.57159183673469 8.60152284263959 0
"2787" 71 12 3.49718367346939 9.32741116751269 0
"2788" 71 13 5.13416326530612 8.60152284263959 0
"2789" 71 14 10.8635918367347 6.87055837563452 0
"2790" 71 15 12.0541224489796 6.31218274111675 0
"2791" 71 16 9.7474693877551 6.256345177766497 0
"2792" 71 17 9.7474693877551 6.92639593908629 0
"2793" 71 18 10.938 6.81472081218274 0
"2794" 71 19 9.6730612244898 6.92639593908629 0
"2795" 71 20 9.7474693877551 6.98223350253807 0
"2796" 71 21 9.6730612244898 7.54060913705584 0
"2797" 71 22 8.18489795918367 8.15482233502538 0
"2798" 71 23 8.11048979591837 8.15482233502538 0
"2799" 71 24 8.18489795918367 8.21065989847716 0
"2800" 71 25 9.89628571428572 8.21065989847716 0
"2801" 71 26 8.25930612244898 8.21065989847716 0
"2802" 71 27 9.7474693877551 8.21065989847716 0
"2803" 71 28 9.7474693877551 8.21065989847716 0
"2804" 71 29 9.7474693877551 8.15482233502538 0
"2805" 71 30 9.7474693877551 8.15482233502538 0
"2806" 71 31 9.7474693877551 8.15482233502538 0
"2807" 71 32 8.18489795918367 8.21065989847716 0
"2808" 71 33 8.18489795918367 8.21065989847716 1
```

```
"2809" 71 34 8.18489795918367 8.21065989847716 0
"2810" 71 35 8.18489795918367 8.21065989847716 0
"2811" 71 36 8.25930612244898 8.21065989847716 0
"2812" 71 37 8.11048979591837 8.21065989847716 0
"2813" 71 38 6.84555102040816 8.21065989847716 0
"2814" 71 39 6.84555102040816 8.21065989847716 0
"2815" 71 40 8.11048979591837 8.21065989847716 0
"2816" 71 41 8.11048979591837 8.21065989847716 0
"2817" 71 42 6.77114285714286 8.21065989847716 0
"2818" 71 43 6.77114285714286 8.0989847715736 0
"2819" 72 11 3.57159183673469 10.3324873096447 0
"2820" 72 12 5.05975510204082 9.27157360406091 1
"2821" 72 13 8.25930612244898 7.48477157360406 0
"2822" 72 14 10.938 6.25634517766497 0
"2823" 72 15 12.0541224489796 5.47461928934011 0
"2824" 72 16 9.7474693877551 6.31218274111675 0
"2825" 72 17 9.7474693877551 6.87055837563452 0
"2826" 72 18 9.7474693877551 6.81472081218274 0
"2827" 72 19 9.7474693877551 6.92639593908629 0
"2828" 72 20 9.67306122448980 8.04314720812183 0
"2829" 72 21 9.82187755102041 6.98223350253807 0
"2830" 72 22 9.7474693877551 8.04314720812183 1
"2831" 72 23 8.18489795918367 8.15482233502538 1
"2832" 72 24 8.25930612244898 8.0989847715736 1
"2833" 72 25 9.7474693877551 8.0989847715736 0
"2834" 72 26 9.7474693877551 8.15482233502538 0
"2835" 72 27 9.7474693877551 8.04314720812183 0
"2836" 72 28 9.7474693877551 7.48477157360406 0
"2837" 72 29 9.7474693877551 6.92639593908629 0
"2838" 72 30 9.7474693877551 7.48477157360406 0
"2839" 72 31 9.7474693877551 8.0989847715736 0
"2840" 72 32 8.18489795918367 8.0989847715736 0
"2841" 72 33 9.7474693877551 8.0989847715736 0
"2842" 72 34 9.7474693877551 8.0989847715736 0
"2843" 72 35 9.7474693877551 7.98730964467005 0
"2844" 72 36 9.7474693877551 8.21065989847716 0
"2845" 72 37 8.18489795918367 8.0989847715736 0
"2846" 72 38 6.77114285714286 8.0989847715736 0
"2847" 72 39 8.18489795918367 8.0989847715736 0
"2848" 72 40 8.18489795918367 8.0989847715736 1
"2849" 72 41 8.25930612244898 8.0989847715736 0
"2850" 72 42 6.77114285714286 8.0989847715736 0
"2851" 72 43 6.77114285714286 8.0989847715736 0
"2852" 73 11 3.57159183673469 9.27157360406091 0
"2853" 73 12 8.18489795918367 8.0989847715736 0
"2854" 73 13 12.1285306122449 5.58629441624366 0
"2855" 73 14 13.3934693877551 4.91624365482233 0
"2856" 73 15 10.938 6.25634517766497 0
"2857" 73 16 9.7474693877551 7.42893401015228 0
"2858" 73 17 9.7474693877551 6.75888324873097 0
"2859" 73 18 8.18489795918367 7.48477157360406 0
"2860" 73 19 8.18489795918367 8.04314720812183 0
"2861" 73 20 8.18489795918367 7.98730964467005 0
"2862" 73 21 9.7474693877551 7.48477157360406 0
"2863" 73 22 9.7474693877551 7.59644670050761 0
"2864" 73 23 8.25930612244898 8.0989847715736 1
"2865" 73 24 9.7474693877551 8.0989847715736 0
"2866" 73 25 9.7474693877551 8.0989847715736 0
"2867" 73 26 9.7474693877551 8.0989847715736 0
"2868" 73 27 9.7474693877551 7.48477157360406 0
"2869" 73 28 9.7474693877551 6.2005076142132 0
"2870" 73 29 9.7474693877551 6.92639593908629 0
"2871" 73 30 9.7474693877551 6.87055837563452 0
"2872" 73 31 9.7474693877551 8.0989847715736 0
"2873" 73 32 8.18489795918367 8.0989847715736 0
"2874" 73 33 9.7474693877551 8.0989847715736 0
"2875" 73 34 9.7474693877551 8.15482233502538 0
"2876" 73 35 9.7474693877551 6.87055837563452 1
"2877" 73 36 9.7474693877551 7.42893401015228 0
"2878" 73 37 8.18489795918367 8.04314720812183 0
"2879" 73 38 6.77114285714286 8.0989847715736 1
"2880" 73 39 8.18489795918367 8.0989847715736 1
"2881" 73 40 8.18489795918367 8.0989847715736 1
"2882" 73 41 8.18489795918367 8.0989847715736 0
"2883" 73 42 9.7474693877551 7.48477157360406 0
```

```
"2884" 73 43 6.69673469387755 8.04314720812183 0
"2885" 73 44 8.25930612244898 0.393401015228429 0
"2886" 74 11 5.13416326530612 8.60152284263959 0
"2887" 74 12 9.6730612244898 7.54060913705584 0
"2888" 74 13 10.938 6.25634517766497 0
"2889" 74 14 12.1285306122449 5.64213197969543 0
"2890" 74 15 10.938 6.81472081218274 0
"2891" 74 16 8.18489795918367 8.0989847715736 0
"2892" 74 17 8.18489795918367 7.42893401015228 0
"2893" 74 18 6.77114285714286 8.15482233502538 0
"2894" 74 19 6.77114285714286 8.15482233502538 1
"2895" 74 20 6.77114285714286 8.15482233502538 1
"2896" 74 21 6.84555102040816 8.15482233502538 1
"2897" 74 22 6.77114285714286 8.21065989847716 0
"2898" 74 23 6.99436734693878 8.21065989847716 0
"2899" 74 24 9.6730612244898 8.21065989847716 0
"2900" 74 25 9.7474693877551 8.0989847715736 0
"2901" 74 26 10.8635918367347 6.98223350253807 0
"2902" 74 27 10.938 6.92639593908629 0
"2903" 74 28 9.7474693877551 6.92639593908629 0
"2904" 74 29 9.82187755102041 6.98223350253807 0
"2905" 74 30 9.7474693877551 6.98223350253807 0
"2906" 74 31 9.7474693877551 7.48477157360406 0
"2907" 74 32 9.7474693877551 8.21065989847716 0
"2908" 74 33 9.6730612244898 8.15482233502538 0
"2909" 74 34 9.7474693877551 7.48477157360406 1
"2910" 74 35 10.8635918367347 6.98223350253807 1
"2911" 74 36 10.938 6.98223350253807 0
"2912" 74 37 11.0124081632653 6.98223350253807 0
"2913" 74 38 9.82187755102041 7.54060913705584 0
"2914" 74 39 12.0541224489796 6.31218274111675 0
"2915" 74 40 11.0124081632653 6.2005076142132 1
"2916" 74 41 12.2029387755102 5.58629441624366 0
"2917" 74 42 12.1285306122449 6.25634517766497 0
"2918" 74 43 9.7474693877551 6.81472081218274 0
"2919" 74 44 8.25930612244898 6.98223350253807 0
"2920" 75 11 5.28297959183674 8.60152284263959 0
"2921" 75 12 6.84555102040816 8.21065989847716 0
"2922" 75 13 8.18489795918367 7.48477157360406 0
"2923" 75 14 9.6730612244898 7.48477157360406 0
"2924" 75 15 6.77114285714286 8.0989847715736 0
"2925" 75 16 6.77114285714286 8.0989847715736 0
"2926" 75 17 6.77114285714286 8.0989847715736 0
"2927" 75 18 6.84555102040816 8.0989847715736 1
"2928" 75 19 6.77114285714286 8.0989847715736 0
"2929" 75 20 6.77114285714286 8.60152284263959 0
"2930" 75 21 6.69673469387755 8.60152284263959 0
"2931" 75 22 5.13416326530612 8.21065989847716 0
"2932" 75 23 6.77114285714286 8.15482233502538 1
"2933" 75 24 8.11048979591837 8.15482233502538 1
"2934" 75 25 9.7474693877551 8.0989847715736 0
"2935" 75 26 11.0124081632653 6.31218274111675 1
"2936" 75 27 10.8635918367347 6.98223350253807 0
"2937" 75 28 10.938 6.92639593908629 0
"2938" 75 29 10.8635918367347 6.87055837563452 0
"2939" 75 30 9.7474693877551 6.25634517766497 0
"2940" 75 31 12.1285306122449 6.31218274111675 0
"2941" 75 32 10.938 6.98223350253807 0
"2942" 75 33 11.0124081632653 7.48477157360406 1
"2943" 75 34 12.0541224489796 6.25634517766497 0
"2944" 75 35 12.2029387755102 6.25634517766497 0
"2945" 75 36 13.3934693877551 5.64213197969543 0
"2946" 75 37 14.7328163265306 5.08375634517767 0
"2947" 75 38 12.2029387755102 6.2005076142132 0
"2948" 75 39 10.8635918367347 7.42893401015228 0
"2949" 75 40 12.1285306122449 5.58629441624366 0
"2950" 75 41 14.584 4.86040609137056 0
"2951" 75 42 13.3934693877551 4.97208121827411 0
"2952" 75 43 12.2773469387755 5.53045685279188 0
"2953" 75 44 11.0124081632653 5.64213197969543 0
"2954" 76 11 5.13416326530612 8.60152284263959 0
"2955" 76 12 5.20857142857143 8.60152284263959 0
"2956" 76 13 6.77114285714286 8.60152284263959 0
"2957" 76 14 6.84555102040816 8.60152284263959 0
"2958" 76 15 6.54791836734694 8.60152284263959 0
```

```
"2959" 76 16 6.77114285714286 7.98730964467005 0
"2960" 76 17 6.77114285714286 0.337563451776651 0
"2961" 76 23 5.13416326530612 8.04314720812183 1
"2962" 76 24 8.25930612244898 8.0989847715736 1
"2963" 76 25 9.82187755102041 6.98223350253807 0
"2964" 76 26 12.0541224489796 5.58629441624366 0
"2965" 76 27 12.2029387755102 6.25634517766497 0
"2966" 76 28 14.4351836734694 5.64213197969543 0
"2967" 76 29 14.3607755102041 6.25634517766497 0
"2968" 76 30 13.3934693877551 5.58629441624366 0
"2969" 76 31 15.4768979591837 4.19035532994924 0
"2970" 76 32 13.3934693877551 5.58629441624366 0
"2971" 76 33 13.3934693877551 5.64213197969543 0
"2972" 76 34 13.3934693877551 5.53045685279188 0
"2973" 76 35 14.2863673469388 5.58629441624366 0
"2974" 76 36 14.4351836734694 4.13451776649746 0
"2975" 76 37 13.3934693877551 5.53045685279188 0
"2976" 76 42 14.3607755102041 3.46446700507614 0
"2977" 76 43 16.3697959183673 3.52030456852792 0
"2978" 76 44 16.295387755102 1.78934010152284 0
"2979" 77 25 10.8635918367347 4.02284263959391 0
"2980" 77 26 14.584 4.13451776649746 0
"2981" 77 27 13.4678775510204 4.97208121827411 0
"2982" 77 28 16.3697959183673 3.52030456852792 0
"2983" 77 31 15.4024897959184 4.13451776649746 0
"2984" 77 32 15.4768979591837 4.19035532994924 0
"2985" 77 33 15.4024897959184 4.91624365482233 0
"2986" 77 34 17.7835510204082 3.40862944162437 0
"2987" 77 43 16.3697959183673 0.337563451776651 0
"2988" 77 44 17.7835510204082 1.17512690355533 0
"2989" 78 26 16.4442040816327 2.62690355329949 0
"2990" 78 27 15.3280816326531 3.52030456852792 0
"2991" 78 28 17.6347346938776 1.90101522842640
"2992" 78 31 16.3697959183673 -0.5 0
"2993" 78 32 15.4024897959184 -0.5 0
"2994" 78 33 16.3697959183673 0.281725888324875 0
"2995" 79 27 18.23 1.956852791878170
```